\pdfoutput=1
\documentclass[epj,nopacs]{svjour}
\usepackage{graphics}
\usepackage{microtype}
\usepackage[hyperindex=true,colorlinks=true]{hyperref}
\hypersetup{citecolor=[rgb]{0.125,0.679,0.196}} 
\usepackage{url}
\usepackage{doi}
\usepackage{slashed}
\usepackage{physics}
\usepackage{feynmp-auto}
\usepackage{multirow}
\usepackage{enumitem}
\usepackage{amsmath}
\usepackage{amsfonts}
\usepackage{mathrsfs}

\usepackage{amssymb}
\usepackage{epsfig}
\usepackage{comment}
\usepackage{subfigure}
\usepackage{booktabs}

\usepackage[numbers]{natbib}

\renewcommand{\arraystretch}{1.2}

\def\b{\beta}

\def\tt{\hat \tau}

\def\lnl{$NN$+$3N\text{(lnl)}$}
\def\sat{NNLO$_{\text{sat}}$}

\allowdisplaybreaks

\begin{document}
\title{Moving away from singly-magic nuclei with Gorkov Green's function theory}

\author{V. Som\`a\inst{1}%
        \and C. Barbieri\inst{2}\fnmsep\inst{3}\fnmsep\inst{4}                     %
        \and T. Duguet\inst{1}\fnmsep\inst{5}
        \and P. Navr\'atil\inst{6}}                     %
%
%
\institute{IRFU, CEA, Universit\'e Paris-Saclay, 91191 Gif-sur-Yvette, France
          \and Department of Physics, University of Surrey, Guildford GU2 7XH, United Kingdom
          \and Dipartimento di Fisica, Universit\`a degli Studi di Milano, Via Celoria 16, 20133 Milano, Italy
          \and INFN, Sezione di Milano, Via Celoria 16, 20133 Milano, Italy
          \and KU Leuven, Institut voor Kern- en Stralingsfysica, 3001 Leuven, Belgium
          \and TRIUMF, 4004 Westbrook Mall, Vancouver, BC, V6T 2A3, Canada}
\date{}
%
\abstract{
\textit{Ab initio} calculations of bulk nuclear properties (ground-state energies, root-mean-square charge radii and charge density distributions) are presented for seven complete isotopic chains around calcium, from argon to chromium.
Calculations are performed within the Gorkov self-consistent Green's function approach at second order and make use of two state-of-the-art two- plus three-nucleon Hamiltonians, \lnl{} and \sat.
An overall good agreement with available experimental data is found, in particular for differential energies (charge radii) when the former (latter) interaction is employed.
Remarkably, neutron magic numbers $N=28,32,34$ emerge and evolve following experimental trends.
In contrast, pairing gaps are systematically underestimated.
General features of the isotopic dependence of charge radii are also reproduced, as well as charge density distributions.
A deterioration of the theoretical description is observed for certain nuclei and ascribed to the inefficient account of (static) quadrupole correlation in the present many-body truncation scheme.
In order to resolve these limitations, we advocate the extension of the formalism towards incorporating breaking of rotational symmetry or, alternatively, performing a stochastic sampling of the self-energy.
}

\maketitle

\section{Introduction}
\label{sec_intro}

A leap forward in ab initio calculations of atomic nuclei occurred about 15 years ago with the (re)introduction, in nuclear structure theory, of so-called \textit{correlation expansion} methods~\cite{Dickhoff04, Kowalski04}.
As opposed to virtually exact approaches, which do not impose any formal approximation on the solution of the many-body Schr{\"o}dinger equation and scale exponentially or factorially with the system size, correlation expansion techniques achieve a polynomial scaling at the price of an approximate, yet controlled and systematically improvable, solution.
Combined with the availability of ``softer'' Hamiltonians, obtained via similarity renormalisation group (SRG) transformations~\cite{Bogner10}, such a favourable scaling progressively enabled the extension of first-principle calculations beyond the region of light nuclei traditionally targeted by ab initio practitioners.
Nowadays, systems up to mass number $A \sim 70$ can be routinely accessed~\cite{Binder14, Hergert14, Hagen16, Taniuchi19, Soma20a}, with a few attempts reaching out to neutron-deficient tin ($A \sim 100$)~\cite{Morris18, Gysbers19} or even neutron-rich tin and xenon ($A \sim 140$)~\cite{Arthuis20} nuclei.

Many-body theories accessing mid-mass nuclei standardly expand the exact ground-state wave function with respect to a reference Slater determinant and can thus efficiently access doubly closed-shell systems dominated by dynamical, i.e. weak, correlations. 
However, the restriction to a single symmetry-restricted reference product state is too limiting to generate a meaningful expansion in open-shell nuclei due to the associated degeneracy with respect to elementary particle-hole excitations. 
The use of more general reference states must be contemplated to lift the degeneracy and tackle, from the outset, strong static correlations characterising open-shell systems.
To this purpose, three different strategies have been explored so far.
The first option relies on reference states mixing a set of appropriately chosen Slater determinants~\cite{Rolik03, Surjan04}. 
Those multi-configurational reference states can, for example, be obtained from a prior no-core shell model calculation~\cite{Barrett13} in a small basis or under the form of a particle-number-projected Hartree-Fock-Bogolyubov (HFB) state. 
Such reference states have been successfully employed in the multi-reference extension of the in-medium similarity renormalisation group method~\cite{Hergert14, Gebrerufael17, Hergert17} or within a perturbative framework yielding multi-configurational perturbation theory ~\cite{Tichai18b}. 
The second possibility consists in using a doubly closed-shell nucleus as a core, and computing the valence-space interaction at play on top of it through a polynomially-scaling method. 
Subsequently, a factorially-scaling diagonalisation is performed to solve the Schr{\"o}dinger equation within the valence space to a high degree of precision~\cite{Bogner14, Jansen14, Stroberg19a}. 
While this method benefits from the maturity of the shell-model technology, its hybrid numerical scaling limits its applicability to nuclei traditionally accessible by the shell model, i.e. $A \lesssim 100$.

A third route, followed here, relies on the use of a reference product state breaking one or several symmetries of the underlying Hamiltonian.
In doing so, one can trade the degeneracy with respect to particle-hole excitations characterising open-shell systems for a degeneracy with respect to transformations of the associated symmetry group.
As a result, the particle-hole degeneracy is lifted and a well-defined many-body expansion on top of a ``deformed'' reference product state can be designed.
This trade-off allows one to access open-shell systems while maintaining a polynomial cost and the intrinsic simplicity of single-reference expansion methods. The handling of the pseudo Goldstone mode associated with the manifold of degenerate states, necessary to restore the broken symmetry, can be safely postponed to a later stage~\cite{Duguet14, Duguet17b, Qiu18}.

Largely employed in the context of nuclear energy density functional~\cite{Bender03}, this approach was imported in ab initio nuclear structure about a decade ago.
First, Gorkov self-consistent Green's function (GSCGF) theory was developed~\cite{Soma11}. 
Few years later, coupled cluster theory was extended to the use of a Bogolyubov reference state~\cite{Signoracci15}.
More recently, Bogolyubov many-body perturbation theory (BMBPT) was introduced as a generalisation of standard M{\o}ller-Plesset theory~\cite{Tichai18a, Tichai20}.
All these techniques rely solely on the breaking of the U(1) symmetry related to particle-number conservation and are thus designed to efficiently account for static pairing correlations.
In order to deal with the other source of strong correlations in nuclei, i.e. the quadrupole correlations typically associated with nuclear deformation, one would need to correspondingly break rotational SU(2) symmetry.
Although work in this direction is in progress (see e.g.~\cite{Yao20, Novario20, Hergert20, Frosini21}), the latter feature is currently unavailable in nearly all state-of-the-art implementations.
As a consequence, the above methods are preferentially applied to singly open-shell (i.e., semi-magic) nuclei, where the role of quadrupole correlations is not predominant.
Indeed, GSCGF and BMBPT have successfully addressed complete semi-magic isotopic chains, e.g. oxygen, calcium or nickel~\cite{Soma13, Soma14b, Lapoux16, Tichai18a, Soma20a, Tichai20, Soma20b}.
The limits of applicability of U(1)-breaking, SU(2)-conserving correlation expansion methods, however, have never been systematically probed.
Therefore, it is worthwhile to push such calculations away from semi-magic nuclei in order to empirically  identify if and where such a strategy eventually fails, i.e. the point beyond which an explicit breaking of SU(2) symmetry will become mandatory.

Recently, specific medium-mass doubly open-shell systems, e.g. some titanium~\cite{Leistenschneider18} or sulfur and argon~\cite{Barbieri19, Mougeot20} isotopes, have been computed within the GSCGF approach.
In this paper, we extend these works to a systematic study of several isotopic chains around semi-magic calcium for which results were not available before.
In particular, we compute ground-state energies, charge radii and selected charge density distributions for chains ranging from argon ($Z=18$) to chromium ($Z=24$) and compare to available experimental data.
Calculations were performed using the recently introduced \lnl{} Hamiltonian~\cite{Soma20a}.
For charge radii and densities, additionally, the \sat{}~\cite{Ekstrom15} Hamiltonian was employed.
Overall, the goals of the present study can be summarised as follows:
\begin{enumerate}
\item \textit{Assess the performance of state-of-the-art ab initio calculations on bulk properties of medium-mass nuclei.}\\
In this respect, the present work follows up on the results of Ref.~\cite{Soma20a}, in which the novel \lnl{} interaction was benchmarked on semi-magic oxygen, calcium and nickel isotopes.
Here it is shown that the global satisfactory agreement with experimental data found in Ref.~\cite{Soma20a} extends to doubly open-shell isotopes around calcium.
Remarkably, neutron magic numbers $N=28,32,34$ emerge and evolve following experimental trends.
While the neutron dripline is not addressed here, the proton dripline is found at or near the experimental one.
As already remarked in Ref.~\cite{Soma20a} for calcium, charge radii computed with \lnl{} are too small compared to the experimental values.
In contrast, \sat{} provides a good overall description of existing data.
Nevertheless, even our best calculations fail to reproduce some finer details, e.g. the steep rise between $N=28$ and $N=32$ and the parabolic-like behaviour between $N=20$ and $N=28$. The latter can be in part ascribed to many-body truncations.
Interestingly, for both interactions, a second, smaller, kink is observed at $N=34$.\\
\item \textit{Analyse pairing properties in nuclei within a first-principle description.}\\
The ability of accessing ground-state energies of odd-even nuclei enables the investigation of pairing effects e.g. by considering three-point mass differences in even-$Z$ isotopic chains. 
The resulting pairing strength turns out to be underestimated compared to experimental observations, which possibly points to missing many-body correlations.\\
\item \textit{Probe the limits of SU(2)-conserving correlation expansion methods in the description of doubly open-shell nuclei.}\\
It is observed that the description of experimental data deteriorates for certain sets of nuclei away from singly-magic calcium. 
It is conjectured that this might signal the onset of significant quadrupole correlations, i.e. static deformation.
A careful scrutiny indeed reveals a correlation between the inaccuracy of the results (quantified in terms of deviation from experimental data) and an estimate of the deformation.\\
\end{enumerate}
Developing the above points, the manuscript is organised as follows.
First, the theoretical and computational scheme is briefly recalled in Sec.~\ref{sec_setup}. 
Section~\ref{sec_energies} is devoted to the study of ground-state energies, in the form of either total (Sec.~\ref{sec_total}) or differential (Secs.~\ref{sec_s2n} and \ref{sec_gaps}) binding energies. 
Further, a discussion of three-point mass differences is presented in Sec.~\ref{sec_TPMD}.
The impact of (expected) nuclear deformation on calculated ground-state energies is investigated in Sec.~\ref{sec_deformation}.
Finally, a systematic survey of nuclear radii and a selection of representative charge density distributions are presented in Sec.~\ref{sec_radii}.
Conclusions and perspectives follow in Sec.~\ref{sec_conclusions}.

\section{Computational set-up}
\label{sec_setup}

All calculations presented here were performed within the Gorkov self-consistent Green's function approach at second order in the algebraic diagrammatic construction expansion [ADC(2)]~\cite{Soma11, Soma14a}.
An extensive study of oxygen, calcium and nickel isotopes has been recently carried out in the same computational scheme and published in Ref.~\cite{Soma20a}.
Hence, only the most salient features are recalled here and the reader is referred to~\cite{Soma20a} for more computational and technical details.

Two different two- plus three-nucleon (2N+3N) Hamiltonians were employed in the present study.
The first one, labelled \lnl, is based on the next-to-next-to-next-to-leading order (N$^3$LO) nucleon-nucleon potential from Entem and Machleidt~\cite{Entem03, Machleidt11} complemented with the N$^2$LO $3N$ interaction for which a combination of local and nonlocal regulators is used~\cite{Soma20a}.
Low-energy constants were fitted to $A=2,3,4$ systems.
This Hamiltonian is further SRG-evolved to a low-momentum scale of $\lambda = 2 \text{ fm}^{-1}$.
The second one, labelled \sat, was introduced in Ref.~\cite{Ekstrom15} with the explicit goal of providing an improved description of saturation properties.
To achieve this, in contrast to \lnl, low-energy constants were simultaneously fitted to few-body systems as well as selected ground-state energies and radii of carbon and oxygen isotopes.
This Hamiltonian is SRG-unevolved.

Three-nucleon forces are treated following the formalism developed in Ref.~\cite{Carbone13}.
In practice, the three-body Hamilton operator is self-consistently convoluted with the correlated one-body density matrix and contributes to one- and two-body effective interactions~\cite{Cipollone15}. The contributions resulting from contracting two- and many-body density matrices were seen to be negligible for our purposes~\cite{Cipollone13,Barbieri14}. Note that
we discard interaction-irreducible diagrams containing three-body vertices. The 
formalism needed to include these at the ADC(3) level was presented in Ref.~\cite{Raimondi18} and their contribution is estimated to be comparable, in terms of both importance and required computing resources, to ADC(5) computations with only two-nucleon interactions.
The procedure used in this paper generates an $A$-dependent symmetry-conserving Hamiltonian that can be viewed as a generalisation of the particle-number-conserving normal-ordered two-body approximation discussed in Ref.~\cite{Ripoche20}.

As for the $k$-body basis used to expand $k$-body operators, a $k$-fold tensor product of one-body harmonic oscillator (HO) bases is presently employed.
The latter include states up to $e_{\text{{\rm max}}} \equiv$~max~$(2n+l)$ = 13. 
While the basis used for two-body operators is consistently truncated at $e_{2\text{max}} = 2 \, e_{\text{max}} = 26$,
three-body basis states are further restricted\footnote{This automatically imposes the same restriction on the $e_{2\text{max}}$ at play in \textit{three-body} operators, for which then $(e^{3\text{-body}}_{1\text{max}}, e^{3\text{-body}}_{2\text{max}}, e^{3\text{-body}}_{3\text{max}}) = (13, 16, 16)$.} to $e_{3\text{max}} = 16 < 3 \, e_{\text{max}}$ due to computational requirements.
For some representative (closed-shell, singly open-shell and doubly open-shell) isotopes, a variation of the HO frequency $\hbar \omega$ was performed in order to locate the optimal value for binding energies and radii.
Based on this analysis, $\hbar \omega = 18 \text{ MeV}$ for both energies and radii for \lnl, and $\hbar \omega = 14 \text{ MeV}$ for radii for \sat{} were identified as optimal values for all isotopic chains.
This analysis confirms what was found in Ref.~\cite{Soma20a} for semi-magic nuclei.
All results presented in the following were obtained with these model space parameters.

\begin{table}
\centering
\renewcommand{\arraystretch}{1.4}
\begin{tabular}{l|c|c|c|}
\cline{2-4}
& \multicolumn{2}{|c|}{\lnl{}} & \multicolumn{1}{|c|}{\sat{}} \\
\cline{2-4}
& $E$ & $r_\text{ch}$ & $r_\text{ch}$ \\
\cline{2-4}
\noalign{\vskip\doublerulesep
         \vskip-\arrayrulewidth}
\hline
\multicolumn{1}{|l||}{Model space ($e_{\text{max}}$)} & 0.5\% & $<$0.1\% & 0.5\% \\
\hline
\multicolumn{1}{|l||}{Model space ($e_{3\text{max}}$)} & 0.2\% & 0.2\% & 0.3\% \\
\hline
\multicolumn{1}{|l||}{ADC truncation} & 2\% & 0.5\% & $<$0.1\% \\
\hline
\multicolumn{1}{|l||}{U(1) breaking} & 0.2\% & $<$0.1\% & $<$0.1\% \\
\hline
\multicolumn{1}{|l||}{Neglected induced op.} & 2\% & 1\% & -  \\
\hline
\hline
\multicolumn{1}{|l||}{Total} & 2.9\% & 1.1\% & 0.6\% \\
\hline
\end{tabular}
\vspace{.35cm}
\caption{Breakdown of method uncertainties for the observables considered in the present study.
Errors from breaking of U(1) symmetry were estimated from the particle-number projected HFB calculations of Ref.~\cite{FrosiniPvt}.
Contributions from neglected two- and three-body radius operators were estimated using Ref.~\cite{Miyagi19}.
All remaining estimates derive from the GSCGF calculations of Ref.~\cite{Soma20a} and the present work.
}
\label{errors}
\end{table}
Given an input Hamiltonian, various types of method uncertainties affect the calculation of a given quantity.
These different sources of theoretical error are scrutinised in Tab.~\ref{errors} for the observables considered in this study.
Uncertainties related to model-space truncation were deduced by varying the $e_{\text{max}}$ and $e_{3\text{max}}$ parameters for selected closed- and open-shell isotopes. 
The error due to the second-order truncation in the ADC expansion is computed by comparing ADC(2) and ADC(3) results in closed-shell calcium isotopes~\cite{Soma20a}. 
Although ADC(3) might introduce additional correlations in open-shell systems, the particle-hole excitations already probed in closed-shell nuclei are expected to dominate for bulk properties such as energies and radii.
Hence, the overall ADC(3) shift is assumed to be of the same order of magnitude in closed- and open-shell isotopes.
This conjecture will have to be corroborated by explicit GSCGF-ADC(3) calculations in the future.
To estimate errors originating from the lack of restoration of the broken U(1) symmetry, ab initio particle-number projected HFB results from Ref.~\cite{FrosiniPvt} have been used as an upper limit. 
This is justified given that the variance characterising ADC(2) propagators ($\sigma^2_\text{N,Z} \sim 2$) is always smaller than the one found at the HFB level.
While symmetry restoration has been recently designed for MBPT and coupled-cluster theory~\cite{Duguet14, Duguet17b}, the existing formalism can not be straightforwardly applied to GSCGF theory, for which a dedicated development is yet to be devised.
Finally, discarding four- and higher-body operators induced by the SRG evolution of the \lnl{} Hamiltonian introduces an additional error. 
This has been estimated by performing calculations at different values of the SRG parameter, namely $\lambda = 1.8 \text{ fm}^{-1}$ and $\lambda = 2.2 \text{ fm}^{-1}$, for selected closed- and open-shell isotopes.
In addition, for radii, the uncertainty originating from having neglected induced two- and three-body radius operators has been accounted for based on the findings of Ref.~\cite{Miyagi19}.
One notices that model-space uncertainties from $e_{\text{max}}$ and $e_{3\text{max}}$ truncations are of similar magnitude, 0.5\% or smaller, for all cases.
Errors coming from the lack of symmetry restoration do not exceed 0.2\% for ground-state energies, while they result completely negligible for radii.
For \lnl{} total energies, the overall error is thus dominated by the many-body truncation and neglected SRG-induced many-body operators, both contributing with about 2\%.
Note that these uncertainties cancel out to a good extent when energy differences\footnote{A notable exception is represented by energy differences near a closed shell, where errors related to the breaking of particle-number do not cancel between a closed-shell and and open-shell system.} like two-neutron separation energies of three-point mass differences are computed.
Also for \lnl{} radii the dominating error appears to be related to neglected many-body operators (mostly in the Hamiltonian).
On the contrary, radii computed with \sat{} are characterised by a very good precision, with a total error of around 0.5\%.

\section{Ground-state energies}
\label{sec_energies}

\subsection{Total energies}
\label{sec_total}

\begin{figure}[b]
\centering
\includegraphics[width=8.5cm]{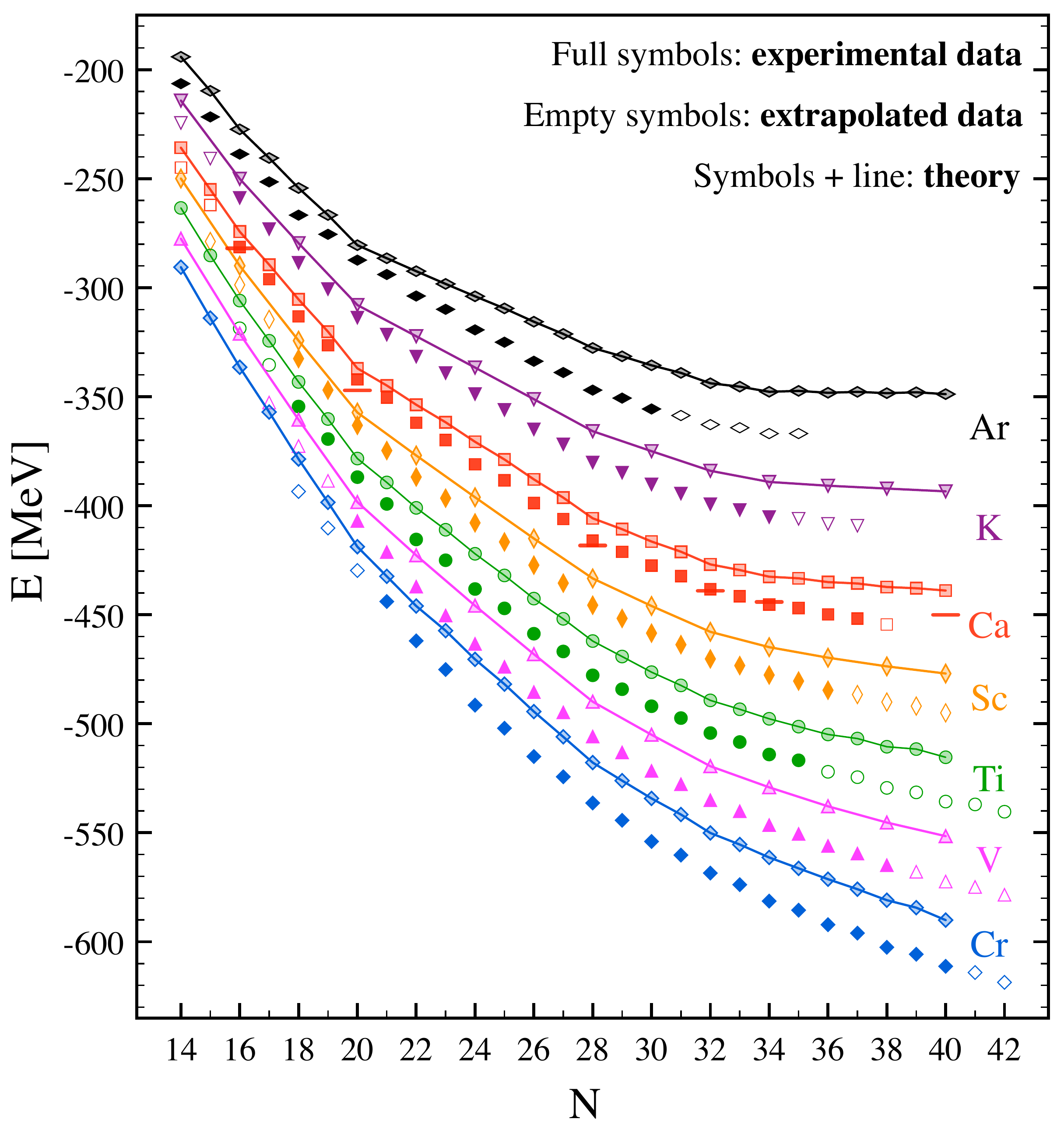}
\caption{Total binding energies along $Z=18-24$ isotopic chains computed at the ADC(2) level with the \lnl{} interaction (symbols joined by solid lines).
For comparison, experimental data (measured~\cite{AME2016, Leistenschneider18, Michimasa18, Xu19, Mougeot20}, full symbols and extrapolated~\cite{AME2016}, empty symbols) are displayed. 
Both calculated and experimental values are shifted by $(20-Z) \times 20$ MeV for a better readability.
For closed-shell calcium isotopes, available ADC(3) results~\cite{Soma20a} are displayed as horizontal lines.}
\label{fig_BE_ArCr}
\end{figure}
Let us start by analysing total ground-state energies along the seven isotopic chains studied in this work, i.e. argon ($Z=18$), potassium ($Z=19$), calcium ($Z=20$), scandium ($Z=21$), titanium ($Z=22$), vanadium ($Z=23$) and chromium ($Z=24$).
The current implementation of GSCGF theory is based on the assumption that $J^\Pi=0^+$ for targeted ground states and is therefore well suited for even-even nuclei. 
The ground-state energy of odd-even systems can be computed via~\cite{Duguet02b}
\begin{equation}
E_\text{odd-even}^A = \tilde{E}^A + \omega_0 \; ,
\end{equation}
where $\tilde{E}^A$ is the ground-state energy of the odd-even nucleus computed as if it had $J^\Pi=0^+$, i.e. as a fully paired even-number-parity state forced to have the right \textit{odd} number of particles on average, and $\omega_0$ is the lowest one-nucleon separation energies in the latter calculation.
Further details can be found in Refs.~\cite{Soma11, Soma14a}.
A more direct but similar approach is to use the addition and separation energies encoded in the spectral function but to recompute the even-even isotope with the center of mass corrections for $A\pm1$, as done in Ref.~\cite{Cipollone15}.
As a result, one can access the ground-state energy of \textit{all} isotopes with \textit{even} $Z$ and that of \textit{odd-even} isotopes with \textit{odd} $Z$.
Other observables, e.g. radii or densities, are instead available only for even-even systems.
Further developments, e.g. involving the use of Hellmann-Feynman theorem, are needed to extend their calculation to odd-even systems.

Computed ground-state energies are presented in Fig.~\ref{fig_BE_ArCr} and compared to experimental (measured and extrapolated) data.
The global behaviour is well captured by the calculated energies across all values of $Z$ and $N$, although underbinding with respect to experiment is observed for all chains.
The deviation \textit{per nucleon} is roughly of the same magnitude for all nuclei, around $0.2-0.3$ MeV (see also Fig.~\ref{fig_errors_ArCr} in the following). 
For calcium isotopes, for which ADC(3) calculations (displayed in Fig.~\ref{fig_BE_ArCr} as horizontal bars) are available~\cite{Soma20a}, the root-mean-square (r.m.s.) deviation of $E/A$ from experiment goes from 0.21 MeV in ADC(2) down to 0.06 MeV in ADC(3) (see also Tab.~\ref{rms}).
This shows that (i) the bulk of the ADC(2) underbinding is due to missing third-order correlations and (ii) the \lnl{} Hamiltonian can reach an excellent agreement with measured total ground-state energies in this mass region once a more refined truncation schemes is used.
A more careful inspection of the absolute r.m.s. deviations also reveals differences between the various isotopic chains, with the ADC(2) inaccuracy increasing when going away from singly-magic calcium.
Specifically, focusing on even-$Z$ isotopes, one goes from an absolute r.m.s. deviation of 10 MeV for calcium to 14 MeV for argon and titanium up to 19 MeV for chromium.
This additionally points to a possible specific deficiency (besides generic third-order terms) related to a poor account of quadrupole correlations, as elaborated on in the following.

\subsection{One- and two-nucleon separation energies}
\label{sec_s2n}

\begin{figure}
\centering
\includegraphics[width=8.5cm]{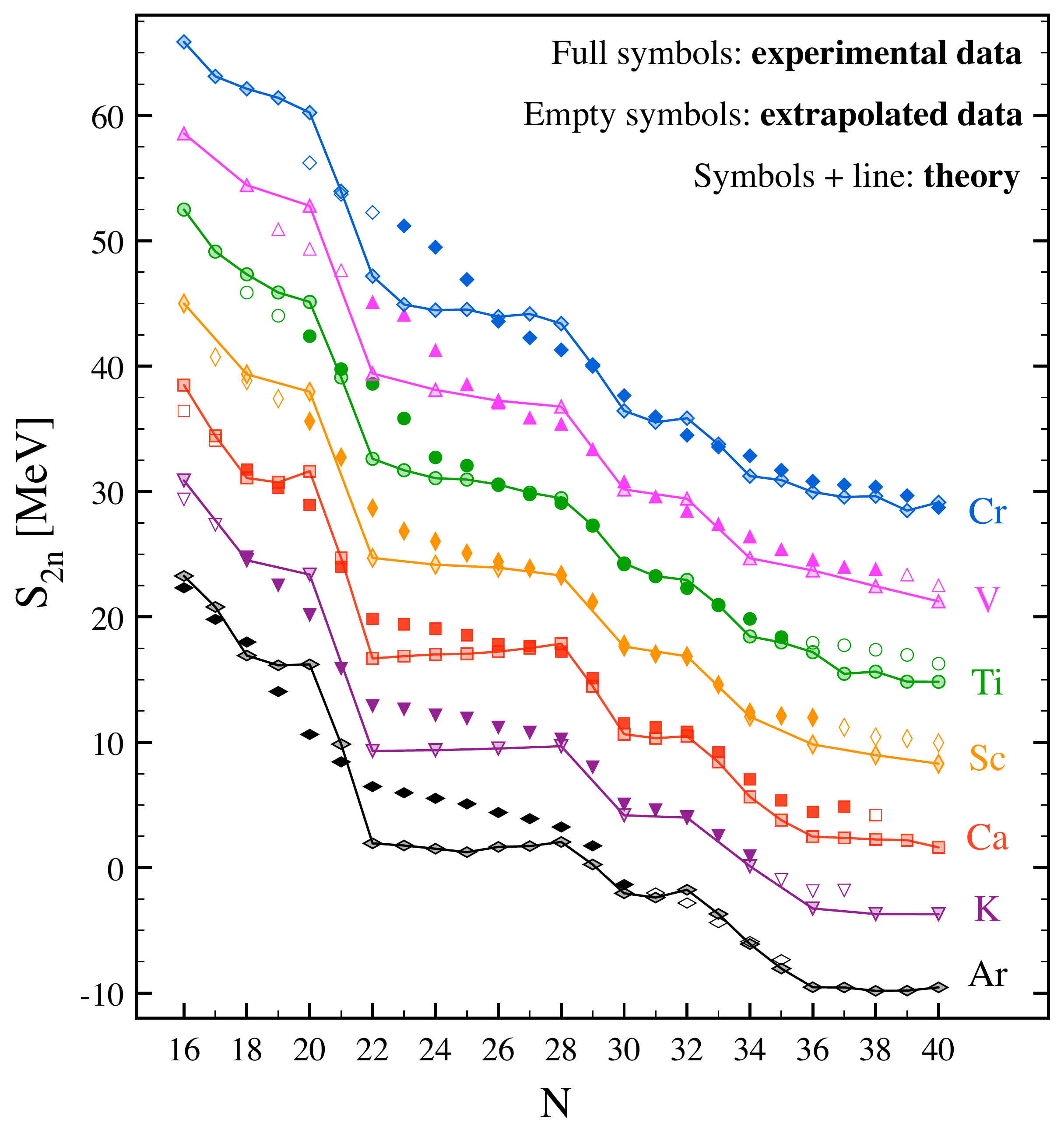}
\caption{Two-neutron separation energies along $Z=18-24$ isotopic chains computed with the \lnl{} interaction (symbols joined by solid lines), compared to experimental (measured, full symbols and extrapolated, empty symbols) data. 
Both calculated and experimental values are shifted by $(Z-20) \times 5$ MeV for a better readability.}
\label{fig_S2n_ArCr}
\end{figure}
Systematically accessing successive nuclides along isotopic or isotonic chains allows to investigate some of the most fundamental properties of atomic nuclei such as the limits of their existence as bound states or the emergence (and evolution) of magic numbers.
Such properties are best studied by looking at total ground-state energy differences. 
Two-neutron separation energies
\begin{equation}
S_\text{2n}(N,Z) \equiv |E(N,Z)| - |E(N-2,Z)|
\end{equation}
are first considered. 
Their values computed from the total energies of Fig.~\ref{fig_BE_ArCr} are shown in Fig.~\ref{fig_S2n_ArCr}, together with available and extrapolated experimental data.
The overall agreement with experiment is remarkable, with computed values following the main trends of measured data.
R.m.s. deviations amount to 2.9, 1.5, 2.0 and 2.2 MeV for argon, calcium, titanium and chromium respectively. 
The two neutron magic numbers $N=20$ and $N=28$, associated with sudden drops of $S_\text{2n}$, are visible in all theoretical curves.
The $N=28$ gap is very well reproduced across all isotopic chains, with the good description carrying over to larger neutron numbers for most chains.
On the contrary, the gap at $N=20$ turns out to be overestimated, with the comparison to experiment worsening when departing from proton magic number $Z=20$.
The description deteriorates also in other regions, e.g. for argon isotopes between $N=20$ and $N=28$ or more generally for chromium isotopes.
The latter observation reflects in the differences between the r.m.s. deviations reported above.
As discussed further below, it might originate in the poorer description of the strong quadrupole correlations characterising doubly open-shell systems.

\begin{figure}[b]
\centering
\includegraphics[width=8.8cm]{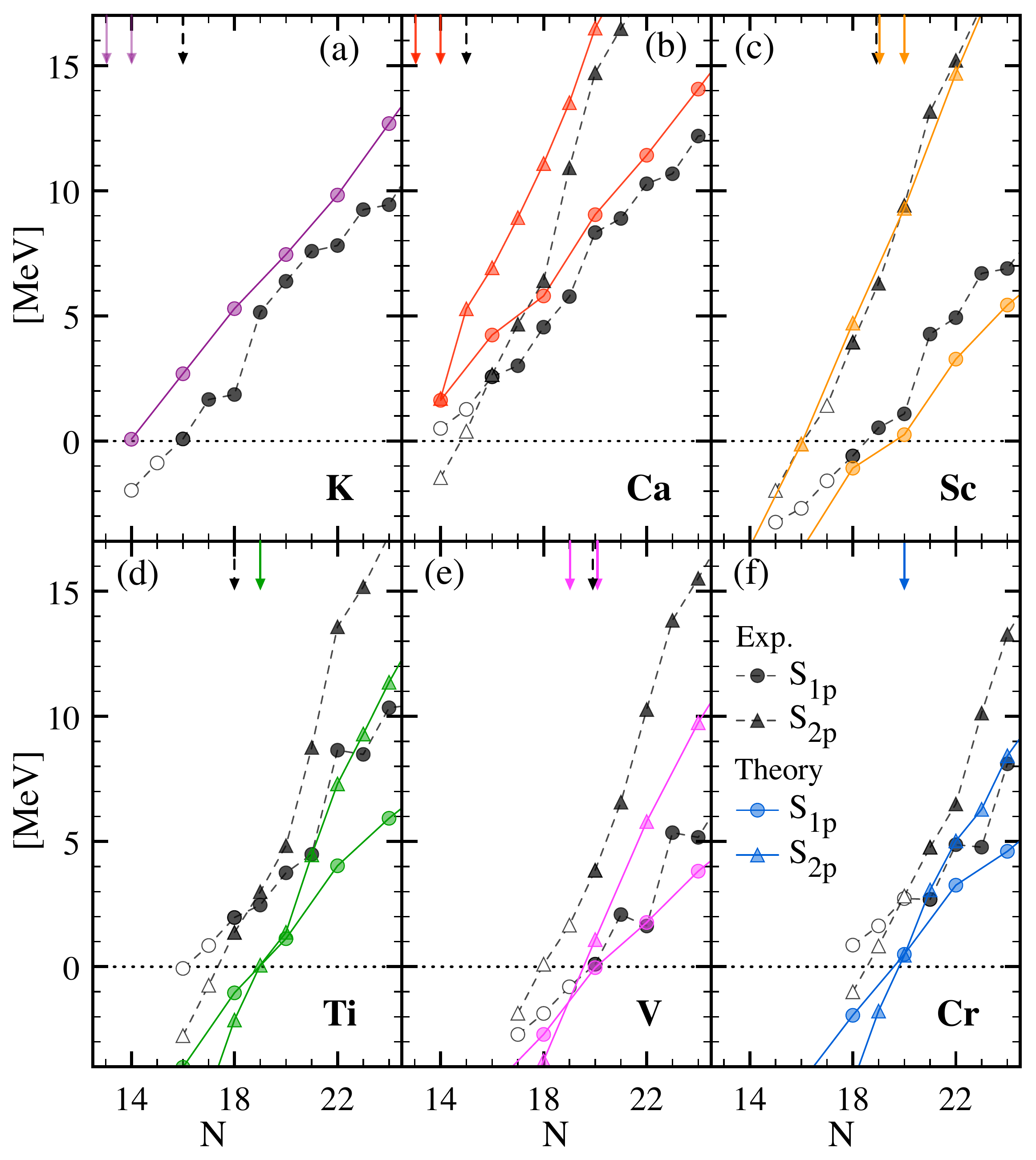}
\caption{One- and two-proton separation energies displayed as a function of neutron number for different $Z$.
Calculations performed with the \lnl{} interaction (symbols joined by solid lines) are compared to existing data (measured, full symbols and extrapolated, empty symbols, all joined by dashed lines). The solid coloured (dashed black) arrows at the top of each panel mark the computed (experimental) driplines. In some cases (K, Ca, Sc, V) the theoretical dripline can not be determined unambiguously from the calculations, hence the two possible values are shown.
}
\label{fig_S12p}
\end{figure}
\begin{figure}
\centering
\includegraphics[width=8cm]{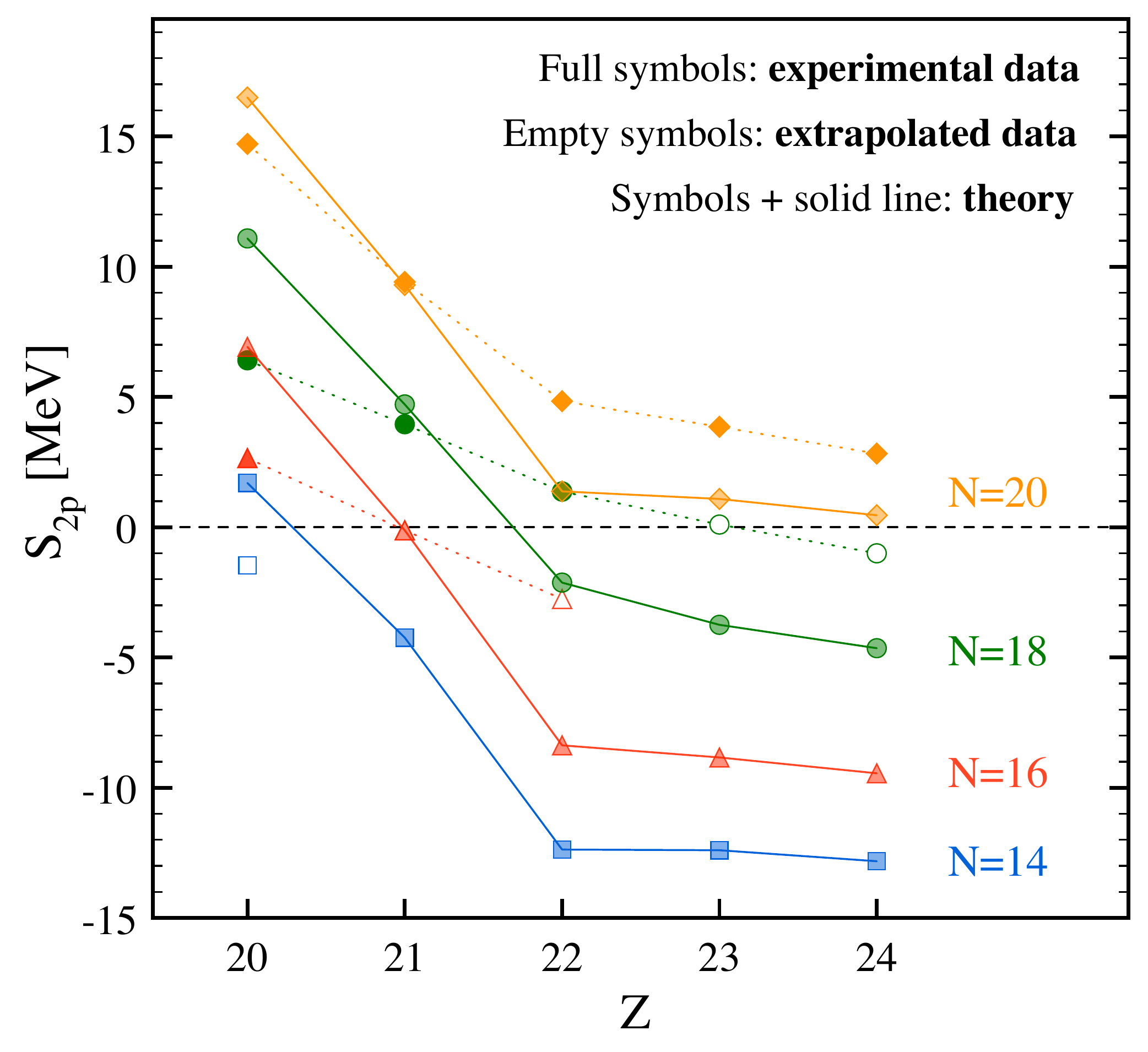}
\caption{Two-proton separation energies displayed as a function of proton number for different isotonic chains.
Calculations performed with the \lnl{} interaction (symbols joined by solid lines) are compared to existing data (measured, full symbols and extrapolated, empty symbols, all joined by dotted lines).}
\label{fig_S2p}
\end{figure}
The neutron dripline, i.e. the position of the last bound system in a given isotopic chain, can be also read from two-neutron separation energies as unbound nuclei are characterised by negative values of $S_\text{2n}$.
None of the computed neutron rich isotopes shown in Fig.~\ref{fig_S2n_ArCr} results unbound, i.e. the dripline is predicted to be located beyond $N=40$ for all considered chains\footnote{Present calculations could not be extended beyond $N=40$ due to convergence issues, see discussion in Ref.~\cite{Soma20a} for more details.}. 
The smallest $S_\text{2n}$ value are reached for $^{56-57}$Ar and are as low as 100 keV.
However, one must remark that continuum coupling is likely to play an important role when binding energies are so close to the neutron emission threshold. 
Presently, the continuum is crudely included via the discretised harmonic oscillator basis, which does not ensure correct asymptotic properties. 
In future studies, in order to reliably determine the position of the neutron dripline, particular care will have to be devoted to a more proper treatment of this aspect.

The coupling to the particle continuum plays a lesser role around the proton dripline because of Coulomb repulsion.
Given that present calculations span several neighbouring chains, the proton dripline can be investigated within this theoretical setting.
Here, the key quantities are one-proton and two-proton separation energies, defined respectively as
\begin{equation}
S_\text{1p}(N,Z) \equiv |E(N,Z)| - |E(N,Z-1)| 
\end{equation}
and
\begin{equation}
S_\text{2p}(N,Z) \equiv |E(N,Z)| - |E(N,Z-2)|  \: .
\end{equation}
For a given element, the most proton-rich isotope for which both $S_\text{1p} > 0$ and $S_\text{2p} > 0$ determines the position of the proton dripline.
In Fig.~\ref{fig_S12p}, measured and computed $S_\text{1p}$ and $S_\text{2p}$ are displayed as a function of neutron number for the isotopic chains considered in this study\footnote{For potassium only $S_\text{1p}$ can be computed, while for argon none of the two separation energies is available in the present calculations.}.
Experimentally, for these elements, the proton dripline has been determined\footnote{Experimentally, the dripline is typically established by means of a void observation of one or several isotopes rather than by determining a negative value of $S_\text{1p}$ or $S_\text{2p}$.} up to vanadium, with the last bound isotopes being $^{35}$K, $^{35}$Ca, $^{40}$Sc, $^{40}$Ti and $^{43}$V.
For chromium, the last known isotope is $^{43}$Cr.
Theoretical curves generally follow the experimental trends yielding an overall correct qualitative description of both $S_\text{1p}$ and $S_\text{2p}$.
Looking more in detail, one observes that calculations tend to overestimate the measured separation energies in potassium and calcium, provide an excellent reproduction of scandium isotopes and underestimate titanium, vanadium and chromium.
As a result, the position of the proton dripline is found at too small $N$ (with a difference of two or three neutrons) for the first two elements.
In scandium, as well as vanadium, the dripline is correctly determined at $N=19$ and $N=20$ respectively.
In titanium and chromium, it is also found respectively at $N=19$ and $N=20$, in this case one neutron away from what observed experimentally.

The cause of this small discrepancy can be traced back to the poor reproduction of the $Z=20$ gap by the \lnl{} Hamiltonian, as evident in Fig.~\ref{fig_S2p}.
Here, two-proton separation energies are plotted as a function of proton number for different isotonic chains.
One notices that, similarly to what observed in Fig.~\ref{fig_S2n_ArCr} for $N=20$, the $Z=20$ gap is overestimated by at least 5 MeV in all considered isotones. 
The disagreement becomes more severe for low neutron numbers, which impacts the determination of the proton dripline in lighter isotopes.
In spite of these shortcomings, this detailed analysis confirms the overall quality of present ab initio calculations, not dissimilar from what emerges from the systematic study reported in Ref.~\cite{Stroberg19b}.

\begin{figure}[t]
\centering
\includegraphics[width=8.5cm]{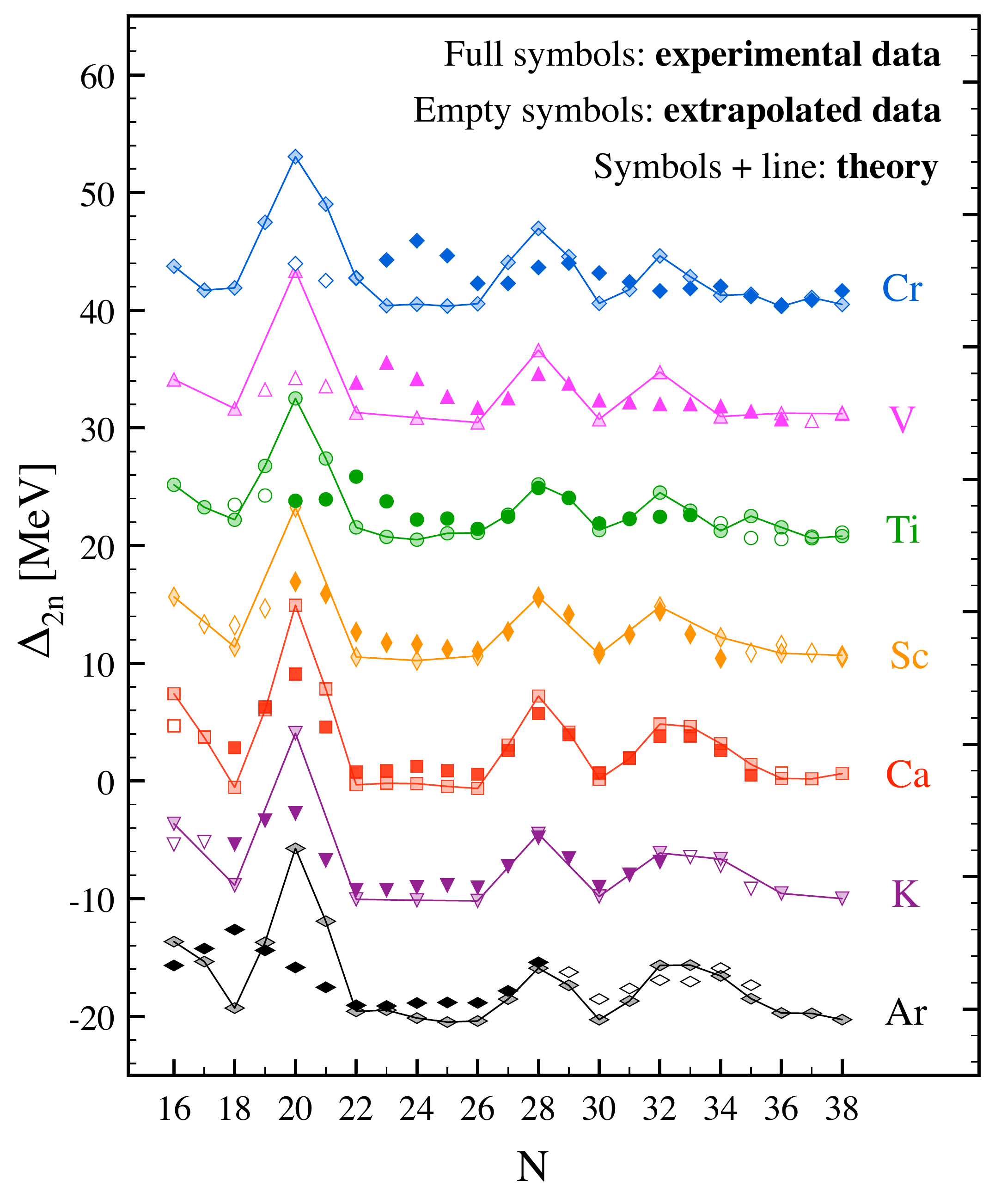}
\caption{Two-neutron shell gaps, Eq.~\eqref{eq_ngaps}, along $Z=18-24$ isotopic chains computed with the \lnl{} interaction (symbols joined by solid lines), compared to experimental (measured, full symbols and extrapolated, empty symbols) data.
Both theoretical and experimental values are shifted by $(Z-20) \times 10$ MeV for a better readability.}
\label{fig_gapsN_ArCr}
\end{figure}

\begin{figure}[t]
\centering
\includegraphics[width=9.5cm]{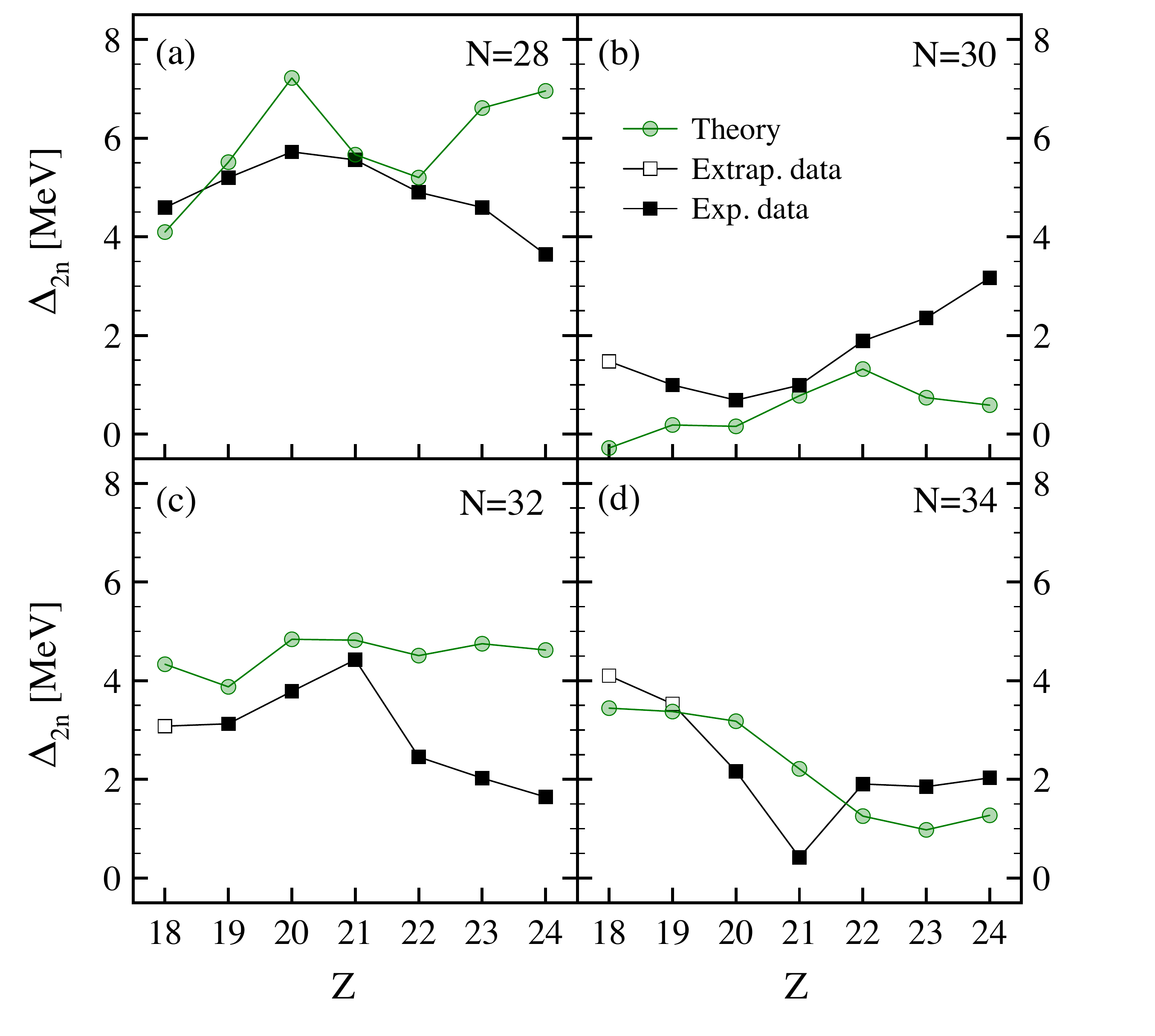}
\caption{Two-neutron shell gaps, Eq.~\eqref{eq_ngaps}, along four isotonic chains computed with the \lnl{} interaction (circles), compared to experimental (measured, full squares, and extrapolated, empty squares) data. 
Results for $N=28,30,32$ and $34$ are shown in panels (a), (b), (c) and (d) respectively.}
\label{fig_magic}
\end{figure}

\begin{figure}[t]
\centering
\includegraphics[width=8.5cm]{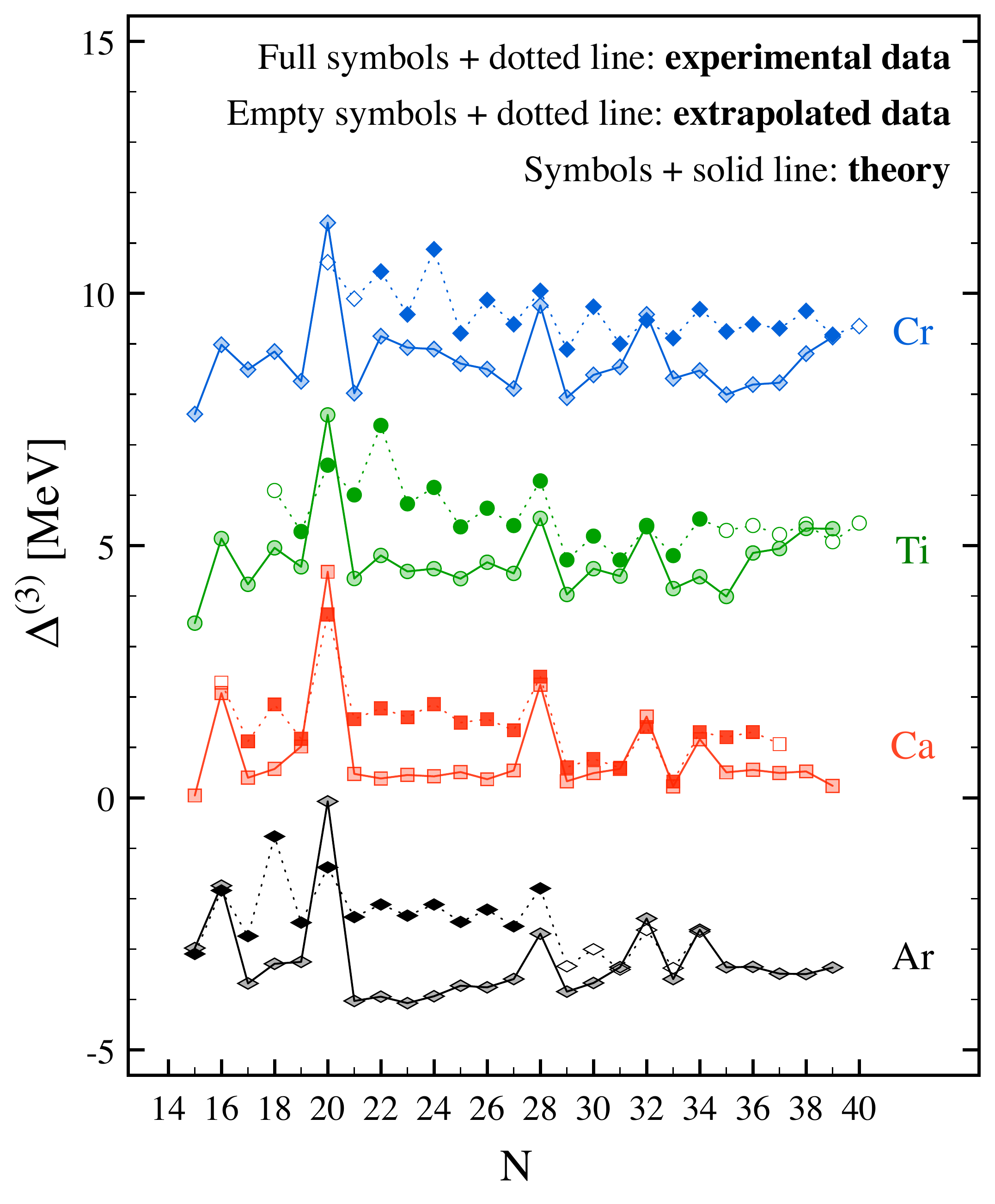}
\caption{Three-point mass differences, Eq.~\eqref{eq_tpmd}, along $Z=18,20,22$ and $24$ isotopic chains computed with the \lnl{} interaction (symbols joined by solid lines), compared to experimental (measured, full symbols and extrapolated, empty symbols) data.
Both calculated and experimental values are shifted by $(Z-20) \times 2$ MeV for a better readability.}
\label{fig_3point_ArCr}
\end{figure}

\subsection{Neutron gaps}
\label{sec_gaps}

A finer insight regarding the magic character of specific neutron numbers can be gained by looking at so-called two-neutron shell gaps, defined  as
\begin{equation}
\Delta_\text{2n}(N,Z) \equiv S_\text{2n}(N,Z) - S_\text{2n}(N+2,Z) 
\label{eq_ngaps}
\end{equation}
and displayed in Fig.~\ref{fig_gapsN_ArCr}. 
As for the $S_\text{2n}$, one first notices an overall very good agreement with experiment, with the clear exception of the $N=20$ peak and its vicinity.
R.m.s. deviations for this quantity are slightly larger to the ones characterising two-neutron separation energies, specifically 3.8, 1.9, 2.9 and 2.4 MeV for argon, calcium, titanium and chromium respectively.
While in semi-magic calcium isotopes calculations only fail to reproduce the height of the peak, experimental data for other isotopes show a displacement of the peak, linked to a possible disappearance of the $N=20$ magic number, which is not reproduced by the present calculations.
In contrast, the $N=28$ peak is very well reproduced up to $Z=22$, with the description only slightly deteriorating for $Z=23$ and $Z=24$.
The emergence of the $N=32$ subclosure is nicely visible in lighter elements, as well as the one at $N=34$ in argon, potassium and calcium.
When going towards higher proton number their evolution is poorly described starting with $N=34$ in scandium and $N=32$ in vanadium. 
The behaviour becomes even more inconsistent for chromium.
Again, this might signal the importance of certain ingredients (e.g. quadrupole correlations) that are missing in the present theoretical framework. 

In spite of these deficiencies, remarkably, all magic numbers as well as their qualitative evolution emerge ``from first principles'', i.e. starting solely from inter-nucleon interactions whose coupling constants have been adjusted only in few-body systems.
Let us stress that, indeed, no ad hoc information about the magic character of these isotopes is inserted at any stage of the calculation.
The emergence of this feature can be better appreciated in Fig.~\ref{fig_magic} where two-neutron gaps are compared to experimental (measured and extrapolated) data along $N=28,30,32$ and $34$ isotonic chains. 
While there is room for improvement in $Z=22,23,24$ isotones for reasons discussed above, the overall description is very reasonable.
In addition, calculations of the $N=28$ gaps were recently extended down to chlorine and sulfur~\cite{Mougeot20} where an excellent agreement with novel precision mass measurement was also found.

\subsection{Three-point mass differences}
\label{sec_TPMD}

One of the longstanding challenges in low-energy nuclear physics relates to the microscopic description of nuclear superfluidity~\cite{Dean03}. 
The microscopic origin of nucleonic pairing, i.e. how it originates in the context of a first-principle calculation and the role played by different types of many-body correlations, remains to be elucidated~\cite{Duguet13}. 
A fundamental, yet unresolved, question relates to how much of the pairing gap in finite nuclei is accounted for at lowest order~\cite{Duguet10,Lesinski11} and how much is due to higher-order processes, i.e. to the induced interaction associated with the exchange of collective medium fluctuations between paired particles~\cite{Barranco04,Gori05,Pastore08,Idini11}. 
By treating normal and anomalous propagators consistently and at the same level of approximation, GSCGF many-body scheme is in an excellent position to contribute to this quest. 
In finite nuclei, the odd-even mass staggering is a good measure of nucleonic, e.g. neutron, pairing.
In particular, the three-point mass difference formula
\begin{equation}
\Delta^{(3)}(N,Z) \equiv \frac{(-1)^N}{2} [ E(N-1,Z) - 2E(N,Z) + E(N+1,Z) ]
\label{eq_tpmd}
\end{equation}
successively evaluated for even and odd $N$ closely encompasses the pairing gap~\cite{Duguet02b,Duguet02a} as long as $N$ does not correspond to a shell closure%
\footnote{Note that $\Delta^{(3)}$ corresponds to half of the energy difference between the lowest unoccupied quasiparticle and the highest occupied quasihole states, that is the particle-hole neutron gap at the Fermi surface. At subshell closures, this is dominated by the gap among different nuclear orbits. However, for open neutron shells only the pairing contribution remains.}. 
Calculated three-point mass differences for argon, calcium, titanium and chromium are compared to available experimental data in Fig.~\ref{fig_3point_ArCr}. 
In spite of a reasonable general trend, the pairing strength generated in the present ab initio calculations is too low compared to experiment.  
This feature is particularly visible for $N \in [21,27]$ isotopes in all considered chains, as well as beyond $N=34$ for calcium and chromium. 
Keeping in mind the possible deficiency of the currently used Hamiltonian, this result likely points to missing higher-order correlations.
The ADC(2) truncation scheme employed here already includes both the lowest-order pairing term and the induced interaction resulting from the exchange of unperturbed particle-hole excitations.
However, it does not account for the collective vibrations that are thought to be responsible for the remaining pairing strength~\cite{Barranco04,Gori05,Pastore08,Idini11}.
Consequently, it does improve on HFB results, e.g., by correcting the odd-even staggering present at the mean-field level (not shown here), but it does not significantly change the amplitude of the pairing gap.
The extension of GSCGF to the ADC(3) level is envisaged in the near future, knowing that such a truncation does indeed seize important features of collective fluctuations and of their effect on superfluidity.

\begin{figure}[b]
\centering
\includegraphics[width=8.5cm]{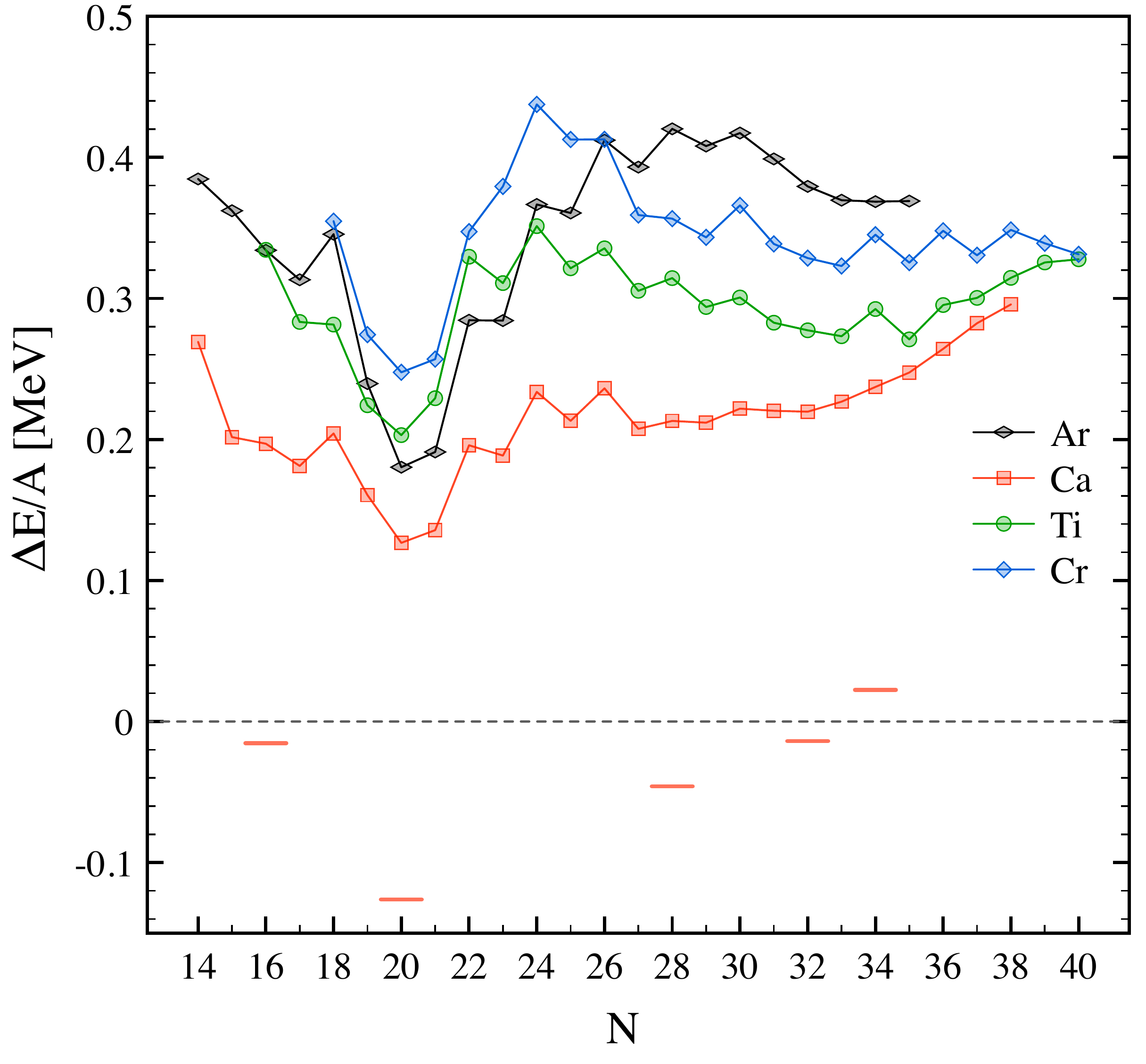}
\caption{Relative ADC(2) errors (theory - experiment) on total binding energies per nucleon along $Z=18,20,22$ and $24$ isotopic chains.
ADC(3) errors are also reported for doubly closed-shell calcium isotopes and displayed as horizontal bars.
Calculations and experimental data are taken from Fig.~\ref{fig_BE_ArCr}.}
\label{fig_errors_ArCr}
\end{figure}
\begin{figure*}
\centering
\includegraphics[width=\textwidth]{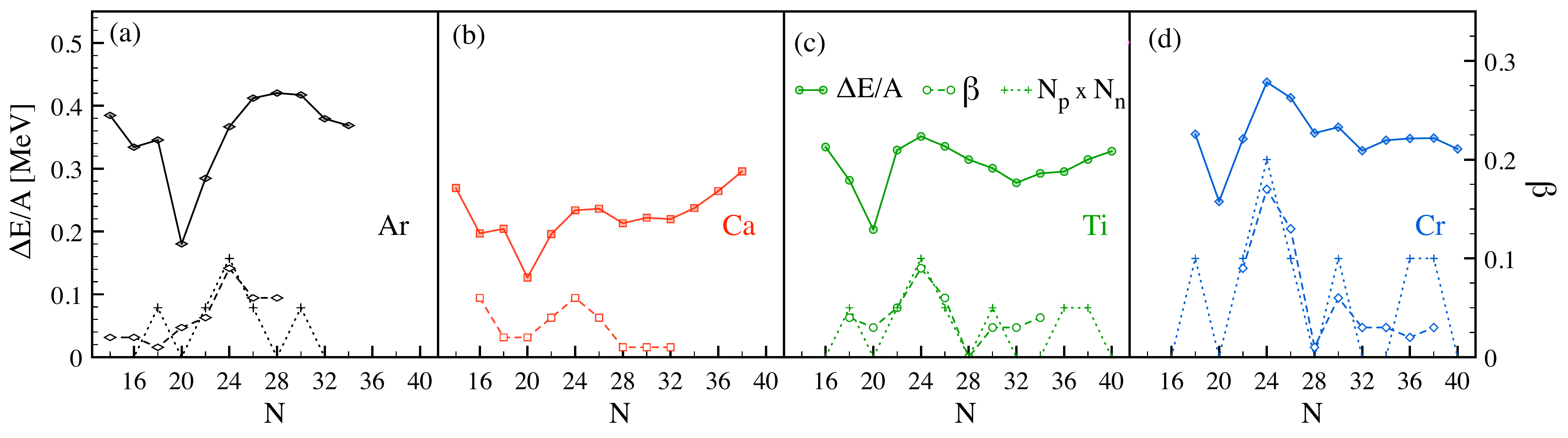}
\caption{Relative errors on total binding energies per nucleon for $Z=18,20,22$ and $24$ isotopes (full symbols and solid lines, taken from Fig.~\ref{fig_errors_ArCr}, referring to the left vertical axis).
Theoretical results are compared to the simple estimate $N_\text{p} \times N_\text{n}$, following Ref.~\cite{Dobaczewski88} (crosses and dotted lines, arbitrarily normalised) and the deformation parameter $\b$ computed via EDF calculations~\cite{Bender06a} (empty symbols and dashed lines, referring to the right vertical axis).}
\label{fig_defomration_ArCr}
\end{figure*}
In titanium and chromium, theoretical and experimental three-point mass differences show further qualitative differences. 
In addition to the average value of $\Delta^{(3)}$ being too low, the increase of its oscillation between $N=20$ and $N=28$ compared to calcium isotopes along with the shell-closure disappearances at $N=28,32,34$ are not captured. 
The oscillation of $\Delta^{(3)}$ around its average is not related to the anomalous part of the self-energy (i.e. the pairing gap) but rather to its normal part (i.e. the effective mean-field)~\cite{Duguet02b,Duguet02a}. 
The qualitative evolution of this staggering from calcium to titanium and chromium pointed out above is thus a fingerprint of increased quadrupole correlations on the normal self-energy.
The absence of this evolution in our theoretical calculation confirms the need to include these correlations consistently in both normal and anomalous channels. 
While extending GSCGF to the ADC(3) level should help better describing the staggering of $\Delta^{(3)}$, an explicit treatment of deformation will probably be the most efficient way to reach a quantitative agreement whenever quadrupole fluctuations become truly collective, i.e. as one moves significantly away from semi-magic systems.

\subsection{Effects of deformation}
\label{sec_deformation}

For several of the quantities discussed above, the poorer agreement with theoretical data when departing from semi-magic calcium has been ascribed to an inefficient description of quadrupole correlations.
To substantiate this observation, differences between computed and experimental ground-state energies per nucleon are displayed in Fig.~\ref{fig_errors_ArCr} for four isotopic chains.
The best agreement with experimental values is found for calcium isotopes. 
Other chains perform generally worse, with the quality of the description deteriorating in particular for neutron-rich argon and chromium isotopes.
In all cases a clear minimum is visible at $N=20$ and a maximum around $N=24$, which suggests a correlation with the closed- or open-shell character of the neutrons and the associated absence or presence of static deformation.
ADC(3) deviations, available for calcium isotopes with sub-shell closures, are also displayed in the figure. 
They illustrate the typical gain achieved by the inclusion of higher-order correlations in semi-magic systems.

The hypothetical correlation with deformation is further examined in Fig.~\ref{fig_defomration_ArCr}, where the four curves of Fig.~\ref{fig_errors_ArCr} are plotted separately and compared to two different quantities measuring the effects of deformation in phenomenological approaches.
First, we consider the simple estimate $N_\text{p} \times N_\text{n}$, where $N_\text{p}$ ($N_\text{n}$) is the number of valence proton (neutron) pairs in a mean-field picture. 
Such a quantity has been shown to provide a good estimate of the so-called \textit{deformation energy} in (single-reference) energy density functional (EDF) calculations~\cite{Dobaczewski88}.
Second, we plot the actual deformation parameter $\beta$ obtained in (multi-reference) EDF calculations~\cite{Bender06a}.
These two estimates of deformation provide a similar picture throughout the four isotopic chains. 
This is consistent with the idea that deformation is mean-field dominated, with beyond-mean-field correlations accounting for additional fluctuations on top.
Turning to our results, one observes that the correlation between the theoretical error $\Delta E / A$ and the two phenomenological estimates is striking for all chains. 
The deformation parameter $\beta$, with smoother variations across sub-shell closures, seems to provide a slightly better account of our theoretical error.
An exception is visible for light argon isotopes, with the mean-field estimate $N_\text{p} \times N_\text{n}$ better capturing the behaviour of $\Delta E / A$ around $N=20$.
This analysis eventually supports the intuition that the collective quadrupole correlations arising in doubly-open shell systems can hardly be captured by present SU(2)-conserving calculations.

Even if in principle all correlations can be accounted for in the current theoretical scheme, one would need to include very high orders in the expansion in order to grasp such quadrupole static correlations.
Indeed, for spherical bases, these are typically associated with the coherent superposition of many particle-many hole excitations that are not included in the low-order many-body truncation schemes currently at reach. 
Extending beyond the ADC(3) approximation involves a factorial increase in the numbers of diagrams and would need a shift of paradigm in which \textit{all} contributions are dealt with at once through stochastic sampling~\cite{VanHoucke12}. 
An alternative solution is the extension of existing expansion methods towards SU(2)-breaking schemes that will enable an efficient description of static deformation from the outset.

\section{Radii}
\label{sec_radii}

Among the basic nuclear properties addressed by ab initio calculations in the past few years, the size of medium-mass nuclei has typically represented (and, to a good extent, still represents) one of the main challenges.
The first sets of calculations that successfully reproduced ground-state energies of oxygen isotopes failed to provide, at the same time, a good description of charge radii~\cite{Lapoux16}.
The \sat{} Hamiltonian, specifically introduced to cure this issue~\cite{Ekstrom15}, very much improved the description of radii although discrepancies for neutron-rich systems have been shown to persist~\cite{Lapoux16, GarciaRuiz16}.
\begin{figure}
\centering
\includegraphics[width=8.5cm]{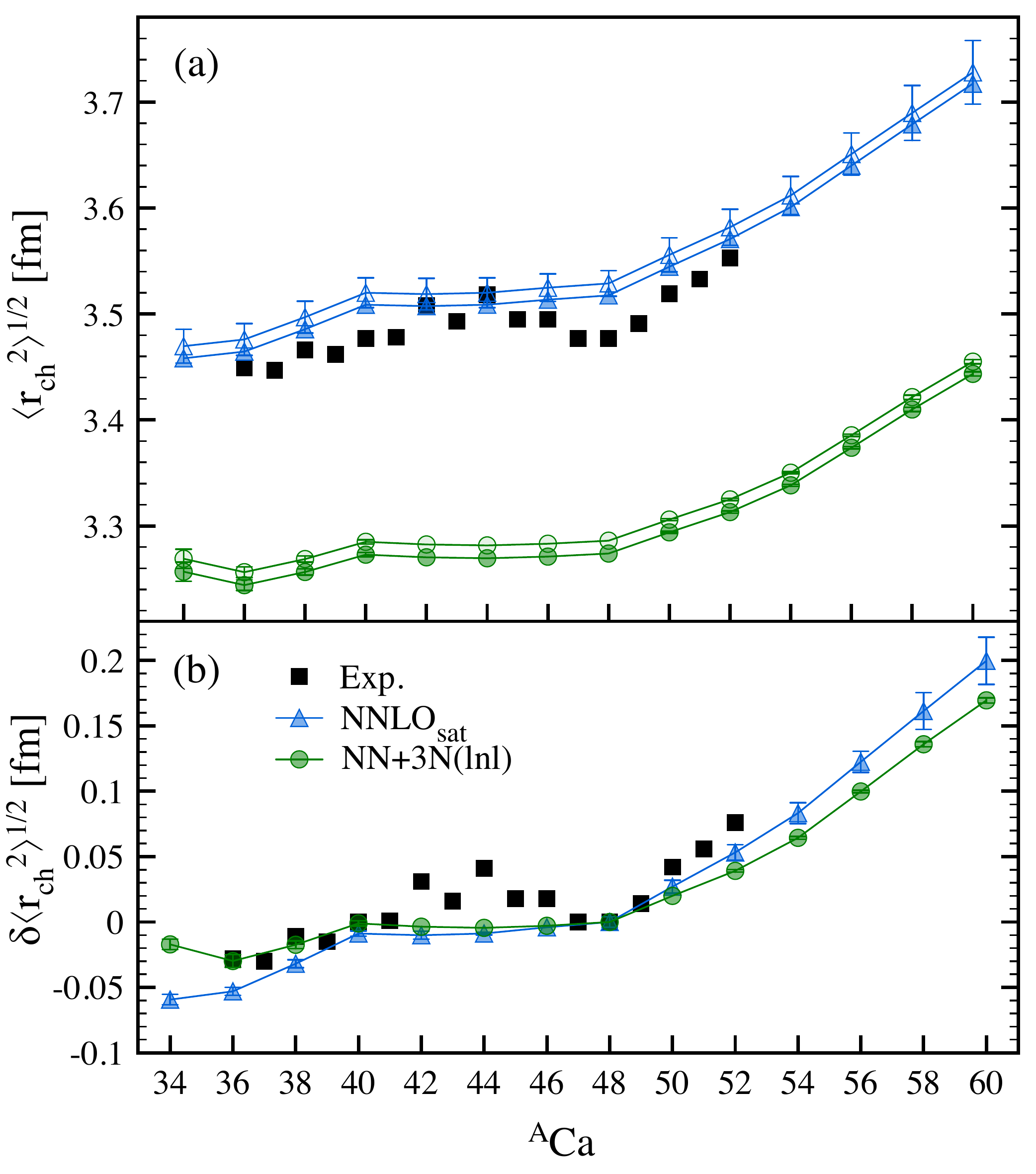}
\caption{(a) Absolute root mean square charge radii of calcium isotopes and (b) differential ones relative to $^{48}$Ca computed with the \lnl{} and \sat{} interactions. 
Available experimental data from Refs.~\cite{Angeli13, GarciaRuiz16, Miller19} are displayed.
Dark (light) symbols were obtained using a value $\langle R^2_{\text{p}} \rangle = 0.6905 \text{ fm}^2$~\cite{Xiong19} ($\langle R^2_{\text{p}} \rangle = 0.770 \text{ fm}^2$~\cite{CODATA2006}) in Eq.~\eqref{eq_r_ch}.
Error bars account for the uncertainty associated to model space truncation (see text for details).
}
\label{fig_radii_Ca}
\end{figure}
An unsatisfactory account of nuclear sizes remains for several Hamiltonians that are currently employed in state-of-the-art calculations~\cite{Simonis17, Soma20a}.
Very recently, new generations of chiral interactions have been proposed and shown to provide promising results for charge radii of closed-shell~\cite{Huther19} as well as some open-shell~\cite{Novario20} medium-mass nuclei. 
The behaviour along isotopic chains around calcium remains however to be investigated.
In Ref.~\cite{Soma20a} charge radii of oxygen, calcium and nickel isotopes have been systematically investigated with the \lnl{} and \sat{} Hamiltonians.
The study confirmed the good performance of \sat{} up to the nickel chains.
Here, in addition to a more refined analysis of calcium isotopes, charge radii along argon, titanium and chromium chains are presented.

Mean square (m.s.) charge radii are computed starting from m.s. point-proton radii $\langle r^2_{\text{p}} \rangle$ as follows
\begin{equation}
\langle
r^2_{\text{ch}}
\rangle
=
\langle
r^2_{\text{p}}
\rangle
+
\langle
R^2_{\text{p}}
\rangle
+
\frac{N}{Z}
\langle
R^2_{\text{n}}
\rangle
+
\langle
r^2
\rangle_{\text{so}}
+
\frac{3 \hbar^2}{4 m_\text{p}^2 c^2} \; .
\label{eq_r_ch}
\end{equation}
The last term corresponds to the relativistic Darwin-Foldy correction~\cite{Friar97} amounting to $0.033 \text{ fm}^2$.
The second to last term, $\langle r^2 \rangle_{\text{so}}$, denotes a spin-orbit correction.
In the present work it has been computed in the mean-field limit following Ref.~\cite{Horowitz12}.
$\langle R^2_{\text{p}} \rangle$ and $\langle R^2_{\text{n}} \rangle$ represent the m.s. charge radius of the proton and the neutron respectively. 
While the latter is relatively well established, $\langle R^2_{\text{n}} \rangle = - 0.1149(27) \text{ fm}^2$~\cite{Angeli13}, the determination of the former has been debated and revised in the past few years.
In the past, the value of $\langle R^2_{\text{p}} \rangle \simeq 0.77 \text{ fm}^2$ inferred from electron scattering experiments was commonly used and included in the CODATA compilation~\cite{CODATA2006}.
Recent experiments, including electronic and muonic hydrogen Lamb shift measurements, favour a lower m.s. radius of $\langle R^2_{\text{p}} \rangle \simeq 0.70 \text{ fm}^2$~\cite{Hammer20}.
As a result, the CODATA value was updated to $\langle R^2_{\text{p}} \rangle \simeq 0.7079 \text{ fm}^2$~\cite{CODATA2018}.
This value is adopted in the present work and used in Eq.~\eqref{eq_r_ch}, unless specified otherwise.
Given the large variation of $\langle R^2_{\text{p}} \rangle$ found in the literature, however, it is worth investigating its impact on computed charge radii, specially in comparison with other sources of theoretical error in the calculation.

Figure~\ref{fig_radii_Ca} shows r.m.s. charge radii along calcium isotopes computed with the \lnl{} and \sat{} Hamiltonians, either as absolute, panel (a), or relative to $^{48}$Ca, panel (b).
For each interaction, the two sets of points (dark and light symbols) were obtained with two different values of the proton radius in Eq.~\eqref{eq_r_ch}, respectively $\langle R^2_{\text{p}} \rangle = 0.6905 \text{ fm}^2$~\cite{Xiong19} and $\langle R^2_{\text{p}} \rangle = 0.770 \text{ fm}^2$~\cite{CODATA2006}.
The two values are representative of the two sets of experimental results discussed above.
For each set of points, error bars conservatively account for the uncertainty coming from truncation of the one-body basis in the calculation.
Specifically, they are obtained from the variation associated to different values of the HO frequency $\hbar \omega$ around the optimal value (itself determined as the closest point to the intersection of the different $e_{\text{{\rm max}}}$ curves, see Fig. 5 of Ref.~\cite{Soma20a}).
While such variation is sizeable for \sat, it is generally smaller than the size of the points for \lnl.
For the latter, however, one should consider also the other uncertainties discussed in Tab.~\ref{errors}, which add up to about 1\%.
In this context, the choice of $\langle R^2_{\text{p}} \rangle$ can lead to a 0.5\% variation, thus comparable with other uncertainties, and should not be overlooked. 
The situation is more favourable for differential radii, as visible in Fig.~\ref{fig_radii_Ca}(b).
For this quantity most of the errors cancel out and one is left with some sizeable model-space truncation uncertainty only for the most neutron-rich isotopes.

In general, while \sat{} only slightly overestimates measured data, results obtained with \lnl{} underestimate the experimental values by about 5\% throughout the calcium chain.
Although the main experimental trend is roughly captured by the theoretical curves (see also Fig.~\ref{fig_radii_all} and associated discussion), two of its peculiar features, namely the parabolic behaviour between $^{40}$Ca and $^{48}$Ca and the steep rise beyond $^{48}$Ca, are missing.
The most important difference between these improved calculations and the ones of Ref.~\cite{Soma20a} relates to the inclusion of the spin-orbit term $\langle r^2 \rangle_{\text{so}}$ in Eq.~\eqref{eq_r_ch}.
This additional correction, not present in Ref.~\cite{Soma20a}, lowers the charge radius of $^{48}$Ca down to roughly the same value as $^{40}$Ca for both interactions, consistently to what observed in experimental data and other ab initio calculations~\cite{GarciaRuiz16}.

\begin{figure*}
\centering
\includegraphics[width=6.6cm]{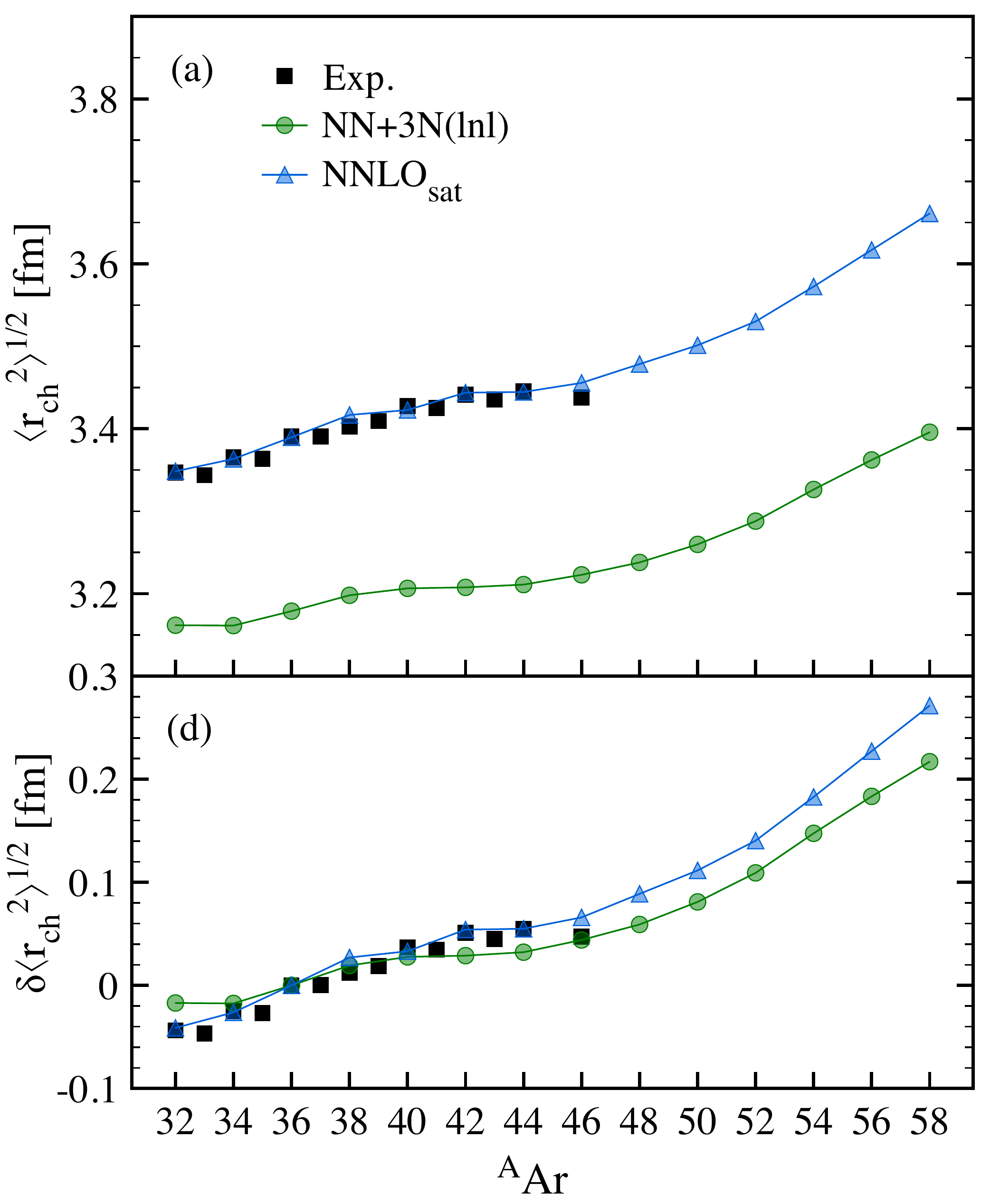}
\hspace{-1.1cm}
\includegraphics[width=6.6cm]{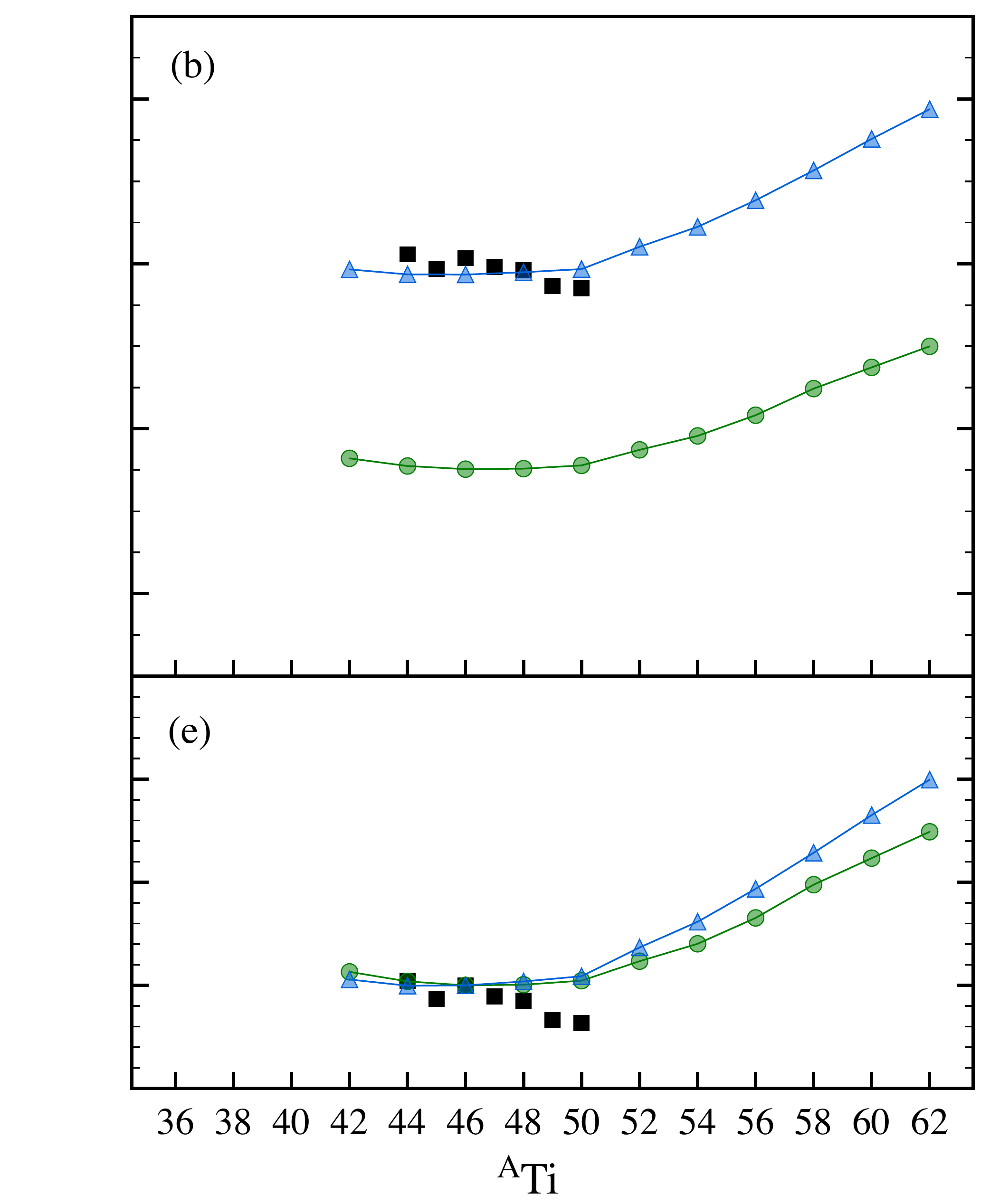}
\hspace{-1.1cm}
\includegraphics[width=6.6cm]{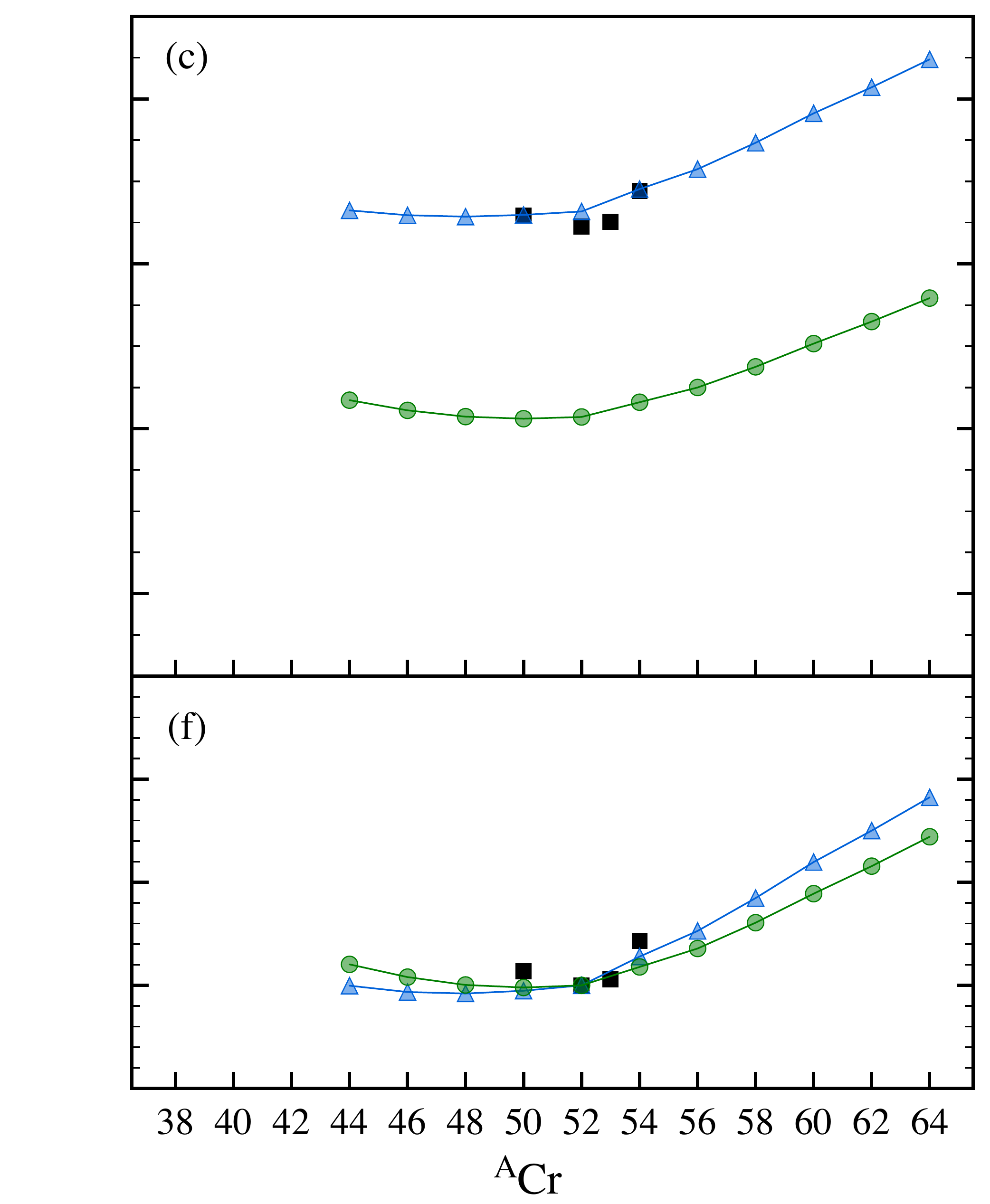}
\caption{Root mean square charge radii of (a) argon, (b) titanium and (c) chromium isotopes computed with the \lnl{} and \sat{} interactions. 
Experimental data are taken from Ref.~\cite{Angeli13}.
Panels (d), (e) and (f) show corresponding differential radii relative to $^{36}$Ar, $^{46}$Ti and $^{52}$Cr respectively.
}
\label{fig_radii_ArTiCr}
\end{figure*}
Let us now move to results for argon, titanium and chromium isotopes, displayed\footnote{Only charge radii of bound isotopes are shown in the following, i.e. in Figs.~\ref{fig_radii_ArTiCr}, \ref{fig_radii_all} and \ref{fig_radii_2832}. While all computed argon and calcium nuclei are found to be bound, titanium and chromium isotopes with $N=14,16,18$ result unbound in the present calculations (with both \lnl{} and \sat{} interactions).} in Fig.~\ref{fig_radii_ArTiCr}.
Globally, the behaviour is similar to the one observed in the calcium chain, with \sat{} calculations very close to experimental data and \lnl{} underestimating experiment by about 5 to 10\%.
In argon, see Fig.~\ref{fig_radii_ArTiCr} (a), charge radii computed with \sat{} reproduce very well existing data, with the notable exception of the most neutron-rich isotope available, $^{46}$Ar.
The trend presents a kink at this nucleus, after which a steady increase with neutron number is observed until $N=34$ where a second kink appears.
Results obtained with \lnl{} follow a similar behaviour past $^{46}$Ar, as one can appreciate by looking at relative charge radii displayed in Fig.~\ref{fig_radii_ArTiCr} (d).
Below $N=28$, however, the \lnl{} slope is somehow different from \sat{} and experimental data.
Experimental points are more scarce for titanium and chromium, with essentially only stable or long-lived isotopes available.
In titanium, Fig.~\ref{fig_radii_ArTiCr} (b) and (e), isotopes with $N=22-26$ are well reproduced by \sat, while $^{50}$Ti is overestimated, similarly to $^{46}$Ar.
\lnl{} follows the same relative trend around stability, with slightly different slopes in the proton- and neutron-rich regions.
Analogous behaviour is observed for chromium, shown in Fig.~\ref{fig_radii_ArTiCr} (c) and (f).
Also in this case the radius of the $N=28$ isotope, $^{52}$Cr, is overestimated by \sat{} calculations, which instead give an excellent reproduction of neighbouring $^{50}$Cr and $^{54}$Cr.
Curves obtained with \lnl{} present the same general features as in the titanium chain.

This qualitative analysis is corroborated by examining in detail the r.m.s. deviations from experiment. 
Absolute r.m.s. deviations for \sat{} are between 0.01 and 0.02 fm, about ten times smaller than the corresponding deviations for \lnl, which amount to 0.2 fm for all isotopic chains.
This clearly confirms the superiority of the \sat{} Hamiltonian for the description of nuclear radii. 
Interestingly, for both interactions r.m.s. deviations are similar in the four isotopic chains.
This indicates that the fingerprints of missing collective correlations might be more subtle for this observable, e.g. showing up in the form of a lack of parabolic behaviour between $N=20$ and $N=28$ even in singly-magic calcium.
R.m.s. deviations of differential radii are substantially independent of the input interaction. 
This points to some global/bulk effect at the origin of the underestimation of nuclear radii when \lnl{} is employed.

To gauge more in detail the quality of the theoretical description, m.s. charge radii along all four isotopic chains are shown together in Fig.~\ref{fig_radii_all}.
By examining available experimental data, one can identify three distinct regions\footnote{Notice that this differentiation also applies to odd-$Z$ chains around calcium and extends up to iron, see Ref.~\cite{GarciaRuiz20}.}:
\begin{enumerate}
\item[a)] Below $N=20$, a steady increase with mild odd-even staggering is observed for calcium and argon.\\
\item[b)] Between $N=20$ and $N=28$, the slope of the experimental trend changes noticeably, going from positive (argon) to null (calcium) and negative (titanium and chromium). Moreover, this is superposed with an inverse parabolic behaviour characterised by a marked odd-even staggering. 
The parabolic trend is weak in argon, but pronounced in calcium and titanium.\\
\item[c)] Above $N=28$, one finds a steep increase with small or even absent signs of odd-even staggering and shell closures.
\end{enumerate}
\begin{figure}
\centering
\includegraphics[width=8.5cm]{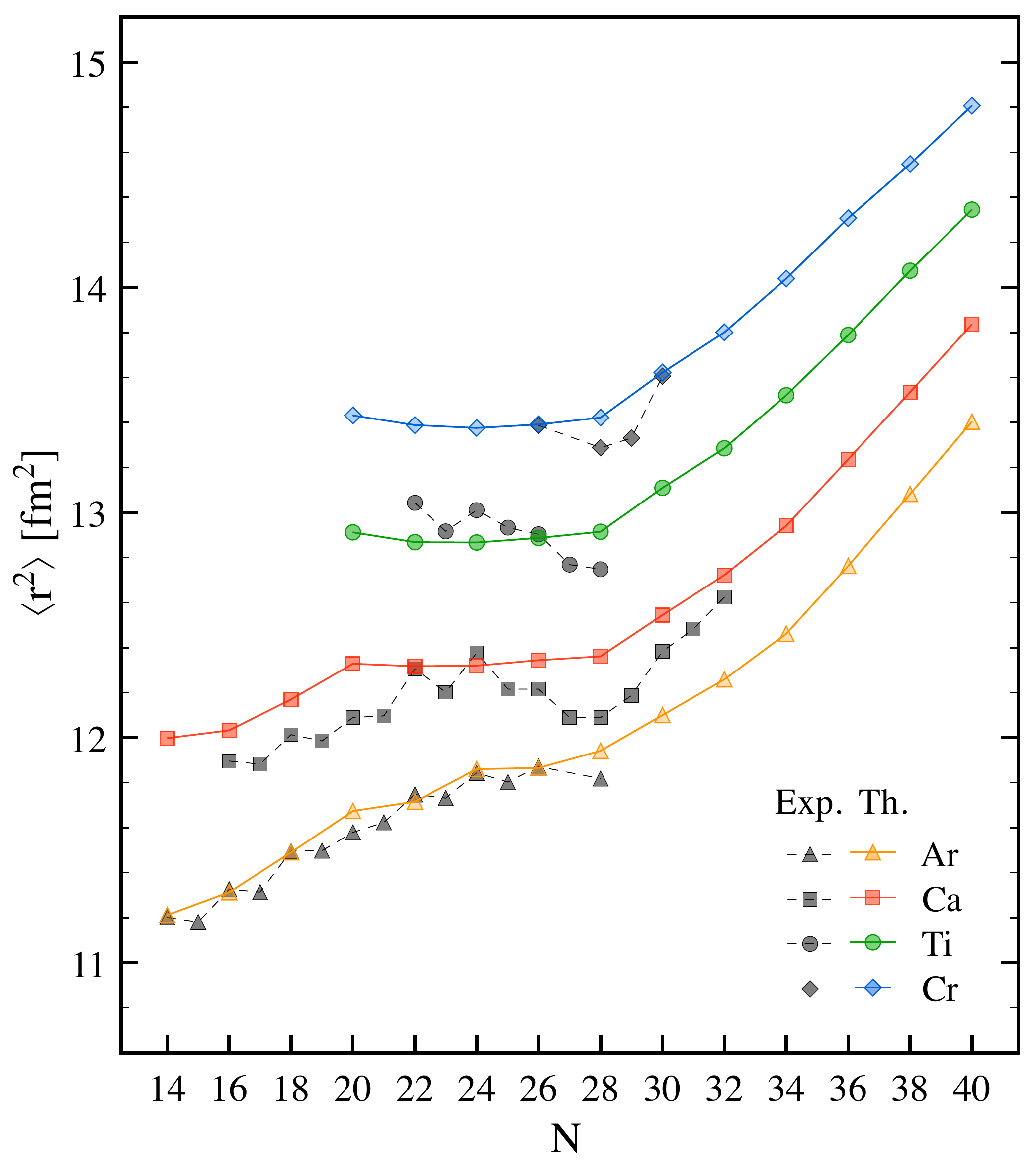}
\caption{Mean square charge radii of argon, calcium, titanium and chromium isotopes computed with the \sat{} interaction (coloured symbols and solid lines) compared to available experimental data~\cite{Angeli13, GarciaRuiz16, Miller19} (dark grey symbols and dashed lines).}
\label{fig_radii_all}
\end{figure}
Computed charge radii do reproduce some but not all of these experimental trends.
Below $N=20$, the steady behaviour is captured by the calculations, although a slight shift is present for calcium.
In the central region, the change in slope from argon to chromium is qualitatively reproduced.
In contrast, the parabolic behaviour is basically absent in all calculated curves. 
Let us note that the charge radii for calcium between $N=20-28$ have been explained in terms of coupling to collective modes in Ref.~\cite{Barranco85} and excitations across the $sd$ and $pf$ orbits using the shell model approach~\cite{Caurier01}. In both cases, quadrupole excitations to (possibly deformed) states are involved. 
The particle-vibration coupling at the origin of this mechanism is encoded in the ADC(3) many-body truncation and in its twin approach, the Faddeev Random phase approximation~\cite{Barbieri07}, which is slightly more sophisticated for collective modes. 
Thus, ADC(3) stands out as the minimum requirement to be able to reproduce the inverted bell behaviour of radii in the central region. However, the above early studies were based on phenomenological interactions. For ab initio applications, it is not clear a priori to what extent the ADC(3) will be sufficient to resolve the low-energy quadrupole deformations with current soft chiral Hamiltonians.
Notice that also EDF calculations have traditionally struggled with the description of this parabolic trend, with the notable exception of Fayans functionals~\cite{Reinhard17, Miller19}.

For all isotopes, the theoretical charge radius at $N=28$ is systematically larger than the measured one. This also affects the slope beyond this point, which results less steep than what observed in experimental data.
This inability to reproduce the pronounced kink at $N=28$ is common to other ab initio calculations as initially discussed in Ref.~\cite{GarciaRuiz16}.
In order to analyse this feature in more details, Fig.~\ref{fig_radii_2832} shows measured and computed m.s. charge radii relative to $N=28$.
The two experimental curves extending beyond $N=28$ do indeed present the same rise towards $N=30$. 
Manganese ($Z=25$) and iron ($Z=26$), for which experimental data are available, also follow this trend.
The same behaviour, with a kink followed by a steep rise essentially independent of $Z$, is found at the $N=50$ and $N=82$ magic numbers~\cite{GarciaRuiz20}.
Remarkably, the theoretical curves capture this basic feature, yielding radii that increase almost independently of $Z$ beyond $N=28$.
As already stressed, however, the slope is less steep than the experimental one, which represents a challenge for most of nuclear structure calculations.
Let us notice that, interestingly, a similar universal behaviour is observed for \lnl{} Hamiltonian, although with a shallower slope than for \sat.
\begin{figure}
\centering
\includegraphics[width=8.5cm]{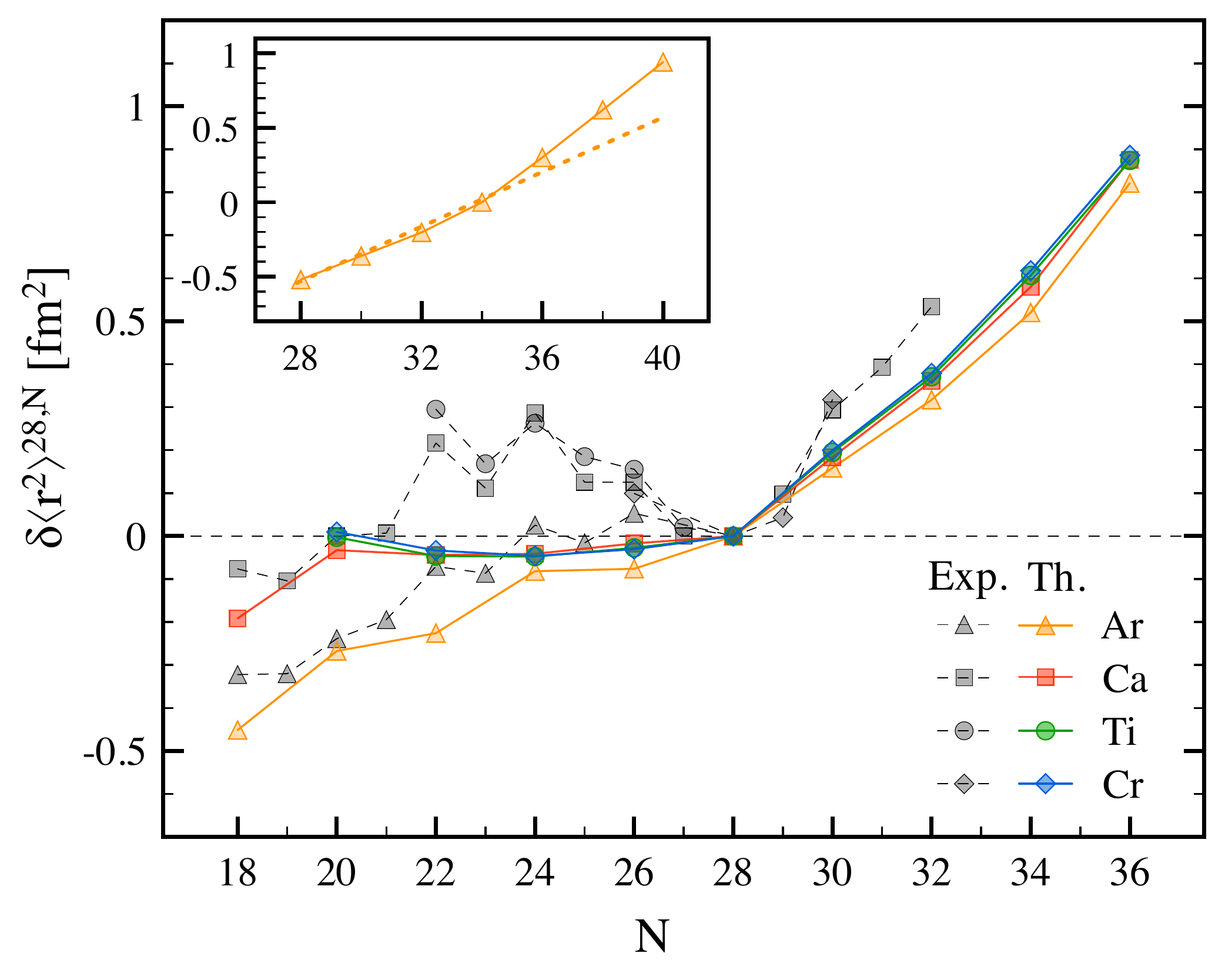}
\caption{Changes in m.s. charge radii for argon, calcium, titanium and chromium relative to $N=28$. 
Results obtained with the \sat{} Hamiltonian (coloured symbols and solid lines) are compared to existing experimental data~\cite{Angeli13, GarciaRuiz16, Miller19} (grey symbols and dashed lines).
In the inset, changes in m.s. charge radii relative to $N=34$ are shown for argon isotopes.
To guide the eye, the linear trend extrapolated from $N=28-32$ is shown as a dashed line.}
\label{fig_radii_2832}
\end{figure}

Furthermore, a second, less pronounced kink is visible at $N=34$, most strongly for argon (see inset of Fig.~\ref{fig_radii_2832}).
The kink fades away with increasing proton number, and is basically absent for chromium that displays a straight linear trend from $N=32$ to $N=40$.
Also in this case, similar features are observed in the charge radii computed with the \lnl{} Hamiltonian.
This behaviour suggests that a (weak) shell closure develops at $N=34$ for neutron-rich nuclei around $Z=20$.
This observation is consistent with the evolution of the $N=34$ neutron gaps computed with \lnl{} and shown in Fig.~\ref{fig_magic}(d).
On the experimental side, the recent measurement of a relatively high value of the $2^+_1$ excitation energy in $^{52}$Ar~\cite{Liu19} and the analysis of quasifree neutron knockout from $^{54}$Ca~\cite{Chen19} also support this picture.

\begin{figure}
\centering
\includegraphics[width=8.5cm]{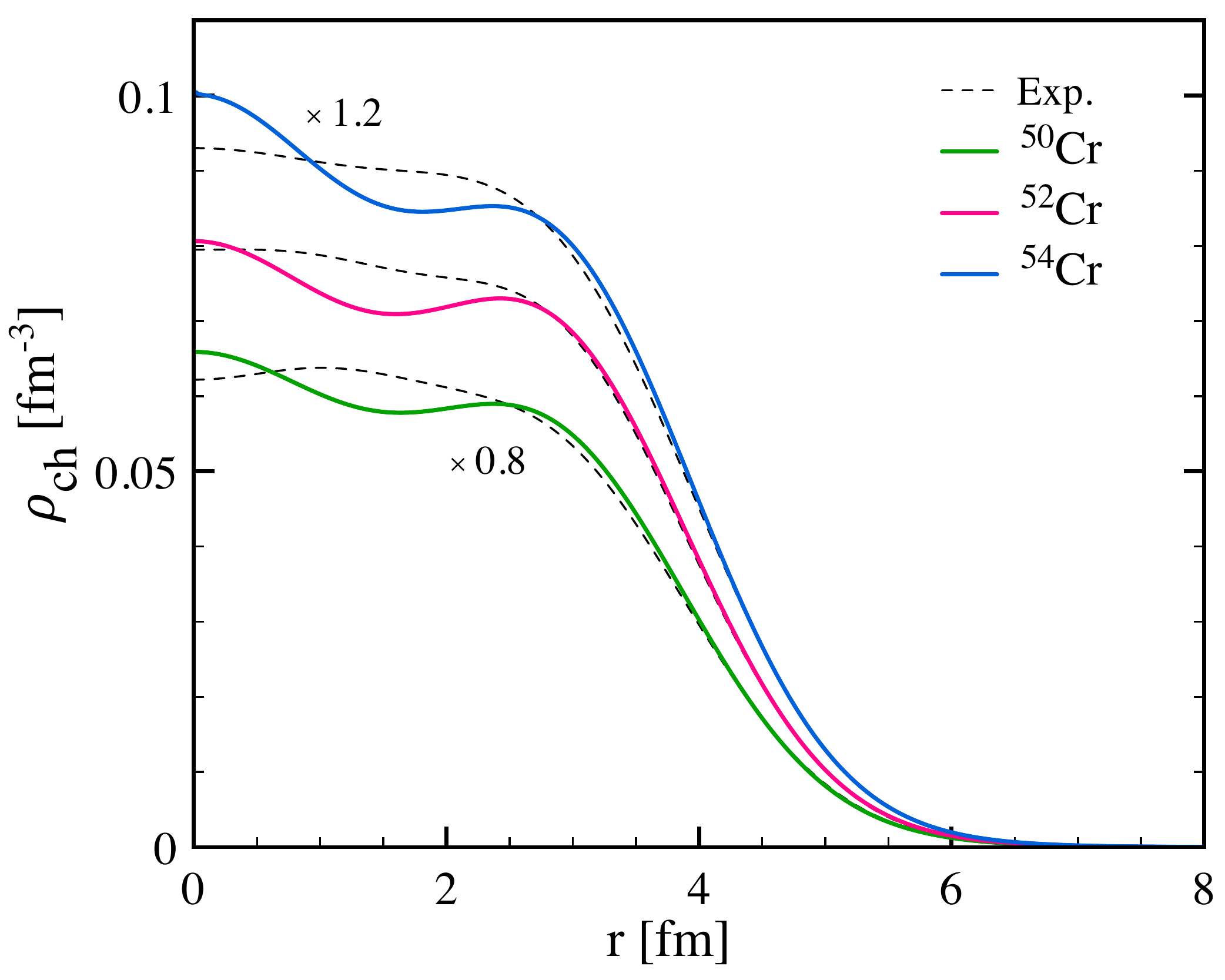}
\caption{Charge density distributions of three chromium isotopes. \sat{} calculations are compared to density profiles determined via electron scattering~\cite{deVries87}. Curves relative to $^{50}$Cr and $^{54}$Cr (both experiment and theory) have been rescaled by a factor 0.8 and 1.2 respectively for better readability.}
\label{fig_densities_Cr}
\end{figure}

To conclude the present section, some examples of charge density distributions in chromium isotopes are shown in Fig.~\ref{fig_densities_Cr}.
Theoretically, the charge distribution is computed as a sum of three terms~\cite{Bertozzi72, Chandra76, Brown79}
\begin{equation}\label{charge_density}
\rho_{{\rm ch}}(r) = \rho_{{\rm ch}}^{{\rm p}}(r) + \rho_{{\rm ch}}^{{\rm n}}(r) + \rho_{{\rm ch}}^{{\rm ls}}(r),
\end{equation}
where $\rho_{{\rm ch}}^{{\rm p}}$ ($\rho_{{\rm ch}}^{{\rm n}}$) is determined by folding the point-proton (point-neutron) density with the finite charge distribution of the proton (neutron) and $\rho_{{\rm ch}}^{{\rm ls}}$ is a relativistic spin-orbit correction.
In addition, centre-of-mass and relativistic Darwin-Foldy corrections are taken into account by employing an effective position variable following Ref.~\cite{Negele70}.
Note that centre-of-mass corrections anyway decrease with increasing mass number. 
Ref.~\cite{Rocco18} used exact Monte Carlo techniques to subtract it from SCGF charge densities obtained with \sat{} and found that it is already under control for $A=16$.
In Fig.~\ref{fig_densities_Cr} distributions of $^{50, 52, 54}$Cr computed with \sat{} are compared to charge profiles determined from electron scattering cross sections~\cite{deVries87}.
Theoretical distributions follow closely the experimental curves in the region around and above $r_{\text{ch}}$.
In contrast, for all three isotopes the behaviour differ in the nuclear interior, with the calculations displaying a dip around 1.5 fm that is not present in the experimental distributions.
The oscillations observed in the theoretical curves are typically interpreted as strong shell effects that have not been washed out by correlations, and are not found in other ADC(2) calculations of spherical nuclei in this mass region (see e.g. $^{40}$Ca in Fig. 15 of Ref.~\cite{Soma20a}).
Therefore this discrepancy could represent another possible signature of missing correlations when the present approach is applied to deformed systems.
Notice that this qualitative behaviour persists for $^{52}$Cr, in spite of the fact that its value of $r_{\text{ch}}$ slightly departs from experiment.

\section{Conclusions}
\label{sec_conclusions}

Correlation expansion methods represent a promising long-term option to simulate the majority, if not all, of atomic nuclei from first principles.
To this purpose, the choice of the reference state, including the use of deformed basis states and the possibility of breaking symmetries, is crucial, notably to account for essential static correlations from the outset.
So far, ab initio approaches have mainly exploited the breaking of U(1) symmetry associated to particle number conservation to account for static pairing correlations.
In the past few years, this strategy has enabled computations of semi-magic, i.e. singly open-shell nuclei, where quadrupole correlations associated to nuclear deformation are typically weak, i.e. predominantly dynamical.
In the present work such U(1)-breaking, SU(2)-conserving calculations are pushed away from semi-magic nuclei in a systematic fashion for the first time.
Results are overall encouraging, with many general experimental features captured by the ab initio simulations.
At the same time, a degradation of the description for certain groups of nuclei signals the inefficient account of (static) quadrupole correlations and calls for a SU(2)-breaking extension of the present theoretical framework.

Bulk nuclear properties, specifically ground-state energies, charge radii and density distributions, were computed along seven isotopic chains around calcium, from argon to chromium. 
Calculations were performed within the Gorkov self-consistent Green's function approach at second order and employed two state-of-the-art two- plus three-nucleon Hamiltonians, \lnl{} and \sat. 
A concise global view of the phenomenological quality of the results can be gained by analysing the r.m.s. deviations from experiment collected in Tab.~\ref{rms}.
\lnl{} results provide a good general description of ground-state energies.
Total binding energies are slightly underestimated, with r.m.s. deviations of about 10-20 MeV depending on the isotopic chain.
For calcium, one finds that going to the next order in the many-body expansion, i.e. ADC(3), brings the r.m.s. deviation from 10.3 MeV down to 2.5 MeV.
This is encouraging in view of being able to provide precise predictions in the future and at the same time underlines the importance of implementing a Gorkov-ADC(3) formalism.

Systematic deviations of ADC(2) on binding energies cancel out to a good extent when computing differential quantities like one- and two-nucleon separation energies and two-neutron shell gaps. 
These energy differences are generally in very good agreement with experiment with r.m.s. deviations of the order of 2 MeV.
In particular, neutron magic numbers $N=28,32,34$ emerge and evolve following experimental trends.
The largest discrepancy with experimental data is found for the $N, Z = 20$ gaps, both overestimated by the calculations.
This impacts the description of the proton dripline, which however remains reasonably reproduced.
In contrast, three-point mass differences along the various isotopic chains evidence that present calculations do not provide sufficient pairing strength. 
The future inclusion of higher-order, e.g. ADC(3), corrections accounting for collective fluctuations might result instrumental for a more accurate description of pairing properties. 
While such computations are routine in Dyson-SCGF (i.e., for closed shells), they become computationally challenging in the Gorkov formalism due to the increase in the number of Bogolyubov mean-field orbits resulting from SU(1)-breaking. 
A full Gorkov-ADC(3) will require improved algorithms, such as importance truncation~\cite{Porro21}, but it is within reach and can be implemented in the foreseeable future.

\begin{table}
\centering
\renewcommand{\arraystretch}{1.4}
\begin{tabular}{|l||c|c|c|c|}
\hline
 & Ar & Ca & Ti & Cr \\
\hline
\hline
\textbf{\textit{NN}+3\textit{N}(lnl)} & \multicolumn{4}{c|}{}  \\
\hline
$E$  [MeV]& 14.1 & 10.3 & 14.2 & 19.2 \\
\hline
$E/A$ [MeV]  & 0.34 & 0.21 & 0.29 & 0.35 \\
\hline
$S_\text{2n}$ [MeV] & 2.90 & 1.56 & 2.05 & 2.22 \\
\hline
$\Delta_\text{2n}$ [MeV] & 3.84 & 1.96 & 2.98 & 2.48 \\
\hline
$\langle r^2_{\text{ch}} \rangle^{1/2}$ [fm] & 0.211 & 0.219 & 0.241 & 0.242 \\
\hline
$\delta \langle r^2_{\text{ch}} \rangle^{1/2}$ [fm] & 0.012 & 0.023 & 0.020 & 0.016 \\
\hline
\hline
\textbf{\sat{}} & \multicolumn{4}{c|}{}  \\
\hline
$\langle r^2_{\text{ch}} \rangle^{1/2}$ [fm] & 0.008 & 0.022 & 0.019 & 0.010 \\
\hline
$\delta \langle r^2_{\text{ch}} \rangle^{1/2}$ [fm] & 0.008 & 0.024 & 0.023 & 0.013 \\
\hline
\end{tabular}
\vspace{.35cm}
\caption{R.m.s. deviation from experimental data of ADC(2) calculations for various observables considered in the present study.
Deviations are reported separately for the four even-$Z$ isotopic chains.
For comparison, the ADC(3) r.m.s. deviations for sub-shell closures amount to 2.5 MeV and 0.06 MeV for $E$ and $E/A$ respectively. 
Differential r.m.s. radii are computed relative to $^{36}$Ar, $^{48}$Ca, $^{46}$Ti and $^{52}$Cr respectively.}
\label{rms}
\end{table}
As remarked in Ref.~\cite{Soma20a}, \lnl{} calculations yield charge radii that underestimate the experimental measurements by about 5 to 10\% throughout all considered chains.
Corresponding infinite-matter calculations, not available at present, would be instrumental to confirm whether this correlates with a poor reproduction of saturation properties.
The deficient description of absolute radii is reflected in r.m.s. deviations that are about ten times larger the ones characterising \sat{} calculations.
Still, relative trends are generally good, which points to some bulk systematic deficiency in the Hamiltonian.
In contrast, \sat{} provides an overall good reproduction of both absolute and relative charge radii with r.m.s. deviations of the order of $0.01-0.02$ fm, of the order of magnitude as the method uncertainties for these observables.
The main experimental trends below $N=20$, between $N=20$ and $N=28$ and above $N=28$ are qualitatively described with the exception of the parabolic behaviour of calcium and titanium.
The largest discrepancy with data is detected for $N=28$ isotopes, whose radius is overestimated in all considered elements.
As a consequence, the steep rise past $N=28$ observed in calcium and chromium is not reproduced to a full extent by the present calculations.
The inability to correctly describe the charge radius difference between $^{48}$Ca and $^{52}$Ca is common to nearly all existing nuclear structure calculations (with the notable exception of Refs.~\cite{Reinhard17, Miller19}) and currently represents a challenge in particular for ab initio approaches.

For some of the doubly open-shell nuclei considered in this study, strong (i.e. static) quadrupole correlations are expected to play an important role and lead to the onset of deformation.
Such correlations are likely to impact the calculated observables but can be hardly accounted for in the current scheme that uses (rotational) symmetry-conserving reference states and incremental extensions of the formalism. 
Indeed, r.m.s. deviations of ground-state energies (both total and per particle) are systematically larger away from singly-magic calcium, reaching their maximum in chromium isotopes.
Moreover, within each isotopic chain, a careful comparison between computed and experimental ground-state energies reveals a clear correlation between the deviation from experiment and the expected degree of deformation (quantified through the deformation parameter $\beta$ obtained via EDF calculations~\cite{Bender06a}).
Remarkably, the same patterns are not observed for charge radii, which display r.m.s. deviations that are roughly independent of the closed/open-shell character.
This points to the fact that effects more complex than the ground-state deformation (e.g. details of the shell structure) play a role in the fine tuning of nuclear sizes.

Different strategies could be envisaged to break through the current limitations of the theoretical method.
Even though the systematic inclusion of higher orders in the ADC($n$) expansion eventually approaches the exact solution of the Schr\"odinger equation, any increase in the ADC($n$) order beyond $n=3$ is, at present, computationally out of reach due to the factorial increase in diagrams and degrees of freedom. 
Besides, such a truncation scheme is unlikely to resolve deformation degrees of freedom until several orders beyond the current capabilities.
Hence alternative routes have to be followed, such as the stochastic sampling of the self-energy or a SU(2)-breaking scheme.
In the first case, one would still work in a standard (spherical or partially deformed) basis but diagrams are summed to very high orders using bold diagrammatic Monte Carlo techniques~\cite{VanHoucke12}. 
This approach is particularly suited to address correlations at medium energies that have been identified as key ingredients to devise ab initio nucleon-nucleus optical potentials~\cite{Idini19}.  
In the second path, the extension towards a SU(2)-breaking scheme would impose nuclear deformation already at the level of the reference state and allow many-body truncations at low ADC($n$) orders, still requiring a final projection on good angular momentum.  
Both approaches will involve sophisticated extensions of the SCGF formalism and will be long-term developments.

\section*{Acknowledgements}

The authors wish to thank M. Frosini for providing the particle-number projected HFB results discussed in Sec.~\ref{sec_setup}, as well as R. Garcia Ruiz and F. Raimondi for useful exchanges.
Calculations were performed by using HPC resources from GENCI-TGCC (Contracts No. A005057392, A007057392) and at the DiRAC Complexity system at the University of Leicester (BIS National E-infrastructure capital grant No. ST/K000373/1 and STFC grant No. ST/K0003259/1). This work was supported by the United Kingdom Science and Technology Facilities Council (STFC) under Grant No. ST/L005816/1 and in part by the NSERC Grant No. SAPIN-2016-00033. TRIUMF receives federal funding via a contribution agreement with the National Research Council of Canada.

\bibliographystyle{apsrev4-1}
\bibliography{newbiblio}

\begin{thebibliography}{94}%
\makeatletter
\providecommand \@ifxundefined [1]{%
 \@ifx{#1\undefined}
}%
\providecommand \@ifnum [1]{%
 \ifnum #1\expandafter \@firstoftwo
 \else \expandafter \@secondoftwo
 \fi
}%
\providecommand \@ifx [1]{%
 \ifx #1\expandafter \@firstoftwo
 \else \expandafter \@secondoftwo
 \fi
}%
\providecommand \natexlab [1]{#1}%
\providecommand \enquote  [1]{``#1''}%
\providecommand \bibnamefont  [1]{#1}%
\providecommand \bibfnamefont [1]{#1}%
\providecommand \citenamefont [1]{#1}%
\providecommand \href@noop [0]{\@secondoftwo}%
\providecommand \href [0]{\begingroup \@sanitize@url \@href}%
\providecommand \@href[1]{\@@startlink{#1}\@@href}%
\providecommand \@@href[1]{\endgroup#1\@@endlink}%
\providecommand \@sanitize@url [0]{\catcode `\\12\catcode `\$12\catcode
  `\&12\catcode `\#12\catcode `\^12\catcode `\_12\catcode `\%12\relax}%
\providecommand \@@startlink[1]{}%
\providecommand \@@endlink[0]{}%
\providecommand \url  [0]{\begingroup\@sanitize@url \@url }%
\providecommand \@url [1]{\endgroup\@href {#1}{\urlprefix }}%
\providecommand \urlprefix  [0]{URL }%
\providecommand \Eprint [0]{\href }%
\providecommand \doibase [0]{http://dx.doi.org/}%
\providecommand \selectlanguage [0]{\@gobble}%
\providecommand \bibinfo  [0]{\@secondoftwo}%
\providecommand \bibfield  [0]{\@secondoftwo}%
\providecommand \translation [1]{[#1]}%
\providecommand \BibitemOpen [0]{}%
\providecommand \bibitemStop [0]{}%
\providecommand \bibitemNoStop [0]{.\EOS\space}%
\providecommand \EOS [0]{\spacefactor3000\relax}%
\providecommand \BibitemShut  [1]{\csname bibitem#1\endcsname}%
\let\auto@bib@innerbib\@empty
\bibitem [{\citenamefont {Dickhoff}\ and\ \citenamefont
  {Barbieri}(2004)}]{Dickhoff04}%
  \BibitemOpen
  \bibfield  {author} {\bibinfo {author} {\bibfnamefont {W.~H.}\ \bibnamefont
  {Dickhoff}}\ and\ \bibinfo {author} {\bibfnamefont {C.}~\bibnamefont
  {Barbieri}},\ }\href {\doibase https://doi.org/10.1016/j.ppnp.2004.02.038}
  {\bibfield  {journal} {\bibinfo  {journal} {Prog. Part. Nucl. Phys.}\
  }\textbf {\bibinfo {volume} {52}},\ \bibinfo {pages} {377} (\bibinfo {year}
  {2004})}\BibitemShut {NoStop}%
\bibitem [{\citenamefont {Kowalski}\ \emph {et~al.}(2004)\citenamefont
  {Kowalski}, \citenamefont {Dean}, \citenamefont {Hjorth-Jensen},
  \citenamefont {Papenbrock},\ and\ \citenamefont {Piecuch}}]{Kowalski04}%
  \BibitemOpen
  \bibfield  {author} {\bibinfo {author} {\bibfnamefont {K.}~\bibnamefont
  {Kowalski}}, \bibinfo {author} {\bibfnamefont {D.~J.}\ \bibnamefont {Dean}},
  \bibinfo {author} {\bibfnamefont {M.}~\bibnamefont {Hjorth-Jensen}}, \bibinfo
  {author} {\bibfnamefont {T.}~\bibnamefont {Papenbrock}}, \ and\ \bibinfo
  {author} {\bibfnamefont {P.}~\bibnamefont {Piecuch}},\ }\href {\doibase
  10.1103/PhysRevLett.92.132501} {\bibfield  {journal} {\bibinfo  {journal}
  {Phys. Rev. Lett.}\ }\textbf {\bibinfo {volume} {92}},\ \bibinfo {pages}
  {132501} (\bibinfo {year} {2004})}\BibitemShut {NoStop}%
\bibitem [{\citenamefont {Bogner}\ \emph {et~al.}(2010)\citenamefont {Bogner},
  \citenamefont {Furnstahl},\ and\ \citenamefont {Schwenk}}]{Bogner10}%
  \BibitemOpen
  \bibfield  {author} {\bibinfo {author} {\bibfnamefont {S.~K.}\ \bibnamefont
  {Bogner}}, \bibinfo {author} {\bibfnamefont {R.~J.}\ \bibnamefont
  {Furnstahl}}, \ and\ \bibinfo {author} {\bibfnamefont {A.}~\bibnamefont
  {Schwenk}},\ }\href {\doibase 10.1016/j.ppnp.2010.03.001} {\bibfield
  {journal} {\bibinfo  {journal} {Prog. Part. Nucl. Phys.}\ }\textbf {\bibinfo
  {volume} {65}},\ \bibinfo {pages} {94} (\bibinfo {year} {2010})}\BibitemShut
  {NoStop}%
\bibitem [{\citenamefont {Binder}\ \emph {et~al.}(2014)\citenamefont {Binder},
  \citenamefont {Langhammer}, \citenamefont {Calci},\ and\ \citenamefont
  {Roth}}]{Binder14}%
  \BibitemOpen
  \bibfield  {author} {\bibinfo {author} {\bibfnamefont {S.}~\bibnamefont
  {Binder}}, \bibinfo {author} {\bibfnamefont {J.}~\bibnamefont {Langhammer}},
  \bibinfo {author} {\bibfnamefont {A.}~\bibnamefont {Calci}}, \ and\ \bibinfo
  {author} {\bibfnamefont {R.}~\bibnamefont {Roth}},\ }\href {\doibase
  https://doi.org/10.1016/j.physletb.2014.07.010} {\bibfield  {journal}
  {\bibinfo  {journal} {Physics Letters B}\ }\textbf {\bibinfo {volume}
  {736}},\ \bibinfo {pages} {119 } (\bibinfo {year} {2014})}\BibitemShut
  {NoStop}%
\bibitem [{\citenamefont {Hergert}\ \emph {et~al.}(2014)\citenamefont
  {Hergert}, \citenamefont {Bogner}, \citenamefont {Morris}, \citenamefont
  {Binder}, \citenamefont {Calci}, \citenamefont {Langhammer},\ and\
  \citenamefont {Roth}}]{Hergert14}%
  \BibitemOpen
  \bibfield  {author} {\bibinfo {author} {\bibfnamefont {H.}~\bibnamefont
  {Hergert}}, \bibinfo {author} {\bibfnamefont {S.~K.}\ \bibnamefont {Bogner}},
  \bibinfo {author} {\bibfnamefont {T.~D.}\ \bibnamefont {Morris}}, \bibinfo
  {author} {\bibfnamefont {S.}~\bibnamefont {Binder}}, \bibinfo {author}
  {\bibfnamefont {A.}~\bibnamefont {Calci}}, \bibinfo {author} {\bibfnamefont
  {J.}~\bibnamefont {Langhammer}}, \ and\ \bibinfo {author} {\bibfnamefont
  {R.}~\bibnamefont {Roth}},\ }\href {\doibase 10.1103/PhysRevC.90.041302}
  {\bibfield  {journal} {\bibinfo  {journal} {Phys. Rev. C}\ }\textbf {\bibinfo
  {volume} {90}},\ \bibinfo {pages} {041302} (\bibinfo {year}
  {2014})}\BibitemShut {NoStop}%
\bibitem [{\citenamefont {Hagen}\ \emph {et~al.}(2016)\citenamefont {Hagen},
  \citenamefont {Jansen},\ and\ \citenamefont {Papenbrock}}]{Hagen16}%
  \BibitemOpen
  \bibfield  {author} {\bibinfo {author} {\bibfnamefont {G.}~\bibnamefont
  {Hagen}}, \bibinfo {author} {\bibfnamefont {G.~R.}\ \bibnamefont {Jansen}}, \
  and\ \bibinfo {author} {\bibfnamefont {T.}~\bibnamefont {Papenbrock}},\
  }\href {\doibase 10.1103/PhysRevLett.117.172501} {\bibfield  {journal}
  {\bibinfo  {journal} {Phys. Rev. Lett.}\ }\textbf {\bibinfo {volume} {117}},\
  \bibinfo {pages} {172501} (\bibinfo {year} {2016})}\BibitemShut {NoStop}%
\bibitem [{\citenamefont {Taniuchi}\ \emph {et~al.}(2019)\citenamefont
  {Taniuchi} \emph {et~al.}}]{Taniuchi19}%
  \BibitemOpen
  \bibfield  {author} {\bibinfo {author} {\bibfnamefont {R.}~\bibnamefont
  {Taniuchi}} \emph {et~al.},\ }\href {\doibase 10.1038/s41586-019-1155-x}
  {\bibfield  {journal} {\bibinfo  {journal} {Nature}\ }\textbf {\bibinfo
  {volume} {569}},\ \bibinfo {pages} {53} (\bibinfo {year} {2019})}\BibitemShut
  {NoStop}%
\bibitem [{\citenamefont {Som\`a}\ \emph {et~al.}(2020)\citenamefont {Som\`a},
  \citenamefont {Navr\'atil}, \citenamefont {Raimondi}, \citenamefont
  {Barbieri},\ and\ \citenamefont {Duguet}}]{Soma20a}%
  \BibitemOpen
  \bibfield  {author} {\bibinfo {author} {\bibfnamefont {V.}~\bibnamefont
  {Som\`a}}, \bibinfo {author} {\bibfnamefont {P.}~\bibnamefont {Navr\'atil}},
  \bibinfo {author} {\bibfnamefont {F.}~\bibnamefont {Raimondi}}, \bibinfo
  {author} {\bibfnamefont {C.}~\bibnamefont {Barbieri}}, \ and\ \bibinfo
  {author} {\bibfnamefont {T.}~\bibnamefont {Duguet}},\ }\href {\doibase
  10.1103/PhysRevC.101.014318} {\bibfield  {journal} {\bibinfo  {journal}
  {Phys. Rev. C}\ }\textbf {\bibinfo {volume} {101}},\ \bibinfo {pages}
  {014318} (\bibinfo {year} {2020})}\BibitemShut {NoStop}%
\bibitem [{\citenamefont {Morris}\ \emph {et~al.}(2018)\citenamefont {Morris},
  \citenamefont {Simonis}, \citenamefont {Stroberg}, \citenamefont {Stumpf},
  \citenamefont {Hagen}, \citenamefont {Holt}, \citenamefont {Jansen},
  \citenamefont {Papenbrock}, \citenamefont {Roth},\ and\ \citenamefont
  {Schwenk}}]{Morris18}%
  \BibitemOpen
  \bibfield  {author} {\bibinfo {author} {\bibfnamefont {T.~D.}\ \bibnamefont
  {Morris}}, \bibinfo {author} {\bibfnamefont {J.}~\bibnamefont {Simonis}},
  \bibinfo {author} {\bibfnamefont {S.~R.}\ \bibnamefont {Stroberg}}, \bibinfo
  {author} {\bibfnamefont {C.}~\bibnamefont {Stumpf}}, \bibinfo {author}
  {\bibfnamefont {G.}~\bibnamefont {Hagen}}, \bibinfo {author} {\bibfnamefont
  {J.~D.}\ \bibnamefont {Holt}}, \bibinfo {author} {\bibfnamefont {G.~R.}\
  \bibnamefont {Jansen}}, \bibinfo {author} {\bibfnamefont {T.}~\bibnamefont
  {Papenbrock}}, \bibinfo {author} {\bibfnamefont {R.}~\bibnamefont {Roth}}, \
  and\ \bibinfo {author} {\bibfnamefont {A.}~\bibnamefont {Schwenk}},\ }\href
  {\doibase 10.1103/PhysRevLett.120.152503} {\bibfield  {journal} {\bibinfo
  {journal} {Phys. Rev. Lett.}\ }\textbf {\bibinfo {volume} {120}},\ \bibinfo
  {pages} {152503} (\bibinfo {year} {2018})}\BibitemShut {NoStop}%
\bibitem [{\citenamefont {Gysbers}\ \emph {et~al.}(2019)\citenamefont {Gysbers}
  \emph {et~al.}}]{Gysbers19}%
  \BibitemOpen
  \bibfield  {author} {\bibinfo {author} {\bibfnamefont {P.}~\bibnamefont
  {Gysbers}} \emph {et~al.},\ }\href {\doibase 10.1038/s41567-019-0450-7}
  {\bibfield  {journal} {\bibinfo  {journal} {Nature Phys.}\ }\textbf {\bibinfo
  {volume} {15}},\ \bibinfo {pages} {428} (\bibinfo {year} {2019})}\BibitemShut
  {NoStop}%
\bibitem [{\citenamefont {Arthuis}\ \emph {et~al.}(2020)\citenamefont
  {Arthuis}, \citenamefont {Barbieri}, \citenamefont {Vorabbi},\ and\
  \citenamefont {Finelli}}]{Arthuis20}%
  \BibitemOpen
  \bibfield  {author} {\bibinfo {author} {\bibfnamefont {P.}~\bibnamefont
  {Arthuis}}, \bibinfo {author} {\bibfnamefont {C.}~\bibnamefont {Barbieri}},
  \bibinfo {author} {\bibfnamefont {M.}~\bibnamefont {Vorabbi}}, \ and\
  \bibinfo {author} {\bibfnamefont {P.}~\bibnamefont {Finelli}},\ }\href
  {\doibase 10.1103/PhysRevLett.125.182501} {\bibfield  {journal} {\bibinfo
  {journal} {Phys. Rev. Lett.}\ }\textbf {\bibinfo {volume} {125}},\ \bibinfo
  {pages} {182501} (\bibinfo {year} {2020})}\BibitemShut {NoStop}%
\bibitem [{\citenamefont {{Rolik}}\ \emph {et~al.}(2003)\citenamefont
  {{Rolik}}, \citenamefont {{Szabados}},\ and\ \citenamefont
  {{Surj{\'a}n}}}]{Rolik03}%
  \BibitemOpen
  \bibfield  {author} {\bibinfo {author} {\bibfnamefont {Z.}~\bibnamefont
  {{Rolik}}}, \bibinfo {author} {\bibfnamefont {A.}~\bibnamefont {{Szabados}}},
  \ and\ \bibinfo {author} {\bibfnamefont {P.~R.}\ \bibnamefont
  {{Surj{\'a}n}}},\ }\href {\doibase 10.1063/1.1584424} {\bibfield  {journal}
  {\bibinfo  {journal} {The Journal of Chemical Physics}\ }\textbf {\bibinfo
  {volume} {119}},\ \bibinfo {pages} {1922} (\bibinfo {year}
  {2003})}\BibitemShut {NoStop}%
\bibitem [{\citenamefont {{Surj{\'a}n}}\ \emph {et~al.}(2004)\citenamefont
  {{Surj{\'a}n}}, \citenamefont {{Rolik}}, \citenamefont {{Szabados}},\ and\
  \citenamefont {{K{\"o}halmi}}}]{Surjan04}%
  \BibitemOpen
  \bibfield  {author} {\bibinfo {author} {\bibfnamefont {P.~R.}\ \bibnamefont
  {{Surj{\'a}n}}}, \bibinfo {author} {\bibfnamefont {Z.}~\bibnamefont
  {{Rolik}}}, \bibinfo {author} {\bibfnamefont {A.}~\bibnamefont {{Szabados}}},
  \ and\ \bibinfo {author} {\bibfnamefont {D.}~\bibnamefont {{K{\"o}halmi}}},\
  }\href {\doibase 10.1002/andp.200310074} {\bibfield  {journal} {\bibinfo
  {journal} {Annalen der Physik}\ }\textbf {\bibinfo {volume} {13}},\ \bibinfo
  {pages} {223} (\bibinfo {year} {2004})}\BibitemShut {NoStop}%
\bibitem [{\citenamefont {Barrett}\ \emph {et~al.}(2013)\citenamefont
  {Barrett}, \citenamefont {Navratil},\ and\ \citenamefont {Vary}}]{Barrett13}%
  \BibitemOpen
  \bibfield  {author} {\bibinfo {author} {\bibfnamefont {B.~R.}\ \bibnamefont
  {Barrett}}, \bibinfo {author} {\bibfnamefont {P.}~\bibnamefont {Navratil}}, \
  and\ \bibinfo {author} {\bibfnamefont {J.~P.}\ \bibnamefont {Vary}},\ }\href
  {\doibase 10.1016/j.ppnp.2012.10.003} {\bibfield  {journal} {\bibinfo
  {journal} {Prog. Part. Nucl. Phys.}\ }\textbf {\bibinfo {volume} {69}},\
  \bibinfo {pages} {131} (\bibinfo {year} {2013})}\BibitemShut {NoStop}%
\bibitem [{\citenamefont {{Gebrerufael}}\ \emph {et~al.}(2017)\citenamefont
  {{Gebrerufael}}, \citenamefont {{Vobig}}, \citenamefont {{Hergert}},\ and\
  \citenamefont {{Roth}}}]{Gebrerufael17}%
  \BibitemOpen
  \bibfield  {author} {\bibinfo {author} {\bibfnamefont {E.}~\bibnamefont
  {{Gebrerufael}}}, \bibinfo {author} {\bibfnamefont {K.}~\bibnamefont
  {{Vobig}}}, \bibinfo {author} {\bibfnamefont {H.}~\bibnamefont {{Hergert}}},
  \ and\ \bibinfo {author} {\bibfnamefont {R.}~\bibnamefont {{Roth}}},\ }\href
  {\doibase 10.1103/PhysRevLett.118.152503} {\bibfield  {journal} {\bibinfo
  {journal} {Phys. Rev. Lett.}\ }\textbf {\bibinfo {volume} {118}},\ \bibinfo
  {eid} {152503} (\bibinfo {year} {2017})}\BibitemShut {NoStop}%
\bibitem [{\citenamefont {Hergert}(2017)}]{Hergert17}%
  \BibitemOpen
  \bibfield  {author} {\bibinfo {author} {\bibfnamefont {H.}~\bibnamefont
  {Hergert}},\ }\href {\doibase 10.1088/1402-4896/92/2/023002} {\bibfield
  {journal} {\bibinfo  {journal} {Phys. Scripta}\ }\textbf {\bibinfo {volume}
  {92}},\ \bibinfo {pages} {023002} (\bibinfo {year} {2017})}\BibitemShut
  {NoStop}%
\bibitem [{\citenamefont {Tichai}\ \emph
  {et~al.}(2018{\natexlab{a}})\citenamefont {Tichai}, \citenamefont
  {Gebrerufael}, \citenamefont {Vobig},\ and\ \citenamefont
  {Roth}}]{Tichai18b}%
  \BibitemOpen
  \bibfield  {author} {\bibinfo {author} {\bibfnamefont {A.}~\bibnamefont
  {Tichai}}, \bibinfo {author} {\bibfnamefont {E.}~\bibnamefont {Gebrerufael}},
  \bibinfo {author} {\bibfnamefont {K.}~\bibnamefont {Vobig}}, \ and\ \bibinfo
  {author} {\bibfnamefont {R.}~\bibnamefont {Roth}},\ }\href {\doibase
  https://doi.org/10.1016/j.physletb.2018.10.029} {\bibfield  {journal}
  {\bibinfo  {journal} {Phys. Lett. B}\ }\textbf {\bibinfo {volume} {786}},\
  \bibinfo {pages} {448 } (\bibinfo {year} {2018}{\natexlab{a}})}\BibitemShut
  {NoStop}%
\bibitem [{\citenamefont {Bogner}\ \emph {et~al.}(2014)\citenamefont {Bogner},
  \citenamefont {Hergert}, \citenamefont {Holt}, \citenamefont {Schwenk},
  \citenamefont {Binder}, \citenamefont {Calci}, \citenamefont {Langhammer},\
  and\ \citenamefont {Roth}}]{Bogner14}%
  \BibitemOpen
  \bibfield  {author} {\bibinfo {author} {\bibfnamefont {S.~K.}\ \bibnamefont
  {Bogner}}, \bibinfo {author} {\bibfnamefont {H.}~\bibnamefont {Hergert}},
  \bibinfo {author} {\bibfnamefont {J.~D.}\ \bibnamefont {Holt}}, \bibinfo
  {author} {\bibfnamefont {A.}~\bibnamefont {Schwenk}}, \bibinfo {author}
  {\bibfnamefont {S.}~\bibnamefont {Binder}}, \bibinfo {author} {\bibfnamefont
  {A.}~\bibnamefont {Calci}}, \bibinfo {author} {\bibfnamefont
  {J.}~\bibnamefont {Langhammer}}, \ and\ \bibinfo {author} {\bibfnamefont
  {R.}~\bibnamefont {Roth}},\ }\href {\doibase 10.1103/PhysRevLett.113.142501}
  {\bibfield  {journal} {\bibinfo  {journal} {Phys. Rev. Lett.}\ }\textbf
  {\bibinfo {volume} {113}},\ \bibinfo {pages} {142501} (\bibinfo {year}
  {2014})}\BibitemShut {NoStop}%
\bibitem [{\citenamefont {Jansen}\ \emph {et~al.}(2014)\citenamefont {Jansen},
  \citenamefont {Engel}, \citenamefont {Hagen}, \citenamefont {Navratil},\ and\
  \citenamefont {Signoracci}}]{Jansen14}%
  \BibitemOpen
  \bibfield  {author} {\bibinfo {author} {\bibfnamefont {G.~R.}\ \bibnamefont
  {Jansen}}, \bibinfo {author} {\bibfnamefont {J.}~\bibnamefont {Engel}},
  \bibinfo {author} {\bibfnamefont {G.}~\bibnamefont {Hagen}}, \bibinfo
  {author} {\bibfnamefont {P.}~\bibnamefont {Navratil}}, \ and\ \bibinfo
  {author} {\bibfnamefont {A.}~\bibnamefont {Signoracci}},\ }\href {\doibase
  10.1103/PhysRevLett.113.142502} {\bibfield  {journal} {\bibinfo  {journal}
  {Phys. Rev. Lett.}\ }\textbf {\bibinfo {volume} {113}},\ \bibinfo {pages}
  {142502} (\bibinfo {year} {2014})}\BibitemShut {NoStop}%
\bibitem [{\citenamefont {Stroberg}\ \emph {et~al.}(2019)\citenamefont
  {Stroberg}, \citenamefont {Bogner}, \citenamefont {Hergert},\ and\
  \citenamefont {Holt}}]{Stroberg19a}%
  \BibitemOpen
  \bibfield  {author} {\bibinfo {author} {\bibfnamefont {S.~R.}\ \bibnamefont
  {Stroberg}}, \bibinfo {author} {\bibfnamefont {S.~K.}\ \bibnamefont
  {Bogner}}, \bibinfo {author} {\bibfnamefont {H.}~\bibnamefont {Hergert}}, \
  and\ \bibinfo {author} {\bibfnamefont {J.~D.}\ \bibnamefont {Holt}},\ }\href
  {\doibase 10.1146/annurev-nucl-101917-021120} {\bibfield  {journal} {\bibinfo
   {journal} {Ann. Rev. Nucl. Part. Sci.}\ }\textbf {\bibinfo {volume} {69}},\
  \bibinfo {pages} {307} (\bibinfo {year} {2019})}\BibitemShut {NoStop}%
\bibitem [{\citenamefont {Duguet}(2014)}]{Duguet14}%
  \BibitemOpen
  \bibfield  {author} {\bibinfo {author} {\bibfnamefont {T.}~\bibnamefont
  {Duguet}},\ }\href {\doibase 10.1088/0954-3899/42/2/025107} {\bibfield
  {journal} {\bibinfo  {journal} {Journal of Physics G: Nuclear and Particle
  Physics}\ }\textbf {\bibinfo {volume} {42}},\ \bibinfo {pages} {025107}
  (\bibinfo {year} {2014})}\BibitemShut {NoStop}%
\bibitem [{\citenamefont {Duguet}\ and\ \citenamefont
  {Signoracci}(2017)}]{Duguet17b}%
  \BibitemOpen
  \bibfield  {author} {\bibinfo {author} {\bibfnamefont {T.}~\bibnamefont
  {Duguet}}\ and\ \bibinfo {author} {\bibfnamefont {A.}~\bibnamefont
  {Signoracci}},\ }\href {http://stacks.iop.org/0954-3899/44/i=1/a=015103}
  {\bibfield  {journal} {\bibinfo  {journal} {J. Phys. G}\ }\textbf {\bibinfo
  {volume} {44}},\ \bibinfo {pages} {015103} (\bibinfo {year}
  {2017})}\BibitemShut {NoStop}%
\bibitem [{\citenamefont {Qiu}\ \emph {et~al.}(2019)\citenamefont {Qiu},
  \citenamefont {Henderson}, \citenamefont {Duguet},\ and\ \citenamefont
  {Scuseria}}]{Qiu18}%
  \BibitemOpen
  \bibfield  {author} {\bibinfo {author} {\bibfnamefont {Y.}~\bibnamefont
  {Qiu}}, \bibinfo {author} {\bibfnamefont {T.~M.}\ \bibnamefont {Henderson}},
  \bibinfo {author} {\bibfnamefont {T.}~\bibnamefont {Duguet}}, \ and\ \bibinfo
  {author} {\bibfnamefont {G.~E.}\ \bibnamefont {Scuseria}},\ }\href {\doibase
  {10.1103/PhysRevC.99.044301}} {\bibfield  {journal} {\bibinfo  {journal}
  {Phys. Rev. C}\ }\textbf {\bibinfo {volume} {99}},\ \bibinfo {pages} {044301}
  (\bibinfo {year} {2019})}\BibitemShut {NoStop}%
\bibitem [{\citenamefont {Bender}\ \emph {et~al.}(2003)\citenamefont {Bender},
  \citenamefont {Heenen},\ and\ \citenamefont {Reinhard}}]{Bender03}%
  \BibitemOpen
  \bibfield  {author} {\bibinfo {author} {\bibfnamefont {M.}~\bibnamefont
  {Bender}}, \bibinfo {author} {\bibfnamefont {P.-H.}\ \bibnamefont {Heenen}},
  \ and\ \bibinfo {author} {\bibfnamefont {P.-G.}\ \bibnamefont {Reinhard}},\
  }\href {\doibase 10.1103/RevModPhys.75.121} {\bibfield  {journal} {\bibinfo
  {journal} {Rev. Mod. Phys.}\ }\textbf {\bibinfo {volume} {75}},\ \bibinfo
  {pages} {121} (\bibinfo {year} {2003})}\BibitemShut {NoStop}%
\bibitem [{\citenamefont {Som\`a}\ \emph {et~al.}(2011)\citenamefont {Som\`a},
  \citenamefont {Duguet},\ and\ \citenamefont {Barbieri}}]{Soma11}%
  \BibitemOpen
  \bibfield  {author} {\bibinfo {author} {\bibfnamefont {V.}~\bibnamefont
  {Som\`a}}, \bibinfo {author} {\bibfnamefont {T.}~\bibnamefont {Duguet}}, \
  and\ \bibinfo {author} {\bibfnamefont {C.}~\bibnamefont {Barbieri}},\ }\href
  {\doibase 10.1103/PhysRevC.84.064317} {\bibfield  {journal} {\bibinfo
  {journal} {Phys. Rev. C}\ }\textbf {\bibinfo {volume} {84}},\ \bibinfo
  {pages} {064317} (\bibinfo {year} {2011})}\BibitemShut {NoStop}%
\bibitem [{\citenamefont {Signoracci}\ \emph {et~al.}(2015)\citenamefont
  {Signoracci}, \citenamefont {Duguet}, \citenamefont {Hagen},\ and\
  \citenamefont {Jansen}}]{Signoracci15}%
  \BibitemOpen
  \bibfield  {author} {\bibinfo {author} {\bibfnamefont {A.}~\bibnamefont
  {Signoracci}}, \bibinfo {author} {\bibfnamefont {T.}~\bibnamefont {Duguet}},
  \bibinfo {author} {\bibfnamefont {G.}~\bibnamefont {Hagen}}, \ and\ \bibinfo
  {author} {\bibfnamefont {G.~R.}\ \bibnamefont {Jansen}},\ }\href {\doibase
  10.1103/PhysRevC.91.064320} {\bibfield  {journal} {\bibinfo  {journal} {Phys.
  Rev. C}\ }\textbf {\bibinfo {volume} {91}},\ \bibinfo {pages} {064320}
  (\bibinfo {year} {2015})}\BibitemShut {NoStop}%
\bibitem [{\citenamefont {Tichai}\ \emph
  {et~al.}(2018{\natexlab{b}})\citenamefont {Tichai}, \citenamefont {Arthuis},
  \citenamefont {Duguet}, \citenamefont {Hergert}, \citenamefont {Som\`a},\
  and\ \citenamefont {Roth}}]{Tichai18a}%
  \BibitemOpen
  \bibfield  {author} {\bibinfo {author} {\bibfnamefont {A.}~\bibnamefont
  {Tichai}}, \bibinfo {author} {\bibfnamefont {P.}~\bibnamefont {Arthuis}},
  \bibinfo {author} {\bibfnamefont {T.}~\bibnamefont {Duguet}}, \bibinfo
  {author} {\bibfnamefont {H.}~\bibnamefont {Hergert}}, \bibinfo {author}
  {\bibfnamefont {V.}~\bibnamefont {Som\`a}}, \ and\ \bibinfo {author}
  {\bibfnamefont {R.}~\bibnamefont {Roth}},\ }\href {\doibase
  10.1016/j.physletb.2018.09.044} {\bibfield  {journal} {\bibinfo  {journal}
  {Phys. Lett. B}\ }\textbf {\bibinfo {volume} {786}},\ \bibinfo {pages} {195}
  (\bibinfo {year} {2018}{\natexlab{b}})}\BibitemShut {NoStop}%
\bibitem [{\citenamefont {Tichai}\ \emph {et~al.}(2020)\citenamefont {Tichai},
  \citenamefont {Roth},\ and\ \citenamefont {Duguet}}]{Tichai20}%
  \BibitemOpen
  \bibfield  {author} {\bibinfo {author} {\bibfnamefont {A.}~\bibnamefont
  {Tichai}}, \bibinfo {author} {\bibfnamefont {R.}~\bibnamefont {Roth}}, \ and\
  \bibinfo {author} {\bibfnamefont {T.}~\bibnamefont {Duguet}},\ }\href
  {\doibase 10.3389/fphy.2020.00164} {\bibfield  {journal} {\bibinfo  {journal}
  {Front. in Phys.}\ }\textbf {\bibinfo {volume} {8}},\ \bibinfo {pages} {164}
  (\bibinfo {year} {2020})}\BibitemShut {NoStop}%
\bibitem [{\citenamefont {Yao}\ \emph {et~al.}(2020)\citenamefont {Yao},
  \citenamefont {Bally}, \citenamefont {Engel}, \citenamefont {Wirth},
  \citenamefont {Rodr\'{\i}guez},\ and\ \citenamefont {Hergert}}]{Yao20}%
  \BibitemOpen
  \bibfield  {author} {\bibinfo {author} {\bibfnamefont {J.~M.}\ \bibnamefont
  {Yao}}, \bibinfo {author} {\bibfnamefont {B.}~\bibnamefont {Bally}}, \bibinfo
  {author} {\bibfnamefont {J.}~\bibnamefont {Engel}}, \bibinfo {author}
  {\bibfnamefont {R.}~\bibnamefont {Wirth}}, \bibinfo {author} {\bibfnamefont
  {T.~R.}\ \bibnamefont {Rodr\'{\i}guez}}, \ and\ \bibinfo {author}
  {\bibfnamefont {H.}~\bibnamefont {Hergert}},\ }\href {\doibase
  10.1103/PhysRevLett.124.232501} {\bibfield  {journal} {\bibinfo  {journal}
  {Phys. Rev. Lett.}\ }\textbf {\bibinfo {volume} {124}},\ \bibinfo {pages}
  {232501} (\bibinfo {year} {2020})}\BibitemShut {NoStop}%
\bibitem [{\citenamefont {Novario}\ \emph {et~al.}(2020)\citenamefont
  {Novario}, \citenamefont {Hagen}, \citenamefont {Jansen},\ and\ \citenamefont
  {Papenbrock}}]{Novario20}%
  \BibitemOpen
  \bibfield  {author} {\bibinfo {author} {\bibfnamefont {S.~J.}\ \bibnamefont
  {Novario}}, \bibinfo {author} {\bibfnamefont {G.}~\bibnamefont {Hagen}},
  \bibinfo {author} {\bibfnamefont {G.~R.}\ \bibnamefont {Jansen}}, \ and\
  \bibinfo {author} {\bibfnamefont {T.}~\bibnamefont {Papenbrock}},\ }\href
  {\doibase 10.1103/PhysRevC.102.051303} {\bibfield  {journal} {\bibinfo
  {journal} {Phys. Rev. C}\ }\textbf {\bibinfo {volume} {102}},\ \bibinfo
  {pages} {051303} (\bibinfo {year} {2020})}\BibitemShut {NoStop}%
\bibitem [{\citenamefont {Hergert}(2020)}]{Hergert20}%
  \BibitemOpen
  \bibfield  {author} {\bibinfo {author} {\bibfnamefont {H.}~\bibnamefont
  {Hergert}},\ }\href {\doibase 10.3389/fphy.2020.00379} {\bibfield  {journal}
  {\bibinfo  {journal} {Front. in Phys.}\ }\textbf {\bibinfo {volume} {8}},\
  \bibinfo {pages} {379} (\bibinfo {year} {2020})}\BibitemShut {NoStop}%
\bibitem [{\citenamefont {Frosini}\ \emph {et~al.}(2021)\citenamefont
  {Frosini}, \citenamefont {Duguet}, \citenamefont {Bally}, \citenamefont
  {Beaujeault-Taudi\`ere}, \citenamefont {Ebran},\ and\ \citenamefont
  {Som\`a}}]{Frosini21}%
  \BibitemOpen
  \bibfield  {author} {\bibinfo {author} {\bibfnamefont {M.}~\bibnamefont
  {Frosini}}, \bibinfo {author} {\bibfnamefont {T.}~\bibnamefont {Duguet}},
  \bibinfo {author} {\bibfnamefont {B.}~\bibnamefont {Bally}}, \bibinfo
  {author} {\bibfnamefont {Y.}~\bibnamefont {Beaujeault-Taudi\`ere}}, \bibinfo
  {author} {\bibfnamefont {J.~P.}\ \bibnamefont {Ebran}}, \ and\ \bibinfo
  {author} {\bibfnamefont {V.}~\bibnamefont {Som\`a}},\ }\href {\doibase
  10.1140/epja/s10050-021-00458-z} {\bibfield  {journal} {\bibinfo  {journal}
  {Eur. Phys. J. A}\ }\textbf {\bibinfo {volume} {57}},\ \bibinfo {pages} {151}
  (\bibinfo {year} {2021})}\BibitemShut {NoStop}%
\bibitem [{\citenamefont {Som\`a}\ \emph {et~al.}(2013)\citenamefont {Som\`a},
  \citenamefont {Barbieri},\ and\ \citenamefont {Duguet}}]{Soma13}%
  \BibitemOpen
  \bibfield  {author} {\bibinfo {author} {\bibfnamefont {V.}~\bibnamefont
  {Som\`a}}, \bibinfo {author} {\bibfnamefont {C.}~\bibnamefont {Barbieri}}, \
  and\ \bibinfo {author} {\bibfnamefont {T.}~\bibnamefont {Duguet}},\ }\href
  {\doibase 10.1103/PhysRevC.87.011303} {\bibfield  {journal} {\bibinfo
  {journal} {Phys. Rev. C}\ }\textbf {\bibinfo {volume} {87}},\ \bibinfo
  {pages} {011303} (\bibinfo {year} {2013})}\BibitemShut {NoStop}%
\bibitem [{\citenamefont {Som\`a}\ \emph
  {et~al.}(2014{\natexlab{a}})\citenamefont {Som\`a}, \citenamefont
  {Cipollone}, \citenamefont {Barbieri}, \citenamefont {Navr\'atil},\ and\
  \citenamefont {Duguet}}]{Soma14b}%
  \BibitemOpen
  \bibfield  {author} {\bibinfo {author} {\bibfnamefont {V.}~\bibnamefont
  {Som\`a}}, \bibinfo {author} {\bibfnamefont {A.}~\bibnamefont {Cipollone}},
  \bibinfo {author} {\bibfnamefont {C.}~\bibnamefont {Barbieri}}, \bibinfo
  {author} {\bibfnamefont {P.}~\bibnamefont {Navr\'atil}}, \ and\ \bibinfo
  {author} {\bibfnamefont {T.}~\bibnamefont {Duguet}},\ }\href {\doibase
  10.1103/PhysRevC.89.061301} {\bibfield  {journal} {\bibinfo  {journal} {Phys.
  Rev. C}\ }\textbf {\bibinfo {volume} {89}},\ \bibinfo {pages} {061301}
  (\bibinfo {year} {2014}{\natexlab{a}})}\BibitemShut {NoStop}%
\bibitem [{\citenamefont {Lapoux}\ \emph {et~al.}(2016)\citenamefont {Lapoux},
  \citenamefont {Som\`a}, \citenamefont {Barbieri}, \citenamefont {Hergert},
  \citenamefont {Holt},\ and\ \citenamefont {Stroberg}}]{Lapoux16}%
  \BibitemOpen
  \bibfield  {author} {\bibinfo {author} {\bibfnamefont {V.}~\bibnamefont
  {Lapoux}}, \bibinfo {author} {\bibfnamefont {V.}~\bibnamefont {Som\`a}},
  \bibinfo {author} {\bibfnamefont {C.}~\bibnamefont {Barbieri}}, \bibinfo
  {author} {\bibfnamefont {H.}~\bibnamefont {Hergert}}, \bibinfo {author}
  {\bibfnamefont {J.~D.}\ \bibnamefont {Holt}}, \ and\ \bibinfo {author}
  {\bibfnamefont {S.~R.}\ \bibnamefont {Stroberg}},\ }\href {\doibase
  10.1103/PhysRevLett.117.052501} {\bibfield  {journal} {\bibinfo  {journal}
  {Phys. Rev. Lett.}\ }\textbf {\bibinfo {volume} {117}},\ \bibinfo {pages}
  {052501} (\bibinfo {year} {2016})}\BibitemShut {NoStop}%
\bibitem [{\citenamefont {Som\`a}(2020)}]{Soma20b}%
  \BibitemOpen
  \bibfield  {author} {\bibinfo {author} {\bibfnamefont {V.}~\bibnamefont
  {Som\`a}},\ }\href {\doibase 10.3389/fphy.2020.00340} {\bibfield  {journal}
  {\bibinfo  {journal} {Front. in Phys.}\ }\textbf {\bibinfo {volume} {8}},\
  \bibinfo {pages} {340} (\bibinfo {year} {2020})}\BibitemShut {NoStop}%
\bibitem [{\citenamefont {Leistenschneider}\ \emph {et~al.}(2018)\citenamefont
  {Leistenschneider}, \citenamefont {Reiter}, \citenamefont {Ayet
  San~Andr\'es}, \citenamefont {Kootte}, \citenamefont {Holt}, \citenamefont
  {Navr\'atil}, \citenamefont {Babcock}, \citenamefont {Barbieri},
  \citenamefont {Barquest}, \citenamefont {Bergmann}, \citenamefont {Bollig},
  \citenamefont {Brunner}, \citenamefont {Dunling}, \citenamefont {Finlay},
  \citenamefont {Geissel}, \citenamefont {Graham}, \citenamefont {Greiner},
  \citenamefont {Hergert}, \citenamefont {Hornung}, \citenamefont {Jesch},
  \citenamefont {Klawitter}, \citenamefont {Lan}, \citenamefont {Lascar},
  \citenamefont {Leach}, \citenamefont {Lippert}, \citenamefont {McKay},
  \citenamefont {Paul}, \citenamefont {Schwenk}, \citenamefont {Short},
  \citenamefont {Simonis}, \citenamefont {Som\`a}, \citenamefont
  {Steinbr\"ugge}, \citenamefont {Stroberg}, \citenamefont {Thompson},
  \citenamefont {Wieser}, \citenamefont {Will}, \citenamefont {Yavor},
  \citenamefont {Andreoiu}, \citenamefont {Dickel}, \citenamefont {Dillmann},
  \citenamefont {Gwinner}, \citenamefont {Pla\ss{}}, \citenamefont
  {Scheidenberger}, \citenamefont {Kwiatkowski},\ and\ \citenamefont
  {Dilling}}]{Leistenschneider18}%
  \BibitemOpen
  \bibfield  {author} {\bibinfo {author} {\bibfnamefont {E.}~\bibnamefont
  {Leistenschneider}}, \bibinfo {author} {\bibfnamefont {M.~P.}\ \bibnamefont
  {Reiter}}, \bibinfo {author} {\bibfnamefont {S.}~\bibnamefont {Ayet
  San~Andr\'es}}, \bibinfo {author} {\bibfnamefont {B.}~\bibnamefont {Kootte}},
  \bibinfo {author} {\bibfnamefont {J.~D.}\ \bibnamefont {Holt}}, \bibinfo
  {author} {\bibfnamefont {P.}~\bibnamefont {Navr\'atil}}, \bibinfo {author}
  {\bibfnamefont {C.}~\bibnamefont {Babcock}}, \bibinfo {author} {\bibfnamefont
  {C.}~\bibnamefont {Barbieri}}, \bibinfo {author} {\bibfnamefont {B.~R.}\
  \bibnamefont {Barquest}}, \bibinfo {author} {\bibfnamefont {J.}~\bibnamefont
  {Bergmann}}, \bibinfo {author} {\bibfnamefont {J.}~\bibnamefont {Bollig}},
  \bibinfo {author} {\bibfnamefont {T.}~\bibnamefont {Brunner}}, \bibinfo
  {author} {\bibfnamefont {E.}~\bibnamefont {Dunling}}, \bibinfo {author}
  {\bibfnamefont {A.}~\bibnamefont {Finlay}}, \bibinfo {author} {\bibfnamefont
  {H.}~\bibnamefont {Geissel}}, \bibinfo {author} {\bibfnamefont
  {L.}~\bibnamefont {Graham}}, \bibinfo {author} {\bibfnamefont
  {F.}~\bibnamefont {Greiner}}, \bibinfo {author} {\bibfnamefont
  {H.}~\bibnamefont {Hergert}}, \bibinfo {author} {\bibfnamefont
  {C.}~\bibnamefont {Hornung}}, \bibinfo {author} {\bibfnamefont
  {C.}~\bibnamefont {Jesch}}, \bibinfo {author} {\bibfnamefont
  {R.}~\bibnamefont {Klawitter}}, \bibinfo {author} {\bibfnamefont
  {Y.}~\bibnamefont {Lan}}, \bibinfo {author} {\bibfnamefont {D.}~\bibnamefont
  {Lascar}}, \bibinfo {author} {\bibfnamefont {K.~G.}\ \bibnamefont {Leach}},
  \bibinfo {author} {\bibfnamefont {W.}~\bibnamefont {Lippert}}, \bibinfo
  {author} {\bibfnamefont {J.~E.}\ \bibnamefont {McKay}}, \bibinfo {author}
  {\bibfnamefont {S.~F.}\ \bibnamefont {Paul}}, \bibinfo {author}
  {\bibfnamefont {A.}~\bibnamefont {Schwenk}}, \bibinfo {author} {\bibfnamefont
  {D.}~\bibnamefont {Short}}, \bibinfo {author} {\bibfnamefont
  {J.}~\bibnamefont {Simonis}}, \bibinfo {author} {\bibfnamefont
  {V.}~\bibnamefont {Som\`a}}, \bibinfo {author} {\bibfnamefont
  {R.}~\bibnamefont {Steinbr\"ugge}}, \bibinfo {author} {\bibfnamefont {S.~R.}\
  \bibnamefont {Stroberg}}, \bibinfo {author} {\bibfnamefont {R.}~\bibnamefont
  {Thompson}}, \bibinfo {author} {\bibfnamefont {M.~E.}\ \bibnamefont
  {Wieser}}, \bibinfo {author} {\bibfnamefont {C.}~\bibnamefont {Will}},
  \bibinfo {author} {\bibfnamefont {M.}~\bibnamefont {Yavor}}, \bibinfo
  {author} {\bibfnamefont {C.}~\bibnamefont {Andreoiu}}, \bibinfo {author}
  {\bibfnamefont {T.}~\bibnamefont {Dickel}}, \bibinfo {author} {\bibfnamefont
  {I.}~\bibnamefont {Dillmann}}, \bibinfo {author} {\bibfnamefont
  {G.}~\bibnamefont {Gwinner}}, \bibinfo {author} {\bibfnamefont {W.~R.}\
  \bibnamefont {Pla\ss{}}}, \bibinfo {author} {\bibfnamefont {C.}~\bibnamefont
  {Scheidenberger}}, \bibinfo {author} {\bibfnamefont {A.~A.}\ \bibnamefont
  {Kwiatkowski}}, \ and\ \bibinfo {author} {\bibfnamefont {J.}~\bibnamefont
  {Dilling}},\ }\href {\doibase 10.1103/PhysRevLett.120.062503} {\bibfield
  {journal} {\bibinfo  {journal} {Phys. Rev. Lett.}\ }\textbf {\bibinfo
  {volume} {120}},\ \bibinfo {pages} {062503} (\bibinfo {year}
  {2018})}\BibitemShut {NoStop}%
\bibitem [{\citenamefont {Barbieri}\ \emph {et~al.}(2019)\citenamefont
  {Barbieri}, \citenamefont {Rocco},\ and\ \citenamefont
  {Som\`a}}]{Barbieri19}%
  \BibitemOpen
  \bibfield  {author} {\bibinfo {author} {\bibfnamefont {C.}~\bibnamefont
  {Barbieri}}, \bibinfo {author} {\bibfnamefont {N.}~\bibnamefont {Rocco}}, \
  and\ \bibinfo {author} {\bibfnamefont {V.}~\bibnamefont {Som\`a}},\ }\href
  {\doibase 10.1103/PhysRevC.100.062501} {\bibfield  {journal} {\bibinfo
  {journal} {Phys. Rev. C}\ }\textbf {\bibinfo {volume} {100}},\ \bibinfo
  {pages} {062501} (\bibinfo {year} {2019})}\BibitemShut {NoStop}%
\bibitem [{\citenamefont {Mougeot}\ \emph {et~al.}(2020)\citenamefont
  {Mougeot}, \citenamefont {Atanasov}, \citenamefont {Barbieri}, \citenamefont
  {Blaum}, \citenamefont {Breitenfeld}, \citenamefont {de~Roubin},
  \citenamefont {Duguet}, \citenamefont {George}, \citenamefont {Herfurth},
  \citenamefont {Herlert}, \citenamefont {Holt}, \citenamefont {Karthein},
  \citenamefont {Lunney}, \citenamefont {Manea}, \citenamefont {Navr\'atil},
  \citenamefont {Neidherr}, \citenamefont {Rosenbusch}, \citenamefont
  {Schweikhard}, \citenamefont {Schwenk}, \citenamefont {Som\`a}, \citenamefont
  {Welker}, \citenamefont {Wienholtz}, \citenamefont {Wolf},\ and\
  \citenamefont {Zuber}}]{Mougeot20}%
  \BibitemOpen
  \bibfield  {author} {\bibinfo {author} {\bibfnamefont {M.}~\bibnamefont
  {Mougeot}}, \bibinfo {author} {\bibfnamefont {D.}~\bibnamefont {Atanasov}},
  \bibinfo {author} {\bibfnamefont {C.}~\bibnamefont {Barbieri}}, \bibinfo
  {author} {\bibfnamefont {K.}~\bibnamefont {Blaum}}, \bibinfo {author}
  {\bibfnamefont {M.}~\bibnamefont {Breitenfeld}}, \bibinfo {author}
  {\bibfnamefont {A.}~\bibnamefont {de~Roubin}}, \bibinfo {author}
  {\bibfnamefont {T.}~\bibnamefont {Duguet}}, \bibinfo {author} {\bibfnamefont
  {S.}~\bibnamefont {George}}, \bibinfo {author} {\bibfnamefont
  {F.}~\bibnamefont {Herfurth}}, \bibinfo {author} {\bibfnamefont
  {A.}~\bibnamefont {Herlert}}, \bibinfo {author} {\bibfnamefont {J.~D.}\
  \bibnamefont {Holt}}, \bibinfo {author} {\bibfnamefont {J.}~\bibnamefont
  {Karthein}}, \bibinfo {author} {\bibfnamefont {D.}~\bibnamefont {Lunney}},
  \bibinfo {author} {\bibfnamefont {V.}~\bibnamefont {Manea}}, \bibinfo
  {author} {\bibfnamefont {P.}~\bibnamefont {Navr\'atil}}, \bibinfo {author}
  {\bibfnamefont {D.}~\bibnamefont {Neidherr}}, \bibinfo {author}
  {\bibfnamefont {M.}~\bibnamefont {Rosenbusch}}, \bibinfo {author}
  {\bibfnamefont {L.}~\bibnamefont {Schweikhard}}, \bibinfo {author}
  {\bibfnamefont {A.}~\bibnamefont {Schwenk}}, \bibinfo {author} {\bibfnamefont
  {V.}~\bibnamefont {Som\`a}}, \bibinfo {author} {\bibfnamefont
  {A.}~\bibnamefont {Welker}}, \bibinfo {author} {\bibfnamefont
  {F.}~\bibnamefont {Wienholtz}}, \bibinfo {author} {\bibfnamefont {R.~N.}\
  \bibnamefont {Wolf}}, \ and\ \bibinfo {author} {\bibfnamefont
  {K.}~\bibnamefont {Zuber}},\ }\href {\doibase 10.1103/PhysRevC.102.014301}
  {\bibfield  {journal} {\bibinfo  {journal} {Phys. Rev. C}\ }\textbf {\bibinfo
  {volume} {102}},\ \bibinfo {pages} {014301} (\bibinfo {year}
  {2020})}\BibitemShut {NoStop}%
\bibitem [{\citenamefont {Ekstr\"om}\ \emph {et~al.}(2015)\citenamefont
  {Ekstr\"om}, \citenamefont {Jansen}, \citenamefont {Wendt}, \citenamefont
  {Hagen}, \citenamefont {Papenbrock}, \citenamefont {Carlsson}, \citenamefont
  {Forss\'en}, \citenamefont {Hjorth-Jensen}, \citenamefont {Navr\'atil},\ and\
  \citenamefont {Nazarewicz}}]{Ekstrom15}%
  \BibitemOpen
  \bibfield  {author} {\bibinfo {author} {\bibfnamefont {A.}~\bibnamefont
  {Ekstr\"om}}, \bibinfo {author} {\bibfnamefont {G.~R.}\ \bibnamefont
  {Jansen}}, \bibinfo {author} {\bibfnamefont {K.~A.}\ \bibnamefont {Wendt}},
  \bibinfo {author} {\bibfnamefont {G.}~\bibnamefont {Hagen}}, \bibinfo
  {author} {\bibfnamefont {T.}~\bibnamefont {Papenbrock}}, \bibinfo {author}
  {\bibfnamefont {B.~D.}\ \bibnamefont {Carlsson}}, \bibinfo {author}
  {\bibfnamefont {C.}~\bibnamefont {Forss\'en}}, \bibinfo {author}
  {\bibfnamefont {M.}~\bibnamefont {Hjorth-Jensen}}, \bibinfo {author}
  {\bibfnamefont {P.}~\bibnamefont {Navr\'atil}}, \ and\ \bibinfo {author}
  {\bibfnamefont {W.}~\bibnamefont {Nazarewicz}},\ }\href {\doibase
  10.1103/PhysRevC.91.051301} {\bibfield  {journal} {\bibinfo  {journal} {Phys.
  Rev. C}\ }\textbf {\bibinfo {volume} {91}},\ \bibinfo {pages} {051301}
  (\bibinfo {year} {2015})}\BibitemShut {NoStop}%
\bibitem [{\citenamefont {Som\`a}\ \emph
  {et~al.}(2014{\natexlab{b}})\citenamefont {Som\`a}, \citenamefont
  {Barbieri},\ and\ \citenamefont {Duguet}}]{Soma14a}%
  \BibitemOpen
  \bibfield  {author} {\bibinfo {author} {\bibfnamefont {V.}~\bibnamefont
  {Som\`a}}, \bibinfo {author} {\bibfnamefont {C.}~\bibnamefont {Barbieri}}, \
  and\ \bibinfo {author} {\bibfnamefont {T.}~\bibnamefont {Duguet}},\ }\href
  {\doibase 10.1103/PhysRevC.89.024323} {\bibfield  {journal} {\bibinfo
  {journal} {Phys. Rev. C}\ }\textbf {\bibinfo {volume} {89}},\ \bibinfo
  {pages} {024323} (\bibinfo {year} {2014}{\natexlab{b}})}\BibitemShut
  {NoStop}%
\bibitem [{\citenamefont {Entem}\ and\ \citenamefont
  {Machleidt}(2003)}]{Entem03}%
  \BibitemOpen
  \bibfield  {author} {\bibinfo {author} {\bibfnamefont {D.~R.}\ \bibnamefont
  {Entem}}\ and\ \bibinfo {author} {\bibfnamefont {R.}~\bibnamefont
  {Machleidt}},\ }\href {\doibase 10.1103/PhysRevC.68.041001} {\bibfield
  {journal} {\bibinfo  {journal} {Phys. Rev. C}\ }\textbf {\bibinfo {volume}
  {68}},\ \bibinfo {pages} {041001} (\bibinfo {year} {2003})}\BibitemShut
  {NoStop}%
\bibitem [{\citenamefont {Machleidt}\ and\ \citenamefont
  {Entem}(2011)}]{Machleidt11}%
  \BibitemOpen
  \bibfield  {author} {\bibinfo {author} {\bibfnamefont {R.}~\bibnamefont
  {Machleidt}}\ and\ \bibinfo {author} {\bibfnamefont {D.}~\bibnamefont
  {Entem}},\ }\href {\doibase https://doi.org/10.1016/j.physrep.2011.02.001}
  {\bibfield  {journal} {\bibinfo  {journal} {Physics Reports}\ }\textbf
  {\bibinfo {volume} {503}},\ \bibinfo {pages} {1 } (\bibinfo {year}
  {2011})}\BibitemShut {NoStop}%
\bibitem [{\citenamefont {Carbone}\ \emph {et~al.}(2013)\citenamefont
  {Carbone}, \citenamefont {Cipollone}, \citenamefont {Barbieri}, \citenamefont
  {Rios},\ and\ \citenamefont {Polls}}]{Carbone13}%
  \BibitemOpen
  \bibfield  {author} {\bibinfo {author} {\bibfnamefont {A.}~\bibnamefont
  {Carbone}}, \bibinfo {author} {\bibfnamefont {A.}~\bibnamefont {Cipollone}},
  \bibinfo {author} {\bibfnamefont {C.}~\bibnamefont {Barbieri}}, \bibinfo
  {author} {\bibfnamefont {A.}~\bibnamefont {Rios}}, \ and\ \bibinfo {author}
  {\bibfnamefont {A.}~\bibnamefont {Polls}},\ }\href {\doibase
  10.1103/PhysRevC.88.054326} {\bibfield  {journal} {\bibinfo  {journal} {Phys.
  Rev. C}\ }\textbf {\bibinfo {volume} {88}},\ \bibinfo {pages} {054326}
  (\bibinfo {year} {2013})}\BibitemShut {NoStop}%
\bibitem [{\citenamefont {Cipollone}\ \emph {et~al.}(2015)\citenamefont
  {Cipollone}, \citenamefont {Barbieri},\ and\ \citenamefont
  {Navr\'atil}}]{Cipollone15}%
  \BibitemOpen
  \bibfield  {author} {\bibinfo {author} {\bibfnamefont {A.}~\bibnamefont
  {Cipollone}}, \bibinfo {author} {\bibfnamefont {C.}~\bibnamefont {Barbieri}},
  \ and\ \bibinfo {author} {\bibfnamefont {P.}~\bibnamefont {Navr\'atil}},\
  }\href {\doibase 10.1103/PhysRevC.92.014306} {\bibfield  {journal} {\bibinfo
  {journal} {Phys. Rev. C}\ }\textbf {\bibinfo {volume} {92}},\ \bibinfo
  {pages} {014306} (\bibinfo {year} {2015})}\BibitemShut {NoStop}%
\bibitem [{\citenamefont {Cipollone}\ \emph {et~al.}(2013)\citenamefont
  {Cipollone}, \citenamefont {Barbieri},\ and\ \citenamefont
  {Navr\'atil}}]{Cipollone13}%
  \BibitemOpen
  \bibfield  {author} {\bibinfo {author} {\bibfnamefont {A.}~\bibnamefont
  {Cipollone}}, \bibinfo {author} {\bibfnamefont {C.}~\bibnamefont {Barbieri}},
  \ and\ \bibinfo {author} {\bibfnamefont {P.}~\bibnamefont {Navr\'atil}},\
  }\href {\doibase 10.1103/PhysRevLett.111.062501} {\bibfield  {journal}
  {\bibinfo  {journal} {Phys. Rev. Lett.}\ }\textbf {\bibinfo {volume} {111}},\
  \bibinfo {pages} {062501} (\bibinfo {year} {2013})}\BibitemShut {NoStop}%
\bibitem [{\citenamefont {Barbieri}(2014)}]{Barbieri14}%
  \BibitemOpen
  \bibfield  {author} {\bibinfo {author} {\bibfnamefont {C.}~\bibnamefont
  {Barbieri}},\ }\href {\doibase 10.1088/1742-6596/529/1/012005} {\bibfield
  {journal} {\bibinfo  {journal} {Journal of Physics: Conference Series}\
  }\textbf {\bibinfo {volume} {529}},\ \bibinfo {pages} {012005} (\bibinfo
  {year} {2014})}\BibitemShut {NoStop}%
\bibitem [{\citenamefont {Raimondi}\ and\ \citenamefont
  {Barbieri}(2018)}]{Raimondi18}%
  \BibitemOpen
  \bibfield  {author} {\bibinfo {author} {\bibfnamefont {F.}~\bibnamefont
  {Raimondi}}\ and\ \bibinfo {author} {\bibfnamefont {C.}~\bibnamefont
  {Barbieri}},\ }\href {\doibase 10.1103/PhysRevC.97.054308} {\bibfield
  {journal} {\bibinfo  {journal} {Phys. Rev. C}\ }\textbf {\bibinfo {volume}
  {97}},\ \bibinfo {pages} {054308} (\bibinfo {year} {2018})}\BibitemShut
  {NoStop}%
\bibitem [{\citenamefont {Ripoche}\ \emph {et~al.}(2020)\citenamefont
  {Ripoche}, \citenamefont {Tichai},\ and\ \citenamefont {Duguet}}]{Ripoche20}%
  \BibitemOpen
  \bibfield  {author} {\bibinfo {author} {\bibfnamefont {J.}~\bibnamefont
  {Ripoche}}, \bibinfo {author} {\bibfnamefont {A.}~\bibnamefont {Tichai}}, \
  and\ \bibinfo {author} {\bibfnamefont {T.}~\bibnamefont {Duguet}},\ }\href
  {\doibase 10.1140/epja/s10050-020-00045-8} {\bibfield  {journal} {\bibinfo
  {journal} {Eur. Phys. J. A}\ }\textbf {\bibinfo {volume} {56}},\ \bibinfo
  {pages} {40} (\bibinfo {year} {2020})}\BibitemShut {NoStop}%
\bibitem [{\citenamefont {Frosini}(2021)}]{FrosiniPvt}%
  \BibitemOpen
  \bibfield  {author} {\bibinfo {author} {\bibfnamefont {M.}~\bibnamefont
  {Frosini}},\ }\href@noop {} {} (\bibinfo {year} {2021}),\ \bibinfo {note}
  {private communication}\BibitemShut {NoStop}%
\bibitem [{\citenamefont {Miyagi}\ \emph {et~al.}(2019)\citenamefont {Miyagi},
  \citenamefont {Abe}, \citenamefont {Kohno}, \citenamefont {Navr\'atil},
  \citenamefont {Okamoto}, \citenamefont {Otsuka}, \citenamefont {Shimizu},\
  and\ \citenamefont {Stroberg}}]{Miyagi19}%
  \BibitemOpen
  \bibfield  {author} {\bibinfo {author} {\bibfnamefont {T.}~\bibnamefont
  {Miyagi}}, \bibinfo {author} {\bibfnamefont {T.}~\bibnamefont {Abe}},
  \bibinfo {author} {\bibfnamefont {M.}~\bibnamefont {Kohno}}, \bibinfo
  {author} {\bibfnamefont {P.}~\bibnamefont {Navr\'atil}}, \bibinfo {author}
  {\bibfnamefont {R.}~\bibnamefont {Okamoto}}, \bibinfo {author} {\bibfnamefont
  {T.}~\bibnamefont {Otsuka}}, \bibinfo {author} {\bibfnamefont
  {N.}~\bibnamefont {Shimizu}}, \ and\ \bibinfo {author} {\bibfnamefont
  {S.~R.}\ \bibnamefont {Stroberg}},\ }\href {\doibase
  10.1103/PhysRevC.100.034310} {\bibfield  {journal} {\bibinfo  {journal}
  {Phys. Rev. C}\ }\textbf {\bibinfo {volume} {100}},\ \bibinfo {pages}
  {034310} (\bibinfo {year} {2019})}\BibitemShut {NoStop}%
\bibitem [{\citenamefont {Huang}\ \emph {et~al.}(2017)\citenamefont {Huang},
  \citenamefont {Audi}, \citenamefont {Wang}, \citenamefont {Kondev},
  \citenamefont {Naimi},\ and\ \citenamefont {Xu}}]{AME2016}%
  \BibitemOpen
  \bibfield  {author} {\bibinfo {author} {\bibfnamefont {W.}~\bibnamefont
  {Huang}}, \bibinfo {author} {\bibfnamefont {G.}~\bibnamefont {Audi}},
  \bibinfo {author} {\bibfnamefont {M.}~\bibnamefont {Wang}}, \bibinfo {author}
  {\bibfnamefont {F.~G.}\ \bibnamefont {Kondev}}, \bibinfo {author}
  {\bibfnamefont {S.}~\bibnamefont {Naimi}}, \ and\ \bibinfo {author}
  {\bibfnamefont {X.}~\bibnamefont {Xu}},\ }\href {\doibase
  10.1088/1674-1137/41/3/030002} {\bibfield  {journal} {\bibinfo  {journal}
  {Chinese Physics C}\ }\textbf {\bibinfo {volume} {41}},\ \bibinfo {pages}
  {030002} (\bibinfo {year} {2017})}\BibitemShut {NoStop}%
\bibitem [{\citenamefont {Michimasa}\ \emph {et~al.}(2018)\citenamefont
  {Michimasa}, \citenamefont {Kobayashi}, \citenamefont {Kiyokawa},
  \citenamefont {Ota}, \citenamefont {Ahn}, \citenamefont {Baba}, \citenamefont
  {Berg}, \citenamefont {Dozono}, \citenamefont {Fukuda}, \citenamefont
  {Furuno}, \citenamefont {Ideguchi}, \citenamefont {Inabe}, \citenamefont
  {Kawabata}, \citenamefont {Kawase}, \citenamefont {Kisamori}, \citenamefont
  {Kobayashi}, \citenamefont {Kubo}, \citenamefont {Kubota}, \citenamefont
  {Lee}, \citenamefont {Matsushita}, \citenamefont {Miya}, \citenamefont
  {Mizukami}, \citenamefont {Nagakura}, \citenamefont {Nishimura},
  \citenamefont {Oikawa}, \citenamefont {Sakai}, \citenamefont {Shimizu},
  \citenamefont {Stolz}, \citenamefont {Suzuki}, \citenamefont {Takaki},
  \citenamefont {Takeda}, \citenamefont {Takeuchi}, \citenamefont {Tokieda},
  \citenamefont {Uesaka}, \citenamefont {Yako}, \citenamefont {Yamaguchi},
  \citenamefont {Yanagisawa}, \citenamefont {Yokoyama}, \citenamefont
  {Yoshida},\ and\ \citenamefont {Shimoura}}]{Michimasa18}%
  \BibitemOpen
  \bibfield  {author} {\bibinfo {author} {\bibfnamefont {S.}~\bibnamefont
  {Michimasa}}, \bibinfo {author} {\bibfnamefont {M.}~\bibnamefont
  {Kobayashi}}, \bibinfo {author} {\bibfnamefont {Y.}~\bibnamefont {Kiyokawa}},
  \bibinfo {author} {\bibfnamefont {S.}~\bibnamefont {Ota}}, \bibinfo {author}
  {\bibfnamefont {D.~S.}\ \bibnamefont {Ahn}}, \bibinfo {author} {\bibfnamefont
  {H.}~\bibnamefont {Baba}}, \bibinfo {author} {\bibfnamefont {G.~P.~A.}\
  \bibnamefont {Berg}}, \bibinfo {author} {\bibfnamefont {M.}~\bibnamefont
  {Dozono}}, \bibinfo {author} {\bibfnamefont {N.}~\bibnamefont {Fukuda}},
  \bibinfo {author} {\bibfnamefont {T.}~\bibnamefont {Furuno}}, \bibinfo
  {author} {\bibfnamefont {E.}~\bibnamefont {Ideguchi}}, \bibinfo {author}
  {\bibfnamefont {N.}~\bibnamefont {Inabe}}, \bibinfo {author} {\bibfnamefont
  {T.}~\bibnamefont {Kawabata}}, \bibinfo {author} {\bibfnamefont
  {S.}~\bibnamefont {Kawase}}, \bibinfo {author} {\bibfnamefont
  {K.}~\bibnamefont {Kisamori}}, \bibinfo {author} {\bibfnamefont
  {K.}~\bibnamefont {Kobayashi}}, \bibinfo {author} {\bibfnamefont
  {T.}~\bibnamefont {Kubo}}, \bibinfo {author} {\bibfnamefont {Y.}~\bibnamefont
  {Kubota}}, \bibinfo {author} {\bibfnamefont {C.~S.}\ \bibnamefont {Lee}},
  \bibinfo {author} {\bibfnamefont {M.}~\bibnamefont {Matsushita}}, \bibinfo
  {author} {\bibfnamefont {H.}~\bibnamefont {Miya}}, \bibinfo {author}
  {\bibfnamefont {A.}~\bibnamefont {Mizukami}}, \bibinfo {author}
  {\bibfnamefont {H.}~\bibnamefont {Nagakura}}, \bibinfo {author}
  {\bibfnamefont {D.}~\bibnamefont {Nishimura}}, \bibinfo {author}
  {\bibfnamefont {H.}~\bibnamefont {Oikawa}}, \bibinfo {author} {\bibfnamefont
  {H.}~\bibnamefont {Sakai}}, \bibinfo {author} {\bibfnamefont
  {Y.}~\bibnamefont {Shimizu}}, \bibinfo {author} {\bibfnamefont
  {A.}~\bibnamefont {Stolz}}, \bibinfo {author} {\bibfnamefont
  {H.}~\bibnamefont {Suzuki}}, \bibinfo {author} {\bibfnamefont
  {M.}~\bibnamefont {Takaki}}, \bibinfo {author} {\bibfnamefont
  {H.}~\bibnamefont {Takeda}}, \bibinfo {author} {\bibfnamefont
  {S.}~\bibnamefont {Takeuchi}}, \bibinfo {author} {\bibfnamefont
  {H.}~\bibnamefont {Tokieda}}, \bibinfo {author} {\bibfnamefont
  {T.}~\bibnamefont {Uesaka}}, \bibinfo {author} {\bibfnamefont
  {K.}~\bibnamefont {Yako}}, \bibinfo {author} {\bibfnamefont {Y.}~\bibnamefont
  {Yamaguchi}}, \bibinfo {author} {\bibfnamefont {Y.}~\bibnamefont
  {Yanagisawa}}, \bibinfo {author} {\bibfnamefont {R.}~\bibnamefont
  {Yokoyama}}, \bibinfo {author} {\bibfnamefont {K.}~\bibnamefont {Yoshida}}, \
  and\ \bibinfo {author} {\bibfnamefont {S.}~\bibnamefont {Shimoura}},\ }\href
  {\doibase 10.1103/PhysRevLett.121.022506} {\bibfield  {journal} {\bibinfo
  {journal} {Phys. Rev. Lett.}\ }\textbf {\bibinfo {volume} {121}},\ \bibinfo
  {pages} {022506} (\bibinfo {year} {2018})}\BibitemShut {NoStop}%
\bibitem [{\citenamefont {Xu}\ \emph {et~al.}(2019)\citenamefont {Xu} \emph
  {et~al.}}]{Xu19}%
  \BibitemOpen
  \bibfield  {author} {\bibinfo {author} {\bibfnamefont {X.}~\bibnamefont {Xu}}
  \emph {et~al.},\ }\href {\doibase 10.1103/PhysRevC.99.064303} {\bibfield
  {journal} {\bibinfo  {journal} {Phys. Rev. C}\ }\textbf {\bibinfo {volume}
  {99}},\ \bibinfo {pages} {064303} (\bibinfo {year} {2019})}\BibitemShut
  {NoStop}%
\bibitem [{\citenamefont {Duguet}\ \emph
  {et~al.}(2001{\natexlab{a}})\citenamefont {Duguet}, \citenamefont {Bonche},
  \citenamefont {Heenen},\ and\ \citenamefont {Meyer}}]{Duguet02b}%
  \BibitemOpen
  \bibfield  {author} {\bibinfo {author} {\bibfnamefont {T.}~\bibnamefont
  {Duguet}}, \bibinfo {author} {\bibfnamefont {P.}~\bibnamefont {Bonche}},
  \bibinfo {author} {\bibfnamefont {P.-H.}\ \bibnamefont {Heenen}}, \ and\
  \bibinfo {author} {\bibfnamefont {J.}~\bibnamefont {Meyer}},\ }\href
  {\doibase 10.1103/PhysRevC.65.014311} {\bibfield  {journal} {\bibinfo
  {journal} {Phys. Rev. C}\ }\textbf {\bibinfo {volume} {65}},\ \bibinfo
  {pages} {014311} (\bibinfo {year} {2001}{\natexlab{a}})}\BibitemShut
  {NoStop}%
\bibitem [{\citenamefont {Stroberg}\ \emph {et~al.}(2021)\citenamefont
  {Stroberg}, \citenamefont {Holt}, \citenamefont {Schwenk},\ and\
  \citenamefont {Simonis}}]{Stroberg19b}%
  \BibitemOpen
  \bibfield  {author} {\bibinfo {author} {\bibfnamefont {S.~R.}\ \bibnamefont
  {Stroberg}}, \bibinfo {author} {\bibfnamefont {J.~D.}\ \bibnamefont {Holt}},
  \bibinfo {author} {\bibfnamefont {A.}~\bibnamefont {Schwenk}}, \ and\
  \bibinfo {author} {\bibfnamefont {J.}~\bibnamefont {Simonis}},\ }\href
  {\doibase 10.1103/PhysRevLett.126.022501} {\bibfield  {journal} {\bibinfo
  {journal} {Phys. Rev. Lett.}\ }\textbf {\bibinfo {volume} {126}},\ \bibinfo
  {pages} {022501} (\bibinfo {year} {2021})}\BibitemShut {NoStop}%
\bibitem [{\citenamefont {Dean}\ and\ \citenamefont
  {Hjorth-Jensen}(2003)}]{Dean03}%
  \BibitemOpen
  \bibfield  {author} {\bibinfo {author} {\bibfnamefont {D.~J.}\ \bibnamefont
  {Dean}}\ and\ \bibinfo {author} {\bibfnamefont {M.}~\bibnamefont
  {Hjorth-Jensen}},\ }\href {\doibase 10.1103/RevModPhys.75.607} {\bibfield
  {journal} {\bibinfo  {journal} {Rev. Mod. Phys.}\ }\textbf {\bibinfo {volume}
  {75}},\ \bibinfo {pages} {607} (\bibinfo {year} {2003})}\BibitemShut
  {NoStop}%
\bibitem [{\citenamefont {Duguet}(2013)}]{Duguet13}%
  \BibitemOpen
  \bibfield  {author} {\bibinfo {author} {\bibfnamefont {T.}~\bibnamefont
  {Duguet}},\ }\enquote {\bibinfo {title} {Pairing in finite nuclei from
  low-momentum two- and three-nucleon interactions},}\ in\ \href {\doibase
  10.1142/9789814412490_0017} {\emph {\bibinfo {booktitle} {Fifty Years of
  Nuclear BCS}}}\ (\bibinfo {year} {2013})\ pp.\ \bibinfo {pages}
  {229--242}\BibitemShut {NoStop}%
\bibitem [{\citenamefont {Duguet}\ \emph {et~al.}(2010)\citenamefont {Duguet},
  \citenamefont {Lesinski}, \citenamefont {Hebeler},\ and\ \citenamefont
  {Schwenk}}]{Duguet10}%
  \BibitemOpen
  \bibfield  {author} {\bibinfo {author} {\bibfnamefont {T.}~\bibnamefont
  {Duguet}}, \bibinfo {author} {\bibfnamefont {T.}~\bibnamefont {Lesinski}},
  \bibinfo {author} {\bibfnamefont {K.}~\bibnamefont {Hebeler}}, \ and\
  \bibinfo {author} {\bibfnamefont {A.}~\bibnamefont {Schwenk}},\ }\href
  {\doibase 10.1142/S0217732310000812} {\bibfield  {journal} {\bibinfo
  {journal} {Mod.\ Phys.\ Lett.\ A}\ }\textbf {\bibinfo {volume} {25}},\
  \bibinfo {pages} {1989} (\bibinfo {year} {2010})}\BibitemShut {NoStop}%
\bibitem [{\citenamefont {Lesinski}\ \emph {et~al.}(2012)\citenamefont
  {Lesinski}, \citenamefont {Hebeler}, \citenamefont {Duguet},\ and\
  \citenamefont {Schwenk}}]{Lesinski11}%
  \BibitemOpen
  \bibfield  {author} {\bibinfo {author} {\bibfnamefont {T.}~\bibnamefont
  {Lesinski}}, \bibinfo {author} {\bibfnamefont {K.}~\bibnamefont {Hebeler}},
  \bibinfo {author} {\bibfnamefont {T.}~\bibnamefont {Duguet}}, \ and\ \bibinfo
  {author} {\bibfnamefont {A.}~\bibnamefont {Schwenk}},\ }\href {\doibase
  10.1088/0954-3899/39/1/015108} {\bibfield  {journal} {\bibinfo  {journal}
  {J.\ Phys.\ G}\ }\textbf {\bibinfo {volume} {39}},\ \bibinfo {pages} {015108}
  (\bibinfo {year} {2012})}\BibitemShut {NoStop}%
\bibitem [{\citenamefont {Barranco}\ \emph {et~al.}(2004)\citenamefont
  {Barranco}, \citenamefont {Broglia}, \citenamefont {Col\`o}, \citenamefont
  {Vigezzi},\ and\ \citenamefont {Bortignon}}]{Barranco04}%
  \BibitemOpen
  \bibfield  {author} {\bibinfo {author} {\bibfnamefont {F.}~\bibnamefont
  {Barranco}}, \bibinfo {author} {\bibfnamefont {R.}~\bibnamefont {Broglia}},
  \bibinfo {author} {\bibfnamefont {G.}~\bibnamefont {Col\`o}}, \bibinfo
  {author} {\bibfnamefont {E.}~\bibnamefont {Vigezzi}}, \ and\ \bibinfo
  {author} {\bibfnamefont {P.}~\bibnamefont {Bortignon}},\ }\href {\doibase
  10.1140/epja/i2003-10185-0} {\bibfield  {journal} {\bibinfo  {journal} {Eur.\
  Phys.\ J.\ A}\ }\textbf {\bibinfo {volume} {21}},\ \bibinfo {pages} {57}
  (\bibinfo {year} {2004})}\BibitemShut {NoStop}%
\bibitem [{\citenamefont {Gori}\ \emph {et~al.}(2005)\citenamefont {Gori},
  \citenamefont {Ramponi}, \citenamefont {Barranco}, \citenamefont {Bortignon},
  \citenamefont {Broglia}, \citenamefont {Col\`o},\ and\ \citenamefont
  {Vigezzi}}]{Gori05}%
  \BibitemOpen
  \bibfield  {author} {\bibinfo {author} {\bibfnamefont {G.}~\bibnamefont
  {Gori}}, \bibinfo {author} {\bibfnamefont {F.}~\bibnamefont {Ramponi}},
  \bibinfo {author} {\bibfnamefont {F.}~\bibnamefont {Barranco}}, \bibinfo
  {author} {\bibfnamefont {P.~F.}\ \bibnamefont {Bortignon}}, \bibinfo {author}
  {\bibfnamefont {R.~A.}\ \bibnamefont {Broglia}}, \bibinfo {author}
  {\bibfnamefont {G.}~\bibnamefont {Col\`o}}, \ and\ \bibinfo {author}
  {\bibfnamefont {E.}~\bibnamefont {Vigezzi}},\ }\href {\doibase
  10.1103/PhysRevC.72.011302} {\bibfield  {journal} {\bibinfo  {journal} {Phys.
  Rev. C}\ }\textbf {\bibinfo {volume} {72}},\ \bibinfo {pages} {011302}
  (\bibinfo {year} {2005})}\BibitemShut {NoStop}%
\bibitem [{\citenamefont {Pastore}\ \emph {et~al.}(2008)\citenamefont
  {Pastore}, \citenamefont {Barranco}, \citenamefont {Broglia},\ and\
  \citenamefont {Vigezzi}}]{Pastore08}%
  \BibitemOpen
  \bibfield  {author} {\bibinfo {author} {\bibfnamefont {A.}~\bibnamefont
  {Pastore}}, \bibinfo {author} {\bibfnamefont {F.}~\bibnamefont {Barranco}},
  \bibinfo {author} {\bibfnamefont {R.~A.}\ \bibnamefont {Broglia}}, \ and\
  \bibinfo {author} {\bibfnamefont {E.}~\bibnamefont {Vigezzi}},\ }\href
  {\doibase 10.1103/PhysRevC.78.024315} {\bibfield  {journal} {\bibinfo
  {journal} {Phys. Rev. C}\ }\textbf {\bibinfo {volume} {78}},\ \bibinfo
  {pages} {024315} (\bibinfo {year} {2008})}\BibitemShut {NoStop}%
\bibitem [{\citenamefont {Idini}\ \emph {et~al.}(2011)\citenamefont {Idini},
  \citenamefont {Barranco}, \citenamefont {Vigezzi},\ and\ \citenamefont
  {Broglia}}]{Idini11}%
  \BibitemOpen
  \bibfield  {author} {\bibinfo {author} {\bibfnamefont {A.}~\bibnamefont
  {Idini}}, \bibinfo {author} {\bibfnamefont {F.}~\bibnamefont {Barranco}},
  \bibinfo {author} {\bibfnamefont {E.}~\bibnamefont {Vigezzi}}, \ and\
  \bibinfo {author} {\bibfnamefont {R.}~\bibnamefont {Broglia}},\ }\href
  {\doibase 10.1088/1742-6596/312/9/092032} {\bibfield  {journal} {\bibinfo
  {journal} {J.\ Phys.\ Conf.\ Ser.}\ }\textbf {\bibinfo {volume} {312}},\
  \bibinfo {pages} {092032} (\bibinfo {year} {2011})}\BibitemShut {NoStop}%
\bibitem [{\citenamefont {Duguet}\ \emph
  {et~al.}(2001{\natexlab{b}})\citenamefont {Duguet}, \citenamefont {Bonche},
  \citenamefont {Heenen},\ and\ \citenamefont {Meyer}}]{Duguet02a}%
  \BibitemOpen
  \bibfield  {author} {\bibinfo {author} {\bibfnamefont {T.}~\bibnamefont
  {Duguet}}, \bibinfo {author} {\bibfnamefont {P.}~\bibnamefont {Bonche}},
  \bibinfo {author} {\bibfnamefont {P.-H.}\ \bibnamefont {Heenen}}, \ and\
  \bibinfo {author} {\bibfnamefont {J.}~\bibnamefont {Meyer}},\ }\href
  {\doibase 10.1103/PhysRevC.65.014310} {\bibfield  {journal} {\bibinfo
  {journal} {Phys. Rev. C}\ }\textbf {\bibinfo {volume} {65}},\ \bibinfo
  {pages} {014310} (\bibinfo {year} {2001}{\natexlab{b}})}\BibitemShut
  {NoStop}%
\bibitem [{\citenamefont {Dobaczewski}\ \emph {et~al.}(1988)\citenamefont
  {Dobaczewski}, \citenamefont {Nazarewicz}, \citenamefont {Skalski},\ and\
  \citenamefont {Werner}}]{Dobaczewski88}%
  \BibitemOpen
  \bibfield  {author} {\bibinfo {author} {\bibfnamefont {J.}~\bibnamefont
  {Dobaczewski}}, \bibinfo {author} {\bibfnamefont {W.}~\bibnamefont
  {Nazarewicz}}, \bibinfo {author} {\bibfnamefont {J.}~\bibnamefont {Skalski}},
  \ and\ \bibinfo {author} {\bibfnamefont {T.}~\bibnamefont {Werner}},\ }\href
  {\doibase 10.1103/PhysRevLett.60.2254} {\bibfield  {journal} {\bibinfo
  {journal} {Phys. Rev. Lett.}\ }\textbf {\bibinfo {volume} {60}},\ \bibinfo
  {pages} {2254} (\bibinfo {year} {1988})}\BibitemShut {NoStop}%
\bibitem [{\citenamefont {Bender}\ \emph {et~al.}(2006)\citenamefont {Bender},
  \citenamefont {Bertsch},\ and\ \citenamefont {Heenen}}]{Bender06a}%
  \BibitemOpen
  \bibfield  {author} {\bibinfo {author} {\bibfnamefont {M.}~\bibnamefont
  {Bender}}, \bibinfo {author} {\bibfnamefont {G.~F.}\ \bibnamefont {Bertsch}},
  \ and\ \bibinfo {author} {\bibfnamefont {P.-H.}\ \bibnamefont {Heenen}},\
  }\href {\doibase 10.1103/PhysRevC.73.034322} {\bibfield  {journal} {\bibinfo
  {journal} {Phys. Rev. C}\ }\textbf {\bibinfo {volume} {73}},\ \bibinfo
  {pages} {034322} (\bibinfo {year} {2006})}\BibitemShut {NoStop}%
\bibitem [{\citenamefont {Van~Houcke}\ \emph {et~al.}(2012)\citenamefont
  {Van~Houcke}, \citenamefont {Werner}, \citenamefont {Kozik}, \citenamefont
  {Prokof'ev}, \citenamefont {Svistunov}, \citenamefont {Ku}, \citenamefont
  {Sommer}, \citenamefont {Cheuk}, \citenamefont {Schirotzek},\ and\
  \citenamefont {Zwierlein}}]{VanHoucke12}%
  \BibitemOpen
  \bibfield  {author} {\bibinfo {author} {\bibfnamefont {K.}~\bibnamefont
  {Van~Houcke}}, \bibinfo {author} {\bibfnamefont {F.}~\bibnamefont {Werner}},
  \bibinfo {author} {\bibfnamefont {E.}~\bibnamefont {Kozik}}, \bibinfo
  {author} {\bibfnamefont {N.}~\bibnamefont {Prokof'ev}}, \bibinfo {author}
  {\bibfnamefont {B.}~\bibnamefont {Svistunov}}, \bibinfo {author}
  {\bibfnamefont {M.~J.~H.}\ \bibnamefont {Ku}}, \bibinfo {author}
  {\bibfnamefont {A.~T.}\ \bibnamefont {Sommer}}, \bibinfo {author}
  {\bibfnamefont {L.~W.}\ \bibnamefont {Cheuk}}, \bibinfo {author}
  {\bibfnamefont {A.}~\bibnamefont {Schirotzek}}, \ and\ \bibinfo {author}
  {\bibfnamefont {M.~W.}\ \bibnamefont {Zwierlein}},\ }\href {\doibase
  10.1038/nphys2273} {\bibfield  {journal} {\bibinfo  {journal} {Nature
  Physics}\ }\textbf {\bibinfo {volume} {8}},\ \bibinfo {pages} {366} (\bibinfo
  {year} {2012})}\BibitemShut {NoStop}%
\bibitem [{\citenamefont {Garcia~Ruiz}\ \emph {et~al.}(2016)\citenamefont
  {Garcia~Ruiz}, \citenamefont {Bissell}, \citenamefont {Blaum}, \citenamefont
  {Ekstr{\"o}m}, \citenamefont {Fr{\"o}mmgen}, \citenamefont {Hagen},
  \citenamefont {Hammen}, \citenamefont {Hebeler}, \citenamefont {Holt},
  \citenamefont {Jansen}, \citenamefont {Kowalska}, \citenamefont {Kreim},
  \citenamefont {Nazarewicz}, \citenamefont {Neugart}, \citenamefont {Neyens},
  \citenamefont {N{\"o}rtersh{\"a}user}, \citenamefont {Papenbrock},
  \citenamefont {Papuga}, \citenamefont {Schwenk}, \citenamefont {Simonis},
  \citenamefont {Wendt},\ and\ \citenamefont {Yordanov}}]{GarciaRuiz16}%
  \BibitemOpen
  \bibfield  {author} {\bibinfo {author} {\bibfnamefont {R.~F.}\ \bibnamefont
  {Garcia~Ruiz}}, \bibinfo {author} {\bibfnamefont {M.~L.}\ \bibnamefont
  {Bissell}}, \bibinfo {author} {\bibfnamefont {K.}~\bibnamefont {Blaum}},
  \bibinfo {author} {\bibfnamefont {A.}~\bibnamefont {Ekstr{\"o}m}}, \bibinfo
  {author} {\bibfnamefont {N.}~\bibnamefont {Fr{\"o}mmgen}}, \bibinfo {author}
  {\bibfnamefont {G.}~\bibnamefont {Hagen}}, \bibinfo {author} {\bibfnamefont
  {M.}~\bibnamefont {Hammen}}, \bibinfo {author} {\bibfnamefont
  {K.}~\bibnamefont {Hebeler}}, \bibinfo {author} {\bibfnamefont {J.~D.}\
  \bibnamefont {Holt}}, \bibinfo {author} {\bibfnamefont {G.~R.}\ \bibnamefont
  {Jansen}}, \bibinfo {author} {\bibfnamefont {M.}~\bibnamefont {Kowalska}},
  \bibinfo {author} {\bibfnamefont {K.}~\bibnamefont {Kreim}}, \bibinfo
  {author} {\bibfnamefont {W.}~\bibnamefont {Nazarewicz}}, \bibinfo {author}
  {\bibfnamefont {R.}~\bibnamefont {Neugart}}, \bibinfo {author} {\bibfnamefont
  {G.}~\bibnamefont {Neyens}}, \bibinfo {author} {\bibfnamefont
  {W.}~\bibnamefont {N{\"o}rtersh{\"a}user}}, \bibinfo {author} {\bibfnamefont
  {T.}~\bibnamefont {Papenbrock}}, \bibinfo {author} {\bibfnamefont
  {J.}~\bibnamefont {Papuga}}, \bibinfo {author} {\bibfnamefont
  {A.}~\bibnamefont {Schwenk}}, \bibinfo {author} {\bibfnamefont
  {J.}~\bibnamefont {Simonis}}, \bibinfo {author} {\bibfnamefont {K.~A.}\
  \bibnamefont {Wendt}}, \ and\ \bibinfo {author} {\bibfnamefont {D.~T.}\
  \bibnamefont {Yordanov}},\ }\href {https://doi.org/10.1038/nphys3645}
  {\bibfield  {journal} {\bibinfo  {journal} {Nature Physics}\ }\textbf
  {\bibinfo {volume} {12}},\ \bibinfo {pages} {594} (\bibinfo {year}
  {2016})}\BibitemShut {NoStop}%
\bibitem [{\citenamefont {Angeli}\ and\ \citenamefont
  {Marinova}(2013)}]{Angeli13}%
  \BibitemOpen
  \bibfield  {author} {\bibinfo {author} {\bibfnamefont {I.}~\bibnamefont
  {Angeli}}\ and\ \bibinfo {author} {\bibfnamefont {K.}~\bibnamefont
  {Marinova}},\ }\href {\doibase http://dx.doi.org/10.1016/j.adt.2011.12.006}
  {\bibfield  {journal} {\bibinfo  {journal} {Atomic Data and Nuclear Data
  Tables}\ }\textbf {\bibinfo {volume} {99}},\ \bibinfo {pages} {69 } (\bibinfo
  {year} {2013})}\BibitemShut {NoStop}%
\bibitem [{\citenamefont {Miller}\ \emph {et~al.}(2019)\citenamefont {Miller},
  \citenamefont {Minamisono}, \citenamefont {Klose}, \citenamefont {Garand},
  \citenamefont {Kujawa}, \citenamefont {Lantis}, \citenamefont {Liu},
  \citenamefont {Maa{\ss}}, \citenamefont {Mantica}, \citenamefont
  {Nazarewicz}, \citenamefont {N{\"o}rtersh{\"a}user}, \citenamefont {Pineda},
  \citenamefont {Reinhard}, \citenamefont {Rossi}, \citenamefont {Sommer},
  \citenamefont {Sumithrarachchi}, \citenamefont {Teigelh{\"o}fer},\ and\
  \citenamefont {Watkins}}]{Miller19}%
  \BibitemOpen
  \bibfield  {author} {\bibinfo {author} {\bibfnamefont {A.~J.}\ \bibnamefont
  {Miller}}, \bibinfo {author} {\bibfnamefont {K.}~\bibnamefont {Minamisono}},
  \bibinfo {author} {\bibfnamefont {A.}~\bibnamefont {Klose}}, \bibinfo
  {author} {\bibfnamefont {D.}~\bibnamefont {Garand}}, \bibinfo {author}
  {\bibfnamefont {C.}~\bibnamefont {Kujawa}}, \bibinfo {author} {\bibfnamefont
  {J.~D.}\ \bibnamefont {Lantis}}, \bibinfo {author} {\bibfnamefont
  {Y.}~\bibnamefont {Liu}}, \bibinfo {author} {\bibfnamefont {B.}~\bibnamefont
  {Maa{\ss}}}, \bibinfo {author} {\bibfnamefont {P.~F.}\ \bibnamefont
  {Mantica}}, \bibinfo {author} {\bibfnamefont {W.}~\bibnamefont {Nazarewicz}},
  \bibinfo {author} {\bibfnamefont {W.}~\bibnamefont {N{\"o}rtersh{\"a}user}},
  \bibinfo {author} {\bibfnamefont {S.~V.}\ \bibnamefont {Pineda}}, \bibinfo
  {author} {\bibfnamefont {P.~G.}\ \bibnamefont {Reinhard}}, \bibinfo {author}
  {\bibfnamefont {D.~M.}\ \bibnamefont {Rossi}}, \bibinfo {author}
  {\bibfnamefont {F.}~\bibnamefont {Sommer}}, \bibinfo {author} {\bibfnamefont
  {C.}~\bibnamefont {Sumithrarachchi}}, \bibinfo {author} {\bibfnamefont
  {A.}~\bibnamefont {Teigelh{\"o}fer}}, \ and\ \bibinfo {author} {\bibfnamefont
  {J.}~\bibnamefont {Watkins}},\ }\href {\doibase 10.1038/s41567-019-0416-9}
  {\bibfield  {journal} {\bibinfo  {journal} {Nature Physics}\ }\textbf
  {\bibinfo {volume} {15}},\ \bibinfo {pages} {432} (\bibinfo {year}
  {2019})}\BibitemShut {NoStop}%
\bibitem [{\citenamefont {Xiong}\ \emph {et~al.}(2019)\citenamefont {Xiong}
  \emph {et~al.}}]{Xiong19}%
  \BibitemOpen
  \bibfield  {author} {\bibinfo {author} {\bibfnamefont {W.}~\bibnamefont
  {Xiong}} \emph {et~al.},\ }\href {\doibase 10.1038/s41586-019-1721-2}
  {\bibfield  {journal} {\bibinfo  {journal} {Nature}\ }\textbf {\bibinfo
  {volume} {575}},\ \bibinfo {pages} {147} (\bibinfo {year}
  {2019})}\BibitemShut {NoStop}%
\bibitem [{\citenamefont {Mohr}\ \emph {et~al.}(2008)\citenamefont {Mohr},
  \citenamefont {Taylor},\ and\ \citenamefont {Newell}}]{CODATA2006}%
  \BibitemOpen
  \bibfield  {author} {\bibinfo {author} {\bibfnamefont {P.~J.}\ \bibnamefont
  {Mohr}}, \bibinfo {author} {\bibfnamefont {B.~N.}\ \bibnamefont {Taylor}}, \
  and\ \bibinfo {author} {\bibfnamefont {D.~B.}\ \bibnamefont {Newell}},\
  }\href {\doibase 10.1103/RevModPhys.80.633} {\bibfield  {journal} {\bibinfo
  {journal} {Rev. Mod. Phys.}\ }\textbf {\bibinfo {volume} {80}},\ \bibinfo
  {pages} {633} (\bibinfo {year} {2008})}\BibitemShut {NoStop}%
\bibitem [{\citenamefont {Simonis}\ \emph {et~al.}(2017)\citenamefont
  {Simonis}, \citenamefont {Stroberg}, \citenamefont {Hebeler}, \citenamefont
  {Holt},\ and\ \citenamefont {Schwenk}}]{Simonis17}%
  \BibitemOpen
  \bibfield  {author} {\bibinfo {author} {\bibfnamefont {J.}~\bibnamefont
  {Simonis}}, \bibinfo {author} {\bibfnamefont {S.~R.}\ \bibnamefont
  {Stroberg}}, \bibinfo {author} {\bibfnamefont {K.}~\bibnamefont {Hebeler}},
  \bibinfo {author} {\bibfnamefont {J.~D.}\ \bibnamefont {Holt}}, \ and\
  \bibinfo {author} {\bibfnamefont {A.}~\bibnamefont {Schwenk}},\ }\href
  {\doibase 10.1103/PhysRevC.96.014303} {\bibfield  {journal} {\bibinfo
  {journal} {Phys. Rev. C}\ }\textbf {\bibinfo {volume} {96}},\ \bibinfo
  {pages} {014303} (\bibinfo {year} {2017})}\BibitemShut {NoStop}%
\bibitem [{\citenamefont {H{\"u}ther}\ \emph {et~al.}(2019)\citenamefont
  {H{\"u}ther}, \citenamefont {Vobig}, \citenamefont {Hebeler}, \citenamefont
  {Machleidt},\ and\ \citenamefont {Roth}}]{Huther19}%
  \BibitemOpen
  \bibfield  {author} {\bibinfo {author} {\bibfnamefont {T.}~\bibnamefont
  {H{\"u}ther}}, \bibinfo {author} {\bibfnamefont {K.}~\bibnamefont {Vobig}},
  \bibinfo {author} {\bibfnamefont {K.}~\bibnamefont {Hebeler}}, \bibinfo
  {author} {\bibfnamefont {R.}~\bibnamefont {Machleidt}}, \ and\ \bibinfo
  {author} {\bibfnamefont {R.}~\bibnamefont {Roth}},\ }\href@noop {} {\
  (\bibinfo {year} {2019})},\ \Eprint {http://arxiv.org/abs/1911.04955}
  {arXiv:1911.04955 [nucl-th]} \BibitemShut {NoStop}%
\bibitem [{\citenamefont {Friar}\ \emph {et~al.}(1997)\citenamefont {Friar},
  \citenamefont {Martorell},\ and\ \citenamefont {Sprung}}]{Friar97}%
  \BibitemOpen
  \bibfield  {author} {\bibinfo {author} {\bibfnamefont {J.~L.}\ \bibnamefont
  {Friar}}, \bibinfo {author} {\bibfnamefont {J.}~\bibnamefont {Martorell}}, \
  and\ \bibinfo {author} {\bibfnamefont {D.~W.~L.}\ \bibnamefont {Sprung}},\
  }\href {\doibase 10.1103/PhysRevA.56.4579} {\bibfield  {journal} {\bibinfo
  {journal} {Phys. Rev. A}\ }\textbf {\bibinfo {volume} {56}},\ \bibinfo
  {pages} {4579} (\bibinfo {year} {1997})}\BibitemShut {NoStop}%
\bibitem [{\citenamefont {Horowitz}\ and\ \citenamefont
  {Piekarewicz}(2012)}]{Horowitz12}%
  \BibitemOpen
  \bibfield  {author} {\bibinfo {author} {\bibfnamefont {C.~J.}\ \bibnamefont
  {Horowitz}}\ and\ \bibinfo {author} {\bibfnamefont {J.}~\bibnamefont
  {Piekarewicz}},\ }\href {\doibase 10.1103/PhysRevC.86.045503} {\bibfield
  {journal} {\bibinfo  {journal} {Phys. Rev. C}\ }\textbf {\bibinfo {volume}
  {86}},\ \bibinfo {pages} {045503} (\bibinfo {year} {2012})}\BibitemShut
  {NoStop}%
\bibitem [{\citenamefont {Hammer}\ and\ \citenamefont
  {Mei\ss{}ner}(2020)}]{Hammer20}%
  \BibitemOpen
  \bibfield  {author} {\bibinfo {author} {\bibfnamefont {H.-W.}\ \bibnamefont
  {Hammer}}\ and\ \bibinfo {author} {\bibfnamefont {U.-G.}\ \bibnamefont
  {Mei\ss{}ner}},\ }\href {\doibase https://doi.org/10.1016/j.scib.2019.12.012}
  {\bibfield  {journal} {\bibinfo  {journal} {Science Bulletin}\ }\textbf
  {\bibinfo {volume} {65}},\ \bibinfo {pages} {257 } (\bibinfo {year}
  {2020})}\BibitemShut {NoStop}%
\bibitem [{COD()}]{CODATA2018}%
  \BibitemOpen
  \href@noop {} {\enquote {\bibinfo {title} {{CODATA recommended values of the
  fundamental physical constants: 2018}},}\ }\bibinfo {howpublished}
  {\url{https://physics.nist.gov/cgi-bin/cuu/Value?rp}}\BibitemShut {NoStop}%
\bibitem [{\citenamefont {Garcia~Ruiz}\ and\ \citenamefont
  {Vernon}(2020)}]{GarciaRuiz20}%
  \BibitemOpen
  \bibfield  {author} {\bibinfo {author} {\bibfnamefont {R.}~\bibnamefont
  {Garcia~Ruiz}}\ and\ \bibinfo {author} {\bibfnamefont {A.}~\bibnamefont
  {Vernon}},\ }\href {\doibase 10.1140/epja/s10050-020-00134-8} {\bibfield
  {journal} {\bibinfo  {journal} {Eur. Phys. J. A}\ }\textbf {\bibinfo {volume}
  {56}},\ \bibinfo {pages} {136} (\bibinfo {year} {2020})}\BibitemShut
  {NoStop}%
\bibitem [{\citenamefont {Barranco}\ and\ \citenamefont
  {Broglia}(1985)}]{Barranco85}%
  \BibitemOpen
  \bibfield  {author} {\bibinfo {author} {\bibfnamefont {F.}~\bibnamefont
  {Barranco}}\ and\ \bibinfo {author} {\bibfnamefont {R.}~\bibnamefont
  {Broglia}},\ }\href {\doibase https://doi.org/10.1016/0370-2693(85)91391-7}
  {\bibfield  {journal} {\bibinfo  {journal} {Physics Letters B}\ }\textbf
  {\bibinfo {volume} {151}},\ \bibinfo {pages} {90 } (\bibinfo {year}
  {1985})}\BibitemShut {NoStop}%
\bibitem [{\citenamefont {Caurier}\ \emph {et~al.}(2001)\citenamefont
  {Caurier}, \citenamefont {Langanke}, \citenamefont {Martínez-Pinedo},
  \citenamefont {Nowacki},\ and\ \citenamefont {Vogel}}]{Caurier01}%
  \BibitemOpen
  \bibfield  {author} {\bibinfo {author} {\bibfnamefont {E.}~\bibnamefont
  {Caurier}}, \bibinfo {author} {\bibfnamefont {K.}~\bibnamefont {Langanke}},
  \bibinfo {author} {\bibfnamefont {G.}~\bibnamefont {Martínez-Pinedo}},
  \bibinfo {author} {\bibfnamefont {F.}~\bibnamefont {Nowacki}}, \ and\
  \bibinfo {author} {\bibfnamefont {P.}~\bibnamefont {Vogel}},\ }\href
  {\doibase https://doi.org/10.1016/S0370-2693(01)01246-1} {\bibfield
  {journal} {\bibinfo  {journal} {Physics Letters B}\ }\textbf {\bibinfo
  {volume} {522}},\ \bibinfo {pages} {240 } (\bibinfo {year}
  {2001})}\BibitemShut {NoStop}%
\bibitem [{\citenamefont {Barbieri}\ \emph {et~al.}(2007)\citenamefont
  {Barbieri}, \citenamefont {Van~Neck},\ and\ \citenamefont
  {Dickhoff}}]{Barbieri07}%
  \BibitemOpen
  \bibfield  {author} {\bibinfo {author} {\bibfnamefont {C.}~\bibnamefont
  {Barbieri}}, \bibinfo {author} {\bibfnamefont {D.}~\bibnamefont {Van~Neck}},
  \ and\ \bibinfo {author} {\bibfnamefont {W.~H.}\ \bibnamefont {Dickhoff}},\
  }\href {\doibase 10.1103/PhysRevA.76.052503} {\bibfield  {journal} {\bibinfo
  {journal} {Phys. Rev. A}\ }\textbf {\bibinfo {volume} {76}},\ \bibinfo
  {pages} {052503} (\bibinfo {year} {2007})}\BibitemShut {NoStop}%
\bibitem [{\citenamefont {Reinhard}\ and\ \citenamefont
  {Nazarewicz}(2017)}]{Reinhard17}%
  \BibitemOpen
  \bibfield  {author} {\bibinfo {author} {\bibfnamefont {P.-G.}\ \bibnamefont
  {Reinhard}}\ and\ \bibinfo {author} {\bibfnamefont {W.}~\bibnamefont
  {Nazarewicz}},\ }\href {\doibase 10.1103/PhysRevC.95.064328} {\bibfield
  {journal} {\bibinfo  {journal} {Phys. Rev. C}\ }\textbf {\bibinfo {volume}
  {95}},\ \bibinfo {pages} {064328} (\bibinfo {year} {2017})}\BibitemShut
  {NoStop}%
\bibitem [{\citenamefont {Liu}\ \emph {et~al.}(2019)\citenamefont {Liu},
  \citenamefont {Obertelli}, \citenamefont {Doornenbal}, \citenamefont
  {Bertulani}, \citenamefont {Hagen}, \citenamefont {Holt}, \citenamefont
  {Jansen}, \citenamefont {Morris}, \citenamefont {Schwenk}, \citenamefont
  {Stroberg}, \citenamefont {Achouri}, \citenamefont {Baba}, \citenamefont
  {Browne}, \citenamefont {Calvet}, \citenamefont {Ch\^ateau}, \citenamefont
  {Chen}, \citenamefont {Chiga}, \citenamefont {Corsi}, \citenamefont
  {Cort\'es}, \citenamefont {Delbart}, \citenamefont {Gheller}, \citenamefont
  {Giganon}, \citenamefont {Gillibert}, \citenamefont {Hilaire}, \citenamefont
  {Isobe}, \citenamefont {Kobayashi}, \citenamefont {Kubota}, \citenamefont
  {Lapoux}, \citenamefont {Motobayashi}, \citenamefont {Murray}, \citenamefont
  {Otsu}, \citenamefont {Panin}, \citenamefont {Paul}, \citenamefont
  {Rodriguez}, \citenamefont {Sakurai}, \citenamefont {Sasano}, \citenamefont
  {Steppenbeck}, \citenamefont {Stuhl}, \citenamefont {Sun}, \citenamefont
  {Togano}, \citenamefont {Uesaka}, \citenamefont {Wimmer}, \citenamefont
  {Yoneda}, \citenamefont {Aktas}, \citenamefont {Aumann}, \citenamefont
  {Chung}, \citenamefont {Flavigny}, \citenamefont {Franchoo}, \citenamefont
  {Ga\ifmmode \check{s}\else \v{s}\fi{}pari\ifmmode~\acute{c}\else \'{c}\fi{}},
  \citenamefont {Gerst}, \citenamefont {Gibelin}, \citenamefont {Hahn},
  \citenamefont {Kim}, \citenamefont {Koiwai}, \citenamefont {Kondo},
  \citenamefont {Koseoglou}, \citenamefont {Lee}, \citenamefont {Lehr},
  \citenamefont {Linh}, \citenamefont {Lokotko}, \citenamefont {MacCormick},
  \citenamefont {Moschner}, \citenamefont {Nakamura}, \citenamefont {Park},
  \citenamefont {Rossi}, \citenamefont {Sahin}, \citenamefont {Sohler},
  \citenamefont {S\"oderstr\"om}, \citenamefont {Takeuchi}, \citenamefont
  {T\"ornqvist}, \citenamefont {Vaquero}, \citenamefont {Wagner}, \citenamefont
  {Wang}, \citenamefont {Werner}, \citenamefont {Xu}, \citenamefont {Yamada},
  \citenamefont {Yan}, \citenamefont {Yang}, \citenamefont {Yasuda},\ and\
  \citenamefont {Zanetti}}]{Liu19}%
  \BibitemOpen
  \bibfield  {author} {\bibinfo {author} {\bibfnamefont {H.~N.}\ \bibnamefont
  {Liu}}, \bibinfo {author} {\bibfnamefont {A.}~\bibnamefont {Obertelli}},
  \bibinfo {author} {\bibfnamefont {P.}~\bibnamefont {Doornenbal}}, \bibinfo
  {author} {\bibfnamefont {C.~A.}\ \bibnamefont {Bertulani}}, \bibinfo {author}
  {\bibfnamefont {G.}~\bibnamefont {Hagen}}, \bibinfo {author} {\bibfnamefont
  {J.~D.}\ \bibnamefont {Holt}}, \bibinfo {author} {\bibfnamefont {G.~R.}\
  \bibnamefont {Jansen}}, \bibinfo {author} {\bibfnamefont {T.~D.}\
  \bibnamefont {Morris}}, \bibinfo {author} {\bibfnamefont {A.}~\bibnamefont
  {Schwenk}}, \bibinfo {author} {\bibfnamefont {R.}~\bibnamefont {Stroberg}},
  \bibinfo {author} {\bibfnamefont {N.}~\bibnamefont {Achouri}}, \bibinfo
  {author} {\bibfnamefont {H.}~\bibnamefont {Baba}}, \bibinfo {author}
  {\bibfnamefont {F.}~\bibnamefont {Browne}}, \bibinfo {author} {\bibfnamefont
  {D.}~\bibnamefont {Calvet}}, \bibinfo {author} {\bibfnamefont
  {F.}~\bibnamefont {Ch\^ateau}}, \bibinfo {author} {\bibfnamefont
  {S.}~\bibnamefont {Chen}}, \bibinfo {author} {\bibfnamefont {N.}~\bibnamefont
  {Chiga}}, \bibinfo {author} {\bibfnamefont {A.}~\bibnamefont {Corsi}},
  \bibinfo {author} {\bibfnamefont {M.~L.}\ \bibnamefont {Cort\'es}}, \bibinfo
  {author} {\bibfnamefont {A.}~\bibnamefont {Delbart}}, \bibinfo {author}
  {\bibfnamefont {J.-M.}\ \bibnamefont {Gheller}}, \bibinfo {author}
  {\bibfnamefont {A.}~\bibnamefont {Giganon}}, \bibinfo {author} {\bibfnamefont
  {A.}~\bibnamefont {Gillibert}}, \bibinfo {author} {\bibfnamefont
  {C.}~\bibnamefont {Hilaire}}, \bibinfo {author} {\bibfnamefont
  {T.}~\bibnamefont {Isobe}}, \bibinfo {author} {\bibfnamefont
  {T.}~\bibnamefont {Kobayashi}}, \bibinfo {author} {\bibfnamefont
  {Y.}~\bibnamefont {Kubota}}, \bibinfo {author} {\bibfnamefont
  {V.}~\bibnamefont {Lapoux}}, \bibinfo {author} {\bibfnamefont
  {T.}~\bibnamefont {Motobayashi}}, \bibinfo {author} {\bibfnamefont
  {I.}~\bibnamefont {Murray}}, \bibinfo {author} {\bibfnamefont
  {H.}~\bibnamefont {Otsu}}, \bibinfo {author} {\bibfnamefont {V.}~\bibnamefont
  {Panin}}, \bibinfo {author} {\bibfnamefont {N.}~\bibnamefont {Paul}},
  \bibinfo {author} {\bibfnamefont {W.}~\bibnamefont {Rodriguez}}, \bibinfo
  {author} {\bibfnamefont {H.}~\bibnamefont {Sakurai}}, \bibinfo {author}
  {\bibfnamefont {M.}~\bibnamefont {Sasano}}, \bibinfo {author} {\bibfnamefont
  {D.}~\bibnamefont {Steppenbeck}}, \bibinfo {author} {\bibfnamefont
  {L.}~\bibnamefont {Stuhl}}, \bibinfo {author} {\bibfnamefont {Y.~L.}\
  \bibnamefont {Sun}}, \bibinfo {author} {\bibfnamefont {Y.}~\bibnamefont
  {Togano}}, \bibinfo {author} {\bibfnamefont {T.}~\bibnamefont {Uesaka}},
  \bibinfo {author} {\bibfnamefont {K.}~\bibnamefont {Wimmer}}, \bibinfo
  {author} {\bibfnamefont {K.}~\bibnamefont {Yoneda}}, \bibinfo {author}
  {\bibfnamefont {O.}~\bibnamefont {Aktas}}, \bibinfo {author} {\bibfnamefont
  {T.}~\bibnamefont {Aumann}}, \bibinfo {author} {\bibfnamefont {L.~X.}\
  \bibnamefont {Chung}}, \bibinfo {author} {\bibfnamefont {F.}~\bibnamefont
  {Flavigny}}, \bibinfo {author} {\bibfnamefont {S.}~\bibnamefont {Franchoo}},
  \bibinfo {author} {\bibfnamefont {I.}~\bibnamefont {Ga\ifmmode \check{s}\else
  \v{s}\fi{}pari\ifmmode~\acute{c}\else \'{c}\fi{}}}, \bibinfo {author}
  {\bibfnamefont {R.-B.}\ \bibnamefont {Gerst}}, \bibinfo {author}
  {\bibfnamefont {J.}~\bibnamefont {Gibelin}}, \bibinfo {author} {\bibfnamefont
  {K.~I.}\ \bibnamefont {Hahn}}, \bibinfo {author} {\bibfnamefont
  {D.}~\bibnamefont {Kim}}, \bibinfo {author} {\bibfnamefont {T.}~\bibnamefont
  {Koiwai}}, \bibinfo {author} {\bibfnamefont {Y.}~\bibnamefont {Kondo}},
  \bibinfo {author} {\bibfnamefont {P.}~\bibnamefont {Koseoglou}}, \bibinfo
  {author} {\bibfnamefont {J.}~\bibnamefont {Lee}}, \bibinfo {author}
  {\bibfnamefont {C.}~\bibnamefont {Lehr}}, \bibinfo {author} {\bibfnamefont
  {B.~D.}\ \bibnamefont {Linh}}, \bibinfo {author} {\bibfnamefont
  {T.}~\bibnamefont {Lokotko}}, \bibinfo {author} {\bibfnamefont
  {M.}~\bibnamefont {MacCormick}}, \bibinfo {author} {\bibfnamefont
  {K.}~\bibnamefont {Moschner}}, \bibinfo {author} {\bibfnamefont
  {T.}~\bibnamefont {Nakamura}}, \bibinfo {author} {\bibfnamefont {S.~Y.}\
  \bibnamefont {Park}}, \bibinfo {author} {\bibfnamefont {D.}~\bibnamefont
  {Rossi}}, \bibinfo {author} {\bibfnamefont {E.}~\bibnamefont {Sahin}},
  \bibinfo {author} {\bibfnamefont {D.}~\bibnamefont {Sohler}}, \bibinfo
  {author} {\bibfnamefont {P.-A.}\ \bibnamefont {S\"oderstr\"om}}, \bibinfo
  {author} {\bibfnamefont {S.}~\bibnamefont {Takeuchi}}, \bibinfo {author}
  {\bibfnamefont {H.}~\bibnamefont {T\"ornqvist}}, \bibinfo {author}
  {\bibfnamefont {V.}~\bibnamefont {Vaquero}}, \bibinfo {author} {\bibfnamefont
  {V.}~\bibnamefont {Wagner}}, \bibinfo {author} {\bibfnamefont
  {S.}~\bibnamefont {Wang}}, \bibinfo {author} {\bibfnamefont {V.}~\bibnamefont
  {Werner}}, \bibinfo {author} {\bibfnamefont {X.}~\bibnamefont {Xu}}, \bibinfo
  {author} {\bibfnamefont {H.}~\bibnamefont {Yamada}}, \bibinfo {author}
  {\bibfnamefont {D.}~\bibnamefont {Yan}}, \bibinfo {author} {\bibfnamefont
  {Z.}~\bibnamefont {Yang}}, \bibinfo {author} {\bibfnamefont {M.}~\bibnamefont
  {Yasuda}}, \ and\ \bibinfo {author} {\bibfnamefont {L.}~\bibnamefont
  {Zanetti}},\ }\href {\doibase 10.1103/PhysRevLett.122.072502} {\bibfield
  {journal} {\bibinfo  {journal} {Phys. Rev. Lett.}\ }\textbf {\bibinfo
  {volume} {122}},\ \bibinfo {pages} {072502} (\bibinfo {year}
  {2019})}\BibitemShut {NoStop}%
\bibitem [{\citenamefont {Chen}\ \emph {et~al.}(2019)\citenamefont {Chen},
  \citenamefont {Lee}, \citenamefont {Doornenbal}, \citenamefont {Obertelli},
  \citenamefont {Barbieri}, \citenamefont {Chazono}, \citenamefont
  {Navr\'atil}, \citenamefont {Ogata}, \citenamefont {Otsuka}, \citenamefont
  {Raimondi}, \citenamefont {Som\`a}, \citenamefont {Utsuno}, \citenamefont
  {Yoshida}, \citenamefont {Baba}, \citenamefont {Browne}, \citenamefont
  {Calvet}, \citenamefont {Ch\^ateau}, \citenamefont {Chiga}, \citenamefont
  {Corsi}, \citenamefont {Cort\'es}, \citenamefont {Delbart}, \citenamefont
  {Gheller}, \citenamefont {Giganon}, \citenamefont {Gillibert}, \citenamefont
  {Hilaire}, \citenamefont {Isobe}, \citenamefont {Kahlbow}, \citenamefont
  {Kobayashi}, \citenamefont {Kubota}, \citenamefont {Lapoux}, \citenamefont
  {Liu}, \citenamefont {Motobayashi}, \citenamefont {Murray}, \citenamefont
  {Otsu}, \citenamefont {Panin}, \citenamefont {Paul}, \citenamefont
  {Rodriguez}, \citenamefont {Sakurai}, \citenamefont {Sasano}, \citenamefont
  {Steppenbeck}, \citenamefont {Stuhl}, \citenamefont {Sun}, \citenamefont
  {Togano}, \citenamefont {Uesaka}, \citenamefont {Wimmer}, \citenamefont
  {Yoneda}, \citenamefont {Achouri}, \citenamefont {Aktas}, \citenamefont
  {Aumann}, \citenamefont {Chung}, \citenamefont {Flavigny}, \citenamefont
  {Franchoo}, \citenamefont {Ga\ifmmode \check{s}\else
  \v{s}\fi{}pari\ifmmode~\acute{c}\else \'{c}\fi{}}, \citenamefont {Gerst},
  \citenamefont {Gibelin}, \citenamefont {Hahn}, \citenamefont {Kim},
  \citenamefont {Koiwai}, \citenamefont {Kondo}, \citenamefont {Koseoglou},
  \citenamefont {Lehr}, \citenamefont {Linh}, \citenamefont {Lokotko},
  \citenamefont {MacCormick}, \citenamefont {Moschner}, \citenamefont
  {Nakamura}, \citenamefont {Park}, \citenamefont {Rossi}, \citenamefont
  {Sahin}, \citenamefont {Sohler}, \citenamefont {S\"oderstr\"om},
  \citenamefont {Takeuchi}, \citenamefont {T\"ornqvist}, \citenamefont
  {Vaquero}, \citenamefont {Wagner}, \citenamefont {Wang}, \citenamefont
  {Werner}, \citenamefont {Xu}, \citenamefont {Yamada}, \citenamefont {Yan},
  \citenamefont {Yang}, \citenamefont {Yasuda},\ and\ \citenamefont
  {Zanetti}}]{Chen19}%
  \BibitemOpen
  \bibfield  {author} {\bibinfo {author} {\bibfnamefont {S.}~\bibnamefont
  {Chen}}, \bibinfo {author} {\bibfnamefont {J.}~\bibnamefont {Lee}}, \bibinfo
  {author} {\bibfnamefont {P.}~\bibnamefont {Doornenbal}}, \bibinfo {author}
  {\bibfnamefont {A.}~\bibnamefont {Obertelli}}, \bibinfo {author}
  {\bibfnamefont {C.}~\bibnamefont {Barbieri}}, \bibinfo {author}
  {\bibfnamefont {Y.}~\bibnamefont {Chazono}}, \bibinfo {author} {\bibfnamefont
  {P.}~\bibnamefont {Navr\'atil}}, \bibinfo {author} {\bibfnamefont
  {K.}~\bibnamefont {Ogata}}, \bibinfo {author} {\bibfnamefont
  {T.}~\bibnamefont {Otsuka}}, \bibinfo {author} {\bibfnamefont
  {F.}~\bibnamefont {Raimondi}}, \bibinfo {author} {\bibfnamefont
  {V.}~\bibnamefont {Som\`a}}, \bibinfo {author} {\bibfnamefont
  {Y.}~\bibnamefont {Utsuno}}, \bibinfo {author} {\bibfnamefont
  {K.}~\bibnamefont {Yoshida}}, \bibinfo {author} {\bibfnamefont
  {H.}~\bibnamefont {Baba}}, \bibinfo {author} {\bibfnamefont {F.}~\bibnamefont
  {Browne}}, \bibinfo {author} {\bibfnamefont {D.}~\bibnamefont {Calvet}},
  \bibinfo {author} {\bibfnamefont {F.}~\bibnamefont {Ch\^ateau}}, \bibinfo
  {author} {\bibfnamefont {N.}~\bibnamefont {Chiga}}, \bibinfo {author}
  {\bibfnamefont {A.}~\bibnamefont {Corsi}}, \bibinfo {author} {\bibfnamefont
  {M.~L.}\ \bibnamefont {Cort\'es}}, \bibinfo {author} {\bibfnamefont
  {A.}~\bibnamefont {Delbart}}, \bibinfo {author} {\bibfnamefont {J.-M.}\
  \bibnamefont {Gheller}}, \bibinfo {author} {\bibfnamefont {A.}~\bibnamefont
  {Giganon}}, \bibinfo {author} {\bibfnamefont {A.}~\bibnamefont {Gillibert}},
  \bibinfo {author} {\bibfnamefont {C.}~\bibnamefont {Hilaire}}, \bibinfo
  {author} {\bibfnamefont {T.}~\bibnamefont {Isobe}}, \bibinfo {author}
  {\bibfnamefont {J.}~\bibnamefont {Kahlbow}}, \bibinfo {author} {\bibfnamefont
  {T.}~\bibnamefont {Kobayashi}}, \bibinfo {author} {\bibfnamefont
  {Y.}~\bibnamefont {Kubota}}, \bibinfo {author} {\bibfnamefont
  {V.}~\bibnamefont {Lapoux}}, \bibinfo {author} {\bibfnamefont {H.~N.}\
  \bibnamefont {Liu}}, \bibinfo {author} {\bibfnamefont {T.}~\bibnamefont
  {Motobayashi}}, \bibinfo {author} {\bibfnamefont {I.}~\bibnamefont {Murray}},
  \bibinfo {author} {\bibfnamefont {H.}~\bibnamefont {Otsu}}, \bibinfo {author}
  {\bibfnamefont {V.}~\bibnamefont {Panin}}, \bibinfo {author} {\bibfnamefont
  {N.}~\bibnamefont {Paul}}, \bibinfo {author} {\bibfnamefont {W.}~\bibnamefont
  {Rodriguez}}, \bibinfo {author} {\bibfnamefont {H.}~\bibnamefont {Sakurai}},
  \bibinfo {author} {\bibfnamefont {M.}~\bibnamefont {Sasano}}, \bibinfo
  {author} {\bibfnamefont {D.}~\bibnamefont {Steppenbeck}}, \bibinfo {author}
  {\bibfnamefont {L.}~\bibnamefont {Stuhl}}, \bibinfo {author} {\bibfnamefont
  {Y.~L.}\ \bibnamefont {Sun}}, \bibinfo {author} {\bibfnamefont
  {Y.}~\bibnamefont {Togano}}, \bibinfo {author} {\bibfnamefont
  {T.}~\bibnamefont {Uesaka}}, \bibinfo {author} {\bibfnamefont
  {K.}~\bibnamefont {Wimmer}}, \bibinfo {author} {\bibfnamefont
  {K.}~\bibnamefont {Yoneda}}, \bibinfo {author} {\bibfnamefont
  {N.}~\bibnamefont {Achouri}}, \bibinfo {author} {\bibfnamefont
  {O.}~\bibnamefont {Aktas}}, \bibinfo {author} {\bibfnamefont
  {T.}~\bibnamefont {Aumann}}, \bibinfo {author} {\bibfnamefont {L.~X.}\
  \bibnamefont {Chung}}, \bibinfo {author} {\bibfnamefont {F.}~\bibnamefont
  {Flavigny}}, \bibinfo {author} {\bibfnamefont {S.}~\bibnamefont {Franchoo}},
  \bibinfo {author} {\bibfnamefont {I.}~\bibnamefont {Ga\ifmmode \check{s}\else
  \v{s}\fi{}pari\ifmmode~\acute{c}\else \'{c}\fi{}}}, \bibinfo {author}
  {\bibfnamefont {R.-B.}\ \bibnamefont {Gerst}}, \bibinfo {author}
  {\bibfnamefont {J.}~\bibnamefont {Gibelin}}, \bibinfo {author} {\bibfnamefont
  {K.~I.}\ \bibnamefont {Hahn}}, \bibinfo {author} {\bibfnamefont
  {D.}~\bibnamefont {Kim}}, \bibinfo {author} {\bibfnamefont {T.}~\bibnamefont
  {Koiwai}}, \bibinfo {author} {\bibfnamefont {Y.}~\bibnamefont {Kondo}},
  \bibinfo {author} {\bibfnamefont {P.}~\bibnamefont {Koseoglou}}, \bibinfo
  {author} {\bibfnamefont {C.}~\bibnamefont {Lehr}}, \bibinfo {author}
  {\bibfnamefont {B.~D.}\ \bibnamefont {Linh}}, \bibinfo {author}
  {\bibfnamefont {T.}~\bibnamefont {Lokotko}}, \bibinfo {author} {\bibfnamefont
  {M.}~\bibnamefont {MacCormick}}, \bibinfo {author} {\bibfnamefont
  {K.}~\bibnamefont {Moschner}}, \bibinfo {author} {\bibfnamefont
  {T.}~\bibnamefont {Nakamura}}, \bibinfo {author} {\bibfnamefont {S.~Y.}\
  \bibnamefont {Park}}, \bibinfo {author} {\bibfnamefont {D.}~\bibnamefont
  {Rossi}}, \bibinfo {author} {\bibfnamefont {E.}~\bibnamefont {Sahin}},
  \bibinfo {author} {\bibfnamefont {D.}~\bibnamefont {Sohler}}, \bibinfo
  {author} {\bibfnamefont {P.-A.}\ \bibnamefont {S\"oderstr\"om}}, \bibinfo
  {author} {\bibfnamefont {S.}~\bibnamefont {Takeuchi}}, \bibinfo {author}
  {\bibfnamefont {H.}~\bibnamefont {T\"ornqvist}}, \bibinfo {author}
  {\bibfnamefont {V.}~\bibnamefont {Vaquero}}, \bibinfo {author} {\bibfnamefont
  {V.}~\bibnamefont {Wagner}}, \bibinfo {author} {\bibfnamefont
  {S.}~\bibnamefont {Wang}}, \bibinfo {author} {\bibfnamefont {V.}~\bibnamefont
  {Werner}}, \bibinfo {author} {\bibfnamefont {X.}~\bibnamefont {Xu}}, \bibinfo
  {author} {\bibfnamefont {H.}~\bibnamefont {Yamada}}, \bibinfo {author}
  {\bibfnamefont {D.}~\bibnamefont {Yan}}, \bibinfo {author} {\bibfnamefont
  {Z.}~\bibnamefont {Yang}}, \bibinfo {author} {\bibfnamefont {M.}~\bibnamefont
  {Yasuda}}, \ and\ \bibinfo {author} {\bibfnamefont {L.}~\bibnamefont
  {Zanetti}},\ }\href {\doibase 10.1103/PhysRevLett.123.142501} {\bibfield
  {journal} {\bibinfo  {journal} {Phys. Rev. Lett.}\ }\textbf {\bibinfo
  {volume} {123}},\ \bibinfo {pages} {142501} (\bibinfo {year}
  {2019})}\BibitemShut {NoStop}%
\bibitem [{\citenamefont {Bertozzi}\ \emph {et~al.}(1972)\citenamefont
  {Bertozzi}, \citenamefont {Friar}, \citenamefont {Heisenberg},\ and\
  \citenamefont {Negele}}]{Bertozzi72}%
  \BibitemOpen
  \bibfield  {author} {\bibinfo {author} {\bibfnamefont {W.}~\bibnamefont
  {Bertozzi}}, \bibinfo {author} {\bibfnamefont {J.}~\bibnamefont {Friar}},
  \bibinfo {author} {\bibfnamefont {J.}~\bibnamefont {Heisenberg}}, \ and\
  \bibinfo {author} {\bibfnamefont {J.}~\bibnamefont {Negele}},\ }\href
  {\doibase 10.1016/0370-2693(72)90662-4} {\bibfield  {journal} {\bibinfo
  {journal} {Phys. Lett. B}\ }\textbf {\bibinfo {volume} {41}},\ \bibinfo
  {pages} {408} (\bibinfo {year} {1972})}\BibitemShut {NoStop}%
\bibitem [{\citenamefont {Chandra}\ and\ \citenamefont
  {Sauer}(1976)}]{Chandra76}%
  \BibitemOpen
  \bibfield  {author} {\bibinfo {author} {\bibfnamefont {H.}~\bibnamefont
  {Chandra}}\ and\ \bibinfo {author} {\bibfnamefont {G.}~\bibnamefont
  {Sauer}},\ }\href {\doibase 10.1103/PhysRevC.13.245} {\bibfield  {journal}
  {\bibinfo  {journal} {Phys. Rev. C}\ }\textbf {\bibinfo {volume} {13}},\
  \bibinfo {pages} {245} (\bibinfo {year} {1976})}\BibitemShut {NoStop}%
\bibitem [{\citenamefont {Brown}\ \emph {et~al.}(1979)\citenamefont {Brown},
  \citenamefont {Massen},\ and\ \citenamefont {Hodgson}}]{Brown79}%
  \BibitemOpen
  \bibfield  {author} {\bibinfo {author} {\bibfnamefont {B.}~\bibnamefont
  {Brown}}, \bibinfo {author} {\bibfnamefont {S.}~\bibnamefont {Massen}}, \
  and\ \bibinfo {author} {\bibfnamefont {P.}~\bibnamefont {Hodgson}},\ }\href
  {\doibase 10.1016/0370-2693(79)90569-0} {\bibfield  {journal} {\bibinfo
  {journal} {Phys. Lett. B}\ }\textbf {\bibinfo {volume} {85}},\ \bibinfo
  {pages} {167} (\bibinfo {year} {1979})}\BibitemShut {NoStop}%
\bibitem [{\citenamefont {Negele}(1970)}]{Negele70}%
  \BibitemOpen
  \bibfield  {author} {\bibinfo {author} {\bibfnamefont {J.~W.}\ \bibnamefont
  {Negele}},\ }\href {\doibase 10.1103/PhysRevC.1.1260} {\bibfield  {journal}
  {\bibinfo  {journal} {Phys. Rev. C}\ }\textbf {\bibinfo {volume} {1}},\
  \bibinfo {pages} {1260} (\bibinfo {year} {1970})}\BibitemShut {NoStop}%
\bibitem [{\citenamefont {Rocco}\ and\ \citenamefont
  {Barbieri}(2018)}]{Rocco18}%
  \BibitemOpen
  \bibfield  {author} {\bibinfo {author} {\bibfnamefont {N.}~\bibnamefont
  {Rocco}}\ and\ \bibinfo {author} {\bibfnamefont {C.}~\bibnamefont
  {Barbieri}},\ }\href {\doibase 10.1103/PhysRevC.98.025501} {\bibfield
  {journal} {\bibinfo  {journal} {Phys. Rev. C}\ }\textbf {\bibinfo {volume}
  {98}},\ \bibinfo {pages} {025501} (\bibinfo {year} {2018})}\BibitemShut
  {NoStop}%
\bibitem [{\citenamefont {De~Vries}\ \emph {et~al.}(1987)\citenamefont
  {De~Vries}, \citenamefont {De~Jager},\ and\ \citenamefont
  {De~Vries}}]{deVries87}%
  \BibitemOpen
  \bibfield  {author} {\bibinfo {author} {\bibfnamefont {H.}~\bibnamefont
  {De~Vries}}, \bibinfo {author} {\bibfnamefont {C.~W.}\ \bibnamefont
  {De~Jager}}, \ and\ \bibinfo {author} {\bibfnamefont {C.}~\bibnamefont
  {De~Vries}},\ }\href {\doibase 10.1016/0092-640X(87)90013-1} {\bibfield
  {journal} {\bibinfo  {journal} {Atom. Data Nucl. Data Tabl.}\ }\textbf
  {\bibinfo {volume} {36}},\ \bibinfo {pages} {495} (\bibinfo {year}
  {1987})}\BibitemShut {NoStop}%
\bibitem [{\citenamefont {Porro}\ \emph {et~al.}(2021)\citenamefont {Porro}
  \emph {et~al.}}]{Porro21}%
  \BibitemOpen
  \bibfield  {author} {\bibinfo {author} {\bibfnamefont {A.}~\bibnamefont
  {Porro}} \emph {et~al.},\ }\href@noop {} {} (\bibinfo {year} {2021}),\
  \bibinfo {note} {unpublished}\BibitemShut {NoStop}%
\bibitem [{\citenamefont {Idini}\ \emph {et~al.}(2019)\citenamefont {Idini},
  \citenamefont {Barbieri},\ and\ \citenamefont {Navr\'atil}}]{Idini19}%
  \BibitemOpen
  \bibfield  {author} {\bibinfo {author} {\bibfnamefont {A.}~\bibnamefont
  {Idini}}, \bibinfo {author} {\bibfnamefont {C.}~\bibnamefont {Barbieri}}, \
  and\ \bibinfo {author} {\bibfnamefont {P.}~\bibnamefont {Navr\'atil}},\
  }\href {\doibase 10.1103/PhysRevLett.123.092501} {\bibfield  {journal}
  {\bibinfo  {journal} {Phys. Rev. Lett.}\ }\textbf {\bibinfo {volume} {123}},\
  \bibinfo {pages} {092501} (\bibinfo {year} {2019})}\BibitemShut {NoStop}%
\end{thebibliography}%

\end{document}